\documentclass[structabstract]{aa}
\usepackage{graphicx,natbib}
\usepackage{txfonts}
\def\id{{\rm d}}
\def\bcK{\mbox{\boldmath${\cal K}$}}
\def\cK{{\cal K}}
\def\bK{\mbox{\boldmath$K$}}
\def\bzero{\mbox{\boldmath$0$}}
\def\bcT{\mbox{\boldmath${\cal T}$}}
\def\cT{{\cal T}}
\def\br{\mbox{\boldmath$r$}}
\def\bx{\mbox{\boldmath$x$}}
\def\bk{\mbox{\boldmath$k$}}
\def\bv{\mbox{\boldmath$v$}}

\newcommand{\mps}{m\,s$^{-1}$}

\def\bvsim{\mbox{\boldmath$v$}}
\def\vsim{v}
\def\bvtarg{\mbox{\boldmath$v$}^{\rm tgt}}
\def\vtarg{v^{\rm tgt}}

\def\vakern{v^{\rm inv}}
\def\bvinv{\mbox{\boldmath$\tilde{v}$}^{\rm inv}}
\def\vinv{\tilde{v}^{\rm inv}}
\def\weights{{\boldmath w}}
\def\foral{\mbox{\rm\ for all\ }}

\begin{document}
   \title{Validated helioseismic inversions for 3-D vector flows}

    \author{M. \v{S}vanda
          \inst{1}\fnmsep\thanks{On leave from Astronomical Institute, Academy of Sciences of the Czech Republic, and Faculty of Mathematics and Physics, Charles University in Prague.}
          \and
          L. Gizon \inst{1,2} 
          \and
          S. M. Hanasoge \inst{1,3}
          \and 
          S. D. Ustyugov \inst{4}
          }
   \offprints{L. Gizon}

   \institute{Max-Planck-Institut f\"ur Sonnensystemforschung, Max-Planck-Stra{\ss}e 2, 37191                                               
              Katlenburg-Lindau, Germany\\                                                                                                  
              \email{[svanda,gizon,hanasoge]@mps.mpg.de}
         \and
             Institut f\"ur Astrophysik, Georg-August-Universit\"at G\"ottingen, 37077 G\"ottingen, Germany
         \and
             Department of Geosciences, Princeton University, Princeton, NJ-08544, USA 
         \and                                                                                                                               
             Keldysh Institute of Applied Mathematics, Russian Academy of Sciences, Miusskaya Sq.~4, RU-125047 Moscow, Russia}       
   \date{Received: 31 Dec 2010; accepted: 20 Apr 2011}

\abstract{According to time--distance helioseismology, information about internal fluid motions is encoded in the travel times of solar waves. The inverse problem consists of inferring three-dimensional vector flows from a set of travel-time measurements. While only few tests of the inversions have been done, it is known that the retrieval of the small-amplitude vertical flow velocities is problematic. A thorough study of biases and noise has not been carried out in realistic conditions.}
{Here we investigate the potential of time--distance helioseismology to infer three-dimensional convective velocities in the near-surface layers of the Sun. We developed a new Subtractive Optimally Localised Averaging (SOLA) code suitable for pipeline pseudo-automatic processing. Compared to its predecessor, the code was improved by accounting for additional constraints in order to get the right answer within a given noise level. The main aim of this study is to validate results obtained by our inversion code. }
{We simulate travel-time maps using a snapshot from a numerical simulation of solar convective flows, realistic Born travel-time sensitivity kernels, and a realistic model of travel-time noise. These synthetic travel times are inverted for flows and the results compared with the known input flow field. Additional constraints are implemented in the inversion: cross-talk minimization between flow components and spatial localization of inversion coefficients.
}
{Using modes $f$, $p_1$ through $p_4$, we show that horizontal convective flow velocities can be inferred without bias, at a signal-to-noise ratio greater than one in the top 3.5 Mm, provided that observations span at least four days. The vertical component of velocity ($v_z$), if it were to be weak, is more difficult to infer and is seriously affected by cross-talk from horizontal velocity components. We emphasise that this cross-talk must be explicitly minimised in order to retrieve $v_z$ in the top 1 Mm. We also show that statistical averaging over many different areas of the Sun allows for reliably measuring of average properties of all three flow components in the top 5.5~Mm of the convection zone.}
{}

   \maketitle
%

\section{Introduction}

The sub-surface of the Sun is optically thick, preventing us from directly observing the interior layers. Understanding the properties of the plasma in these regions has consequences for the theories of convection, stability of sunspots, the dynamics of stratified convection, and others. Most current knowledge about convection comes primarily from computational work \citep[e.g.,][]{2005AA...429..335V,2006ASPC..354...92B,2009ApJ...691..640R}. Helioseismic inversions of the sub-surface flows will play an important role in constraining these theories.  

A powerful way of imaging the solar interior is via inferences gathered from studying the statistics of the acoustic and surface gravity waves at the surface. Solar pressure and surface gravity modes are generated randomly by the vigorous turbulence in the upper convection zone. These oscillations are observed in the solar photosphere by measuring Doppler shifts of photospheric absorption lines. Forward modelling allows us to relate anomalies (like flows, thermal hot/cold spots etc.) to changes in helioseismic observables.  

The aim of helioseismic inversions is to reveal the structure of the subsurface flows (rotation, meridional circulation, convection), magnetic fields, and to measure deviations in the plasma state parameters (temperature, density, pressure) from a quiet Sun average. In this paper, we focus on travel times \citep{1993Natur.362..430D}, i.e., quantities that emerge from fits to cross correlations of observed signals. Time--distance helioseismology is used to measure and interpret changes in travel times of seismic waves caused by inhomogeneities in the structure of the Sun \citep[see review by][]{2010arXiv1001.0930G}. In recent years, time--distance helioseismology has been used to invert for near-surface flows \citep[e.g.,][]{2000JApA...21..339G,2000SoPh..192..177D,2004ApJ...603..776Z,2008SoPh..251..381J}, for flows beneath sunspots \citep[e.g.,][]{1996Natur.379..235D,2001ApJ...557..384Z,2006ApJ...640..516C,2008SoPh..251..291C,2009SSRv..144..249G,2009arXiv0912.4982M} and flows in their vicinity \citep{2000JApA...21..339G}, study the rotational gradient at the base of the convection zone \citep[e.g.,][]{2009ApJ...693.1678H}, etc. 

Helioseismic inversions are performed using two principal methods: The regularised least squares (RLS) and optimally localised averaging (OLA). The RLS method \citep[in time--distance helioseismology used for the first time by][]{1996ApJ...461L..55K} seeks to find the models of the solar interior, which provide the best least-squares fit to the measured travel-time maps, while regularising the solution (e.g., by requiring the smooth solution). The OLA method was developed for geoseismology \citep{1968GeoJ...16..169B,1970RSPTA.266..123B}. A form suitable for use in helioseismology was devised by \cite{1992AA...262L..33P}, who formulated the Subtractive-OLA method. SOLA is based on explicitly constructed spatially confined averaging kernels by taking linear combination of sensitivity kernels, while simultaneously keeping the error magnification small. The resulting coefficients are then used to linearly combine the travel-time maps and obtain the estimate for structure and magnitude of solar plasma perturbations. A SOLA-type inversion is the principal method discussed in the current paper. The SOLA has been used in time--distance local helioseismology in the past by \cite{2007AN....328..234J,2008SoPh..251..381J} who demonstrated the ability of SOLA inversions to reveal the structure of the 3-D internal flows. An efficient approach to solve fully consistent SOLA inversions was introduced by \cite{fastOLA}. In this paper we focus on inversions for three-dimensional vector flows on supergranular scales in the near-surface layers of the solar convection zone. 

\subsection{Validating helioseismic inversions}
Two approaches have been used so far to validate time--distance inversions. The first approach consists of generating synthetic travel-time maps by convolving a (known) arbitrary frozen flow field with travel-time sensitivity kernels and then testing the inversion method using these synthetic travel times \citep[e.g.,][]{1997ASSL..225..241K}. The second approach \citep{2007ApJ...659..848Z,2010AAS...21631905Z} is to use evolving realistic numerical simulations of three-dimensional radiative convection, where the helioseismic waves are naturally excited by the convection. 

The first approach is convenient, but may not represent a realistic situation, in particular because the prescribed subsurface structures are often too idealised. The second approach is preferable, but is limited by computing resources: only simulations of the very near-surface regions of the Sun are available today. Both approaches have shown that inversions are generally able to retrieve the horizontal components of velocity at supergranulation scales in the quiet Sun. However, some problems have been reported. For example, vertical velocities have been measured with the opposite sign near the surface \citep{2007ApJ...659..848Z} and the structure and sign of the flows around sunspots vary with the inversion method used \citep[see, e.g.,][]{2009SSRv..144..249G,2009arXiv0912.4982M}.

The aim of this paper is to validate a particular implementation of SOLA inversions for time--distance helioseismology using a mixed approach: we take a snapshot from a large-box realistic simulation of solar convection \citep{2008ASPC..383...43U} to generate realistic travel-time maps by convolution with Born travel-time sensitivity kernels. A realistic noise component is added to the travel times. The travel-time maps are then inverted using a multichannel SOLA inversion and compared with the known flows. This approach allows us to investigate various types of biases in the results of the inversion and to develop a robust procedure to minimise them. The most serious bias comes from the natural correlations among the components of flows induced by mass conservation that translate into a cross-talk between the components of the inverted flow (as we shall see, this is important when retrieving the vertical flow). Using realistic numerical simulations of convective flow velocities as input is very useful to set the acceptable level of random noise of the inverted flow velocities. The use of travel times with realistic noise properties is important in order to derive realistic estimates of the noise of the inverted flow velocities.

\section{Synthetic travel times}

To construct synthetic travel times, we use a realistic hydrodynamic simulation \citep{2008ASPC..383...43U} of the solar convection. The computational domain is a box 20~Mm in depth and 60~Mm in each horizontal direction. The simulation provides us with a reasonably realistic description of flows in the upper convection zone $\bvsim=(\vsim_x,\vsim_y,\vsim_z)$. Throughout this paper the spatial coordinates are defined as 
\begin{equation}
\bx=(\br, z)\ ,
\end{equation}
where $\br$ is the horizontal position vector and $z$ is the height.

We choose a snapshot ($t=500$~min, see Fig.~\ref{fig:simulation} in the on-line supplement) of the above mentioned simulation for our inversion tests. Following \cite{1997SoPh..170...63D}, we consider different types of travel-time measurements between a surface point at position $\br$ and a concentric annulus or quadrants in order to measure travel times sensitive to flows in inward-outward (denoted by ``oi''), west-east (``we''), and north-south (``ns'') directions. Travel-time maps are denoted by $\tau^a(\br)$, where the superscript $a$ is an integer that uniquely refers to a particular combination of choices in the data analysis: a type of geometry (oi, we, or ns), annulus radius (from 7.3~Mm to 29.2~Mm every 1.46~Mm), and a wave filter (here, ridge filters for one of $f$, $p_1$, $p_2$, $p_3$, or $p_4$ modes). Thus, index $a$ refers to one of $M=3 \times 16 \times 5=240$ possibilities. Additional information describing the measurement procedure in time--distance helioseismology is given by \cite{2005LRSP....2....6G}.

Travel-time maps (travel times as functions of position vector $\br$) are generated by convolving the convection snapshot with sensitivity kernels according to
\begin{equation}
\tau^a(\br) =  \int_{\odot} \bK^a(\br'-\br,z) \cdot \bv(\br',z)\; \id^2\br' \, \id z  + n^a(\br)\ ,
\label{eq:traveltimesdef}
\end{equation}
where $\bK^a = (K_x^a, K_y^a, K_z^a)$ is a vector travel-time sensitivity kernel (see, e.g., Fig.~\ref{fig:sensitivitykernel-ES} in the on-line supplement), $\bv$ is the velocity vector of convecting flows, and the volume integral is taken over the Sun. The noise component is denoted by $n^a$. The original horizontal size of the simulation box of 60~Mm was too small for our purpose. Since the simulated velocities are periodic in the horizontal directions, we copied the simulation box $10\times 10$ times.

Kernels are computed using the single-scattering Born approximation \citep{2007AN....328..228B} and depend on eigenmodes of a background 1-D standard solar model \citep{1996Sci...272.1286C}. The kernels are invariant under horizontal translations; horizontal averages of these kernels for various oscillation modes are displayed in Fig.~\ref{fig:1Dsensitivity} in the on-line supplement to display their sensitivity in depth. All kernels used in this study have sensitivities only in upper-most 10~Mm of the convection zone.

Solar waves are excited by the action of turbulent convection in the Sun and therefore travel times are inherently noisy. Following \cite{2004ApJ...614..472G}, we use a realistic noise covariance matrix (see Fig.~\ref{fig:covariancematrix-ES} in the on-line supplement),
\begin{equation}
\Lambda_{ab}(\br-\br') = E[ n^a(\br) n^b(\br')],
\end{equation}
to generate realizations of the noise. 

\begin{figure}[]
\centering
\includegraphics[width=0.49\textwidth]{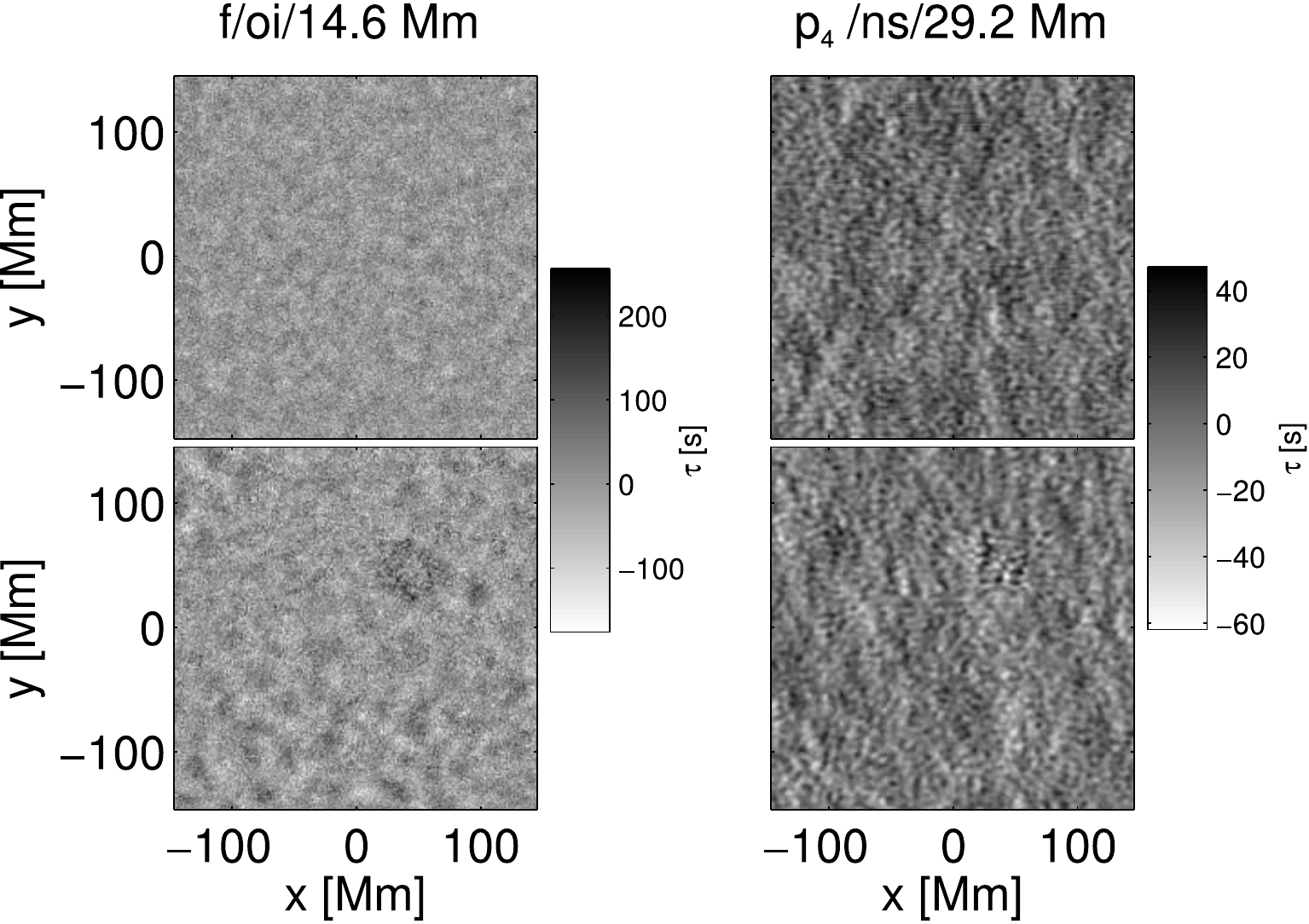}
\caption{Examples of synthetic travel-time maps used in this paper (top row) and observed travel times from SOHO/MDI (bottom row). Temporal length of observation is $T=6$~hours.}
\label{fig:traveltimes-maps}
\end{figure}

\begin{figure}[]
\centering
\includegraphics[width=0.40\textwidth]{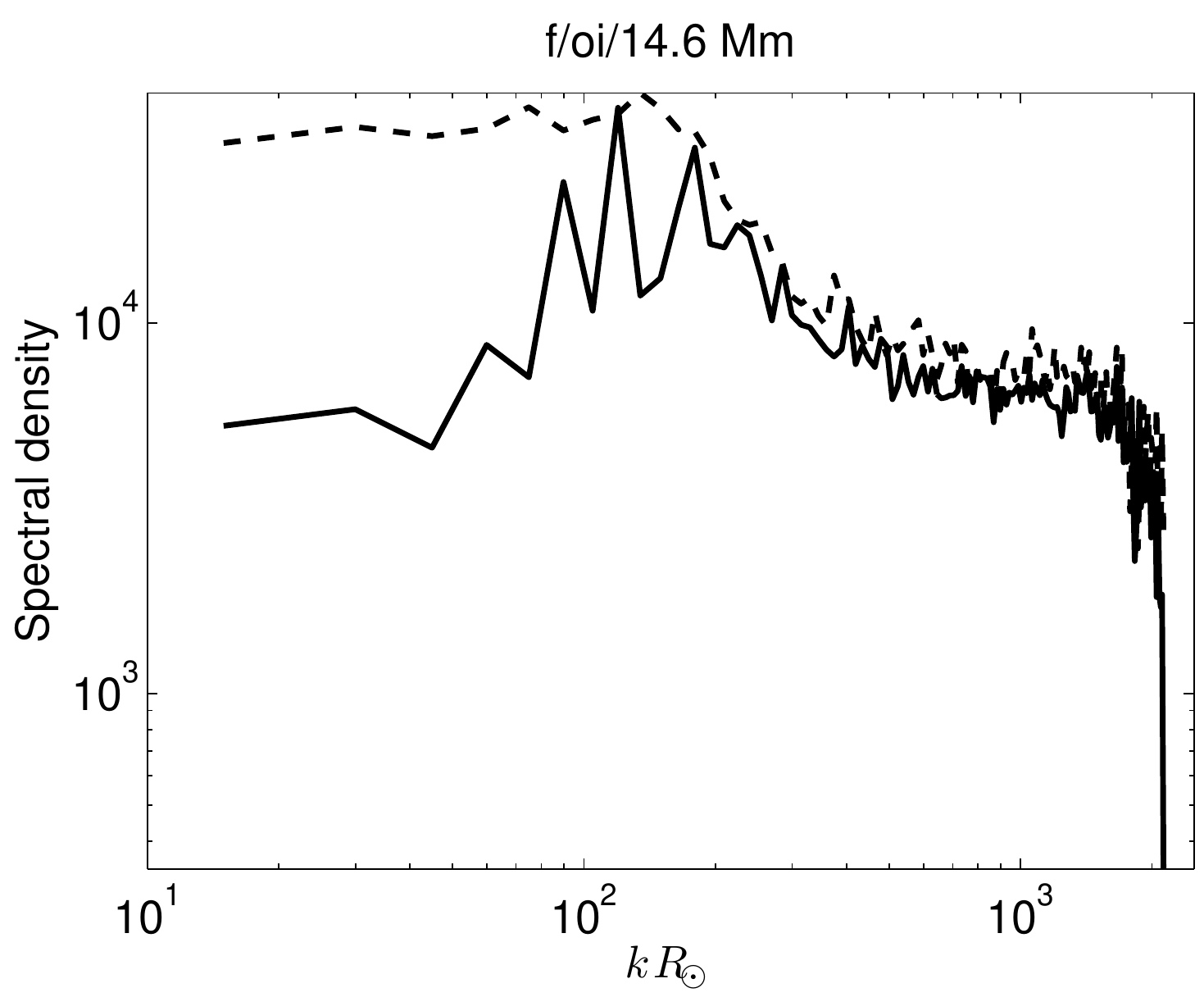}\\
\includegraphics[width=0.40\textwidth]{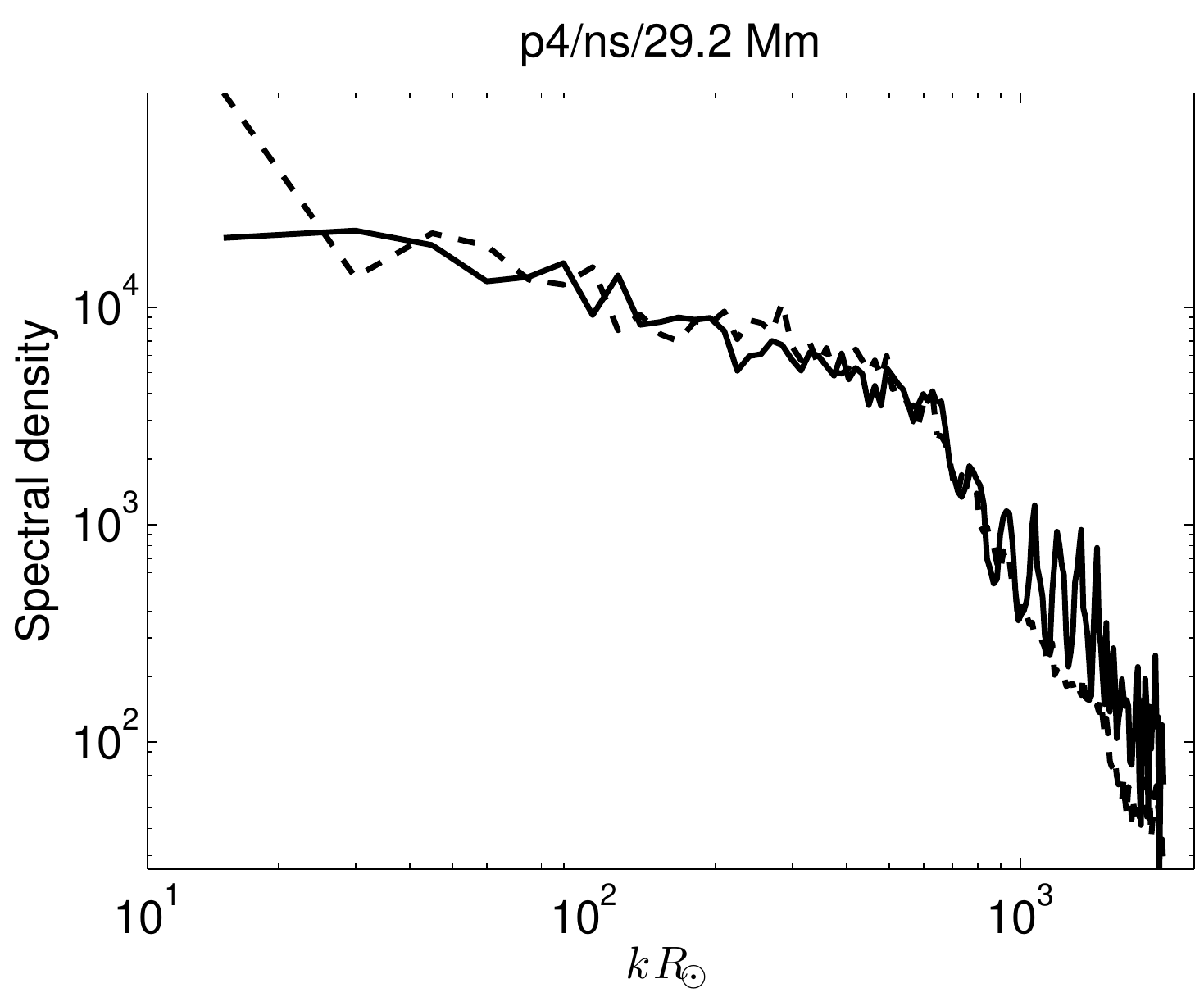}
\caption{Comparison of azimuthally averaged spatial power spectra $| \bar{\tau}^a(k) |^2$ of synthetic (solid) and observed (dashed) travel times displayed in Fig.~\ref{fig:traveltimes-maps} as a function of $k\,R_\odot$, where $k$ is the horizontal wave-number and $R_\odot$ is the radius of the Sun. }
\label{fig:traveltimes-ps}
\end{figure}

Each travel-time map has $400\times400$ pixels with spatial sampling of $1.46$~Mm, corresponding to the pixel size of a Michelson-Doppler-Imager full-disc image \citep[MDI;][]{1995SoPh..162..129S}. In total, we generate 240 different travel-time maps (one for each index $a$, see, e.g., Fig.~\ref{fig:traveltimes-maps}). These travel-time maps have spatial power spectra that are similar to observed travel-time power spectra based on MDI data, as demonstrated in Fig.~\ref{fig:traveltimes-ps}. However, our synthetic travel times have weaker power at low wave-numbers due to the complete lack of signal from the simulated convective velocities below $k\,R_\odot = 60$ (only simulated noise is present).

We check to ensure that point-to-annulus sensitivity kernels and noise covariance matrices obey expected symmetries. Furthermore, the sensitivity kernels and covariance matrices are forced to decay smoothly to zero towards the edge of the computation box by multiplying with a smooth spatial function having zeros far from the region of interest. In addition, we make sure that the horizontal integral of each sensitivity kernel for $v_z$ is zero at all depths. This is a consequence of symmetries associated with point-to-point kernels for $v_z$, which are insensitive to the mean value of vertical velocity \citep{2007AN....328..228B}.


 \section{Improved SOLA inversions}

\begin{table*}[]
\caption{Notations.}
\label{tab:notations}
\begin{tabular}{lp{9cm}l}
\hline
\hline
\rule{0pt}{3mm}Symbol & Meaning & Formula \\
\hline
\rule{0pt}{4mm}$\vsim_\alpha (\bx) $ & $\alpha$-component of flow velocity at $\bx$ (from the simulation) & \\
$\vtarg_\alpha(\bx_0)$ & $\alpha$-component of flow velocity at position $\bx_0$ targeted by the inversion  & $\int_\odot \cT(\br-\br_0, z; z_0) v_\alpha(\bx) \; \id^2\br\,\id z$ \\
${\vakern_\alpha{}^{(\beta)}(\bx_0)}$ & contribution of the $\beta$-component of flow velocity to the inverted $\alpha$-component of flow velocity & $ \int_\odot \cK^\alpha_\beta(\br-\br_0, z; z_0) v_\beta(\bx) \; \id^2\br\,\id z$\\ 
$\vinv_\alpha (\bx_0)$ & inverted $\alpha$-component of flow velocity (contains a noise component) & $\sum_{i, a} w^\alpha_a(\br_i-\br_0; z_0) \tau^a(\br_i) = \sum_{\beta} {\vakern_\alpha}{}^{(\beta)} + {\rm noise}$\\
& & \\
\hline
\end{tabular}
\end{table*}
\subsection{Real-space formulation}
The inversion method we use here is an improved version of the Subtractive Optimally Localised Averages (SOLA) method described by \citeauthor{2007AN....328..234J} (\citeyear{2007AN....328..234J}, \citeyear{2008SoPh..251..381J}, \citeyear{fastOLA}). 

The SOLA algorithm describes how to optimally combine a given set of travel time measurements to infer the underlying properties of medium. In this paper, we wish to retrieve $\vinv_\alpha(\bx_0)$, i.e., an estimate of the $\alpha$-component of the flow velocity in the neighbourhood of position $\bx_0 = (\br_0, z_0)$ in the solar interior. In practice, we search for a linear combination of the travel-time measurements,
\begin{equation}
\vinv_\alpha (\bx_0) = \sum_{i, a} w^\alpha_a(\br_i-\br_0; z_0) \tau^a(\br_i)\ ,
\label{eq:v_tilde}
\end{equation}
where $w^\alpha_a(\br_i)$ are inversion weights to be determined. Combining Eqs. (\ref{eq:traveltimesdef}) and (\ref{eq:v_tilde}), we have
\begin{eqnarray}
\vinv_\alpha (\bx_0) & = & \int_\odot \bcK^\alpha(\br-\br_0, z; z_0) \cdot \bv(\bx) \; \id^2\br\,\id z \nonumber \\
& & + \sum_{i, a} w^\alpha_a(\br_i-\br_0; z_0) n^a(\br_i) , 
\label{eq:tilde_v_vector}
\end{eqnarray}
where $\bcK^\alpha = (\cK_x^\alpha, \cK_y^\alpha, \cK_z^\alpha)$ is a vector averaging kernel with components
\begin{equation}
\cK_\beta^\alpha (\br,z; z_0) \equiv \sum_{i,a} w^\alpha_a(\br_i; z_0) K_\beta^a(\br-\br_i,z).
\label{eq:akerndefinition}
\end{equation}
The expression $\id^2\br\,\id z$ means $\id^3\bx$. 

The expectation value of inverted velocity $\vinv_\alpha (\bx_0)$ may be expanded as 
\begin{equation}
{\rm E}[\vinv_\alpha]= \vakern_\alpha + \sum\limits_{\beta \ne \alpha} {\vakern_\alpha}{}^{(\beta)}\ ,
\label{eq:v_expectation}
\end{equation}
where
\begin{equation}
{\vakern_\alpha}{}^{(\beta)} (\bx_0) \equiv \int_\odot \cK^\alpha_\beta(\br-\br_0, z; z_0) v_\beta(\bx) \; \id^2\br\,\id z\ ,
\label{eq:v_breve}
\end{equation}
and $\vakern_\alpha \equiv {\vakern_\alpha}{}^{(\alpha)}$. The form of equation (\ref{eq:v_expectation}) will be useful later in this paper to optimize the inversion. Notice that the second term on the right side of (\ref{eq:v_expectation}) represents leakage of other flow components into the inverted one. This \emph{cross-talk} may be a significant source of bias and needs to be studied. 

We search for inversion weights $\weights^\alpha$ so that the vector averaging kernel $\bcK^\alpha(\br-\br_0,z;z_0)$ resembles a user-supplied target $\bcT^\alpha(\br-\br_0,z;z_0)$. Because we wish to invert for the $\alpha$-component of the flow, we choose the target with components: 
\begin{equation}
\cT^\alpha_\beta(\br-\br_0,z;z_0)=\cT(\br-\br_0,z;z_0) \delta_{\alpha\beta}\ ,
\end{equation}
where $\cT(\br-\br_0,z;z_0)$ is a function that peaks around $\bx_0=(\br_0,z_0)$,  diminishes rapidly away from that point, and has unit integral. $\delta_{\alpha\beta}$ is a Kronecker $\delta$. Throughout this paper, we choose a simple Gaussian 
\begin{equation}
\cT(\br,z;z_0)= \frac{(4\ln2)^{3/2}}{\pi^{3/2} s^2_h s_z} \exp{\left[-\frac{4\ln{2}}{s_h^2}\|\br\|^2-\frac{4\ln2}{s_z^2}(z-z_0)^2\right]}\ ,
\end{equation}
where $s_h$ and $s_z$ are full widths at half maximum (FWHM) of the Gaussian in horizontal and vertical directions respectively. 

A successful inversion will return a value of $\vakern_\alpha(\bx_0)$ close to the target velocity 
\begin{equation}
\vtarg_\alpha (\bx_0) \equiv \int_\odot \cT(\br-\br_0, z; z_0) v_\alpha(\bx) \; \id^2\br\,\id z\ ,
\label{eq:v_hat}
\end{equation}
and cross-talk velocities ${\vakern_\alpha}{}^{(\beta)}(\bx_0)$ close to zero for $\beta \ne \alpha$. 

A list of the notations used for the various flow velocities referred to in the inversion procedure is given in Table~\ref{tab:notations}.

The problem to be solved to obtain the $w^\alpha$ belongs to the class of constrained regularised optimisations. The terms to be regularised are the following:
\begin{itemize}
\item The \emph{misfit}, i.e., how far the averaging kernel is from the desired target function,
\begin{equation}
\chi^2(\weights^\alpha; z_0)=\int_\odot [\cK_\alpha^\alpha(\br,z; z_0) - \cT(\br,z; z_0)]^2 \; \id^2\br\,\id z\ ,
\label{eq:misfit}
\end{equation}
where $\cK_\alpha^\alpha$ is an implicit function of the weights $\weights^\alpha$.
\item The \emph{cross-talk} quantifying the leakage of the signal from other flow components into the inverted one,
\begin{equation}
XT(\weights^\alpha; z_0)=\sum\limits_{\beta \ne \alpha} \int_\odot [\cK_\beta^\alpha(\br,z; z_0)]^2 \; \id^2\br\,\id z\ .
\label{eq:crosstalk}
\end{equation}
\item The variance of $\vinv_\alpha$, which corresponds to the root-mean-square of noise-related fluctuations in inferred velocities. We refer to this quantity as \emph{predicted error} hereafter,
\begin{equation}
\sigma^2_\alpha(\weights^\alpha; z_0)=\sum\limits_{i,j,a,b} w^\alpha_a(\br_i;z_0) \Lambda_{ab}(\br_i-\br_j) w^\alpha_b(\br_j;z_0)\ .
\label{eq:noise}
\end{equation}
\item The ad-hoc term quantifying the \emph{spread of the weights in space} around $\br_0$,
\begin{equation}
S(\weights^\alpha; z_0)=\sum\limits_{a,i} [w^\alpha_a(\br_i; z_0)]^2\ .
\label{eq:confinement}
\end{equation}
The regularisation based on $S$ ensures that weights decrease towards the edge of the horizontal domain in order to enforce the spatial locality of the inversion and to prevent the weights from oscillating in the spatial domain.
\end{itemize}

\noindent In practice, we search for the weights $\weights^\alpha$ that minimise the cost function
\begin{eqnarray}
F^\alpha(\weights^\alpha) & = &\int_\odot [\cK_\alpha^\alpha(\br,z; z_0) - \cT(\br,z; z_0)]^2 \; \id^2\br\,\id z + \nonumber \\
&+& \nu \,\sum\limits_{\beta \ne \alpha} \int_\odot [\cK_\beta^\alpha(\br,z; z_0)]^2 \; \id^2\br\,\id z \nonumber + \epsilon \sum\limits_{a,i} [w^\alpha_a(\br_i; z_0)]^2 + \\
&+& \mu \sum\limits_{i,j,a,b} w^\alpha_a(\br_i;z_0) \Lambda_{ab}(\br_i-\br_j) w^\alpha_b(\br_j;z_0) .
\label{eq:costfunction}
\end{eqnarray}
Trade-off parameters $\mu$, $\nu$, and $\epsilon$ balance these terms. The strategy to be employed in order to set these values will be described in Section~\ref{sect:tradeoffs}.

The inversion is subject to constraints 
\begin{equation}
\int_\odot \id^2\br\,\id z \, \cK^\alpha_\beta(\br,z;z_0) = \delta_{\alpha \beta}\ \foral \beta\ , 
\label{eq:constraint}
\end{equation}
in order to scale the amplitude of the inverted flow $\bvinv$ appropriately. 

By taking derivative of (\ref{eq:costfunction}) with constraint (\ref{eq:constraint}) added with respect to the weights, the problem can be cast into a linear inverse problem, as explained by \cite{1992AA...262L..33P} or \cite{fastOLA}. In real space, the matrix to be inverted has $(N^2M+P)^2 \simeq (10^7)^2$ elements, where $N=200$ is the number of grid points in one horizontal direction, $P=3$ is the number of physical unknowns (three velocity components), and $M=240$ as already defined earlier.

\subsection{Fourier-space (Multichannel) formulation}
The full problem written in real space is intractable to be solved using nowadays computers, because the matrix to be inverted is too large. \cite{fastOLA} found a solution to this problem by transforming to spatial Fourier space, where the inverse problem decouples as a consequence of the horizontal translation invariance of the sensitivity kernels. Thus, instead of a big linear inverse problem (inverting matrix having $(10^7)^2$ elements), we solve $40\,000$ small linear inverse problems (inverting matrices having $240^2$ elements) in wave-vector space. This approach is called a multichannel inversion \citep{1998ESASP.418..635J}. Here we summarise results of \cite{fastOLA} for the sake of completeness. 

Following \cite{fastOLA}, we use the following definition of the Fourier transform, such that any function $f(\br)$ and its 2-D spatial Fourier transform $\bar{f}(\bk)$ are related according to
\begin{eqnarray}
f(\br) & = & h_k^2 \sum\limits_{\bk} \bar{f}(\bk) \exp{\left( i \bk \cdot \br \right)}\ , \\
\bar{f}(\bk) & = & \frac{h_x^2}{(2\pi)^2} \sum\limits_{\br} f(\br) \exp{\left(-i \bk \cdot \br \right)}\ ,
\end{eqnarray}
where $h_x=1.46$~Mm and $h_k=0.022$~rad/Mm are the grid spacings in the real and Fourier domains respectively. 

For each non-vanishing wave-vector $\bk=(k_x,k_y)$, the vector of weights $W(\bk)=[\bar{w}^\alpha_1(\bk) \;   \bar{w}^\alpha_2(\bk) \;\dots\;  \bar{w}^\alpha_M(\bk)]^{\rm T}$ is the solution to the matrix equation,
\begin{equation}
{h_k^2 N^2\,A(\bk)  W(\bk) }= T(\bk)  \quad  {\rm for\ } \bk \neq \bzero\ .
\label{eq:problem-nonzerok}
\end{equation}
Each matrix $A(\bk)$ has $M\times M$ elements $A_{ab}$ given by
\begin{eqnarray}
A_{ab}(\bk) &=& (2\pi)^2 \int\limits_{-\infty}^{+\infty} dz {\bar{K}_\alpha^{a *}}(\bk,z) \bar{K}^b_\alpha (\bk,z) + \nonumber \\
& + & (2\pi)^2 \nu \sum\limits_{\beta \ne \alpha} \int\limits_{-\infty}^{+\infty} dz {\bar{K}_\beta^{a *}}(\bk,z) \bar{K}^b_\beta (\bk,z) + \nonumber \\
&+& \mu \bar\Lambda_{ab}(\bk) + \epsilon \delta_{ab}\ . \label{eq:a_ca}
\end{eqnarray}
The vector $T(\bk)=[\bar{t}_1(\bk) \;   \bar{t}_2(\bk) \;\dots\;  \bar{t}_M(\bk)]^{\rm T}$ has $M$ elements $\bar{t}_a$ given by 
\begin{equation}
\bar{t}_a (\bk; z_0) = (2\pi)^2 \int\limits_{-\infty}^{+\infty} \id z \, {\bar{K}^{a *}_\alpha}(\bk,z) \bar{\cT}(\bk,z; z_0) \ . \label{eq:t_c}
\end{equation}

In addition, the matrix  equation for the case $\bk=\bzero$ is
\begin{equation}
{\left[ \begin{array}{cc} 
h_k^4 N^2 A(\bzero)  &    C  \\ 
 C^{\rm T} &  0 
\end{array} \right]
\left[ \begin{array}{c}
W(\bzero)  \\ 
L
\end{array} \right]}
=
{\left[ \begin{array}{c}
h_k^2\, T(\bzero)  \\ 
U^\alpha/(h_k^2 N^2)
\end{array} \right]\,,}
\label{eq:problem-zerok}
\end{equation}
where $C$ is an $M\times P$ matrix with elements
\begin{equation}
C_{a \beta} = \int_\odot \id^2 \br\, \id z \, K^a_\beta (\br,z)\ ,\label{eq:c_alpha}
\end{equation}
$L$ is a $1\times P$ vector of Lagrange multipliers, and the vector $U^\alpha$  is a $1 \times P$ unit vector with components $u^\alpha_\beta=\delta_{\alpha\beta}$. 

Solutions to the $N-1$ equations (\ref{eq:problem-nonzerok}) and the equation (\ref{eq:problem-zerok}) give Fourier components of the weights, $\bar{w}^\alpha_a(\bk)$. By taking an inverse Fourier transform we obtain $w^\alpha_a(\br)$. Equation (\ref{eq:v_tilde}) then gives an estimate of the $\alpha$-component of the flow, $\vinv_\alpha$. 

The difference between the above equations and those in \cite{fastOLA} arises from additional constraint terms~(\ref{eq:crosstalk}) and~(\ref{eq:confinement}).

\subsection{Picking trade-off parameters}
\label{sect:tradeoffs}
Trade-off parameters $\mu$, $\nu$, and $\epsilon$ control the balance between various terms in the cost-function~(\ref{eq:costfunction}).

In practice, during the inversion for each flow component, we compute a grid of solutions by varying all three trade-off parameters. For each solution on this grid, we compute the misfit (\ref{eq:misfit}), predicted error of results (\ref{eq:noise}), amount of cross-talk (\ref{eq:crosstalk}), and the spatial power of  weights (\ref{eq:confinement}). All four quantities can be computed in spatial Fourier domain.

Standard optimisation methods to achieve optimality, such as L-curve analysis \citep[see, e.g.,][]{1998Hansen, 2008SoPh..251..381J}, are not particularly useful to our problem. The elbow of the L-curve, which is considered an optimal point in parameter space, was located where the predicted noise level was very large. We therefore developed our own strategy of selecting values for the trade-off parameters.

We start with parameter $\mu$, which controls the trade-off between the misfit and random error. We select $\mu$ such that the noise level of the inverted quantity is less than some target value, chosen by the user, depending on the problem at hand. For example, one may choose a target noise level of 20~m/s to invert for horizontal flows in supergranules. A much lower noise level will be required to invert for the vertical component of the flow \citep{2010ApJ...725L..47D}. The example noise levels $\sigma_\alpha$ are given in Table~\ref{tab:errors}, where we give the predicted inversion error for $\vinv_x$ and $\vinv_z$ and the root-mean-square of the flow averaged with the target function ($\vtarg_x$ and $\vtarg_z$) and averaged with the resulting averaging kernel ($\vakern_x$ and $\vakern_z$). It may be that this selection of $\mu$ leads to an averaging kernel which does not resemble the desired target function at all, in which case one would have to allow for a coarser spatial resolution. 

\begin{table}[!t]
\caption{Predicted inversion errors and the expected magnitude of the velocities at the three different depths discussed throughout this paper.}
\vspace{5mm}
\centering
\label{tab:errors}
\begin{tabular}{l|ccc}
\hline\hline
\rule{0pt}{10pt}Depth [Mm] & 1 & 3.5 & 5.5 \\ 
\hline
\rule{0pt}{5mm}FWHM $s_h$ [Mm] & 15 & 15 & 15 \\
FWHM $s_z$ [Mm] & 1.1 & 2.2 & 3.5 \\
\hline
\multicolumn{4}{c}{\rule{0pt}{5mm}$T=4$~days} \\
\hline
\rule{0pt}{10pt}Noise $\sigma_x$ [m\,s$^{-1}$] & 14 & 20 & 28 \\
Inverted signal $\langle {\vakern_x}^2 \rangle_h^{1/2}$ [m\,s$^{-1}$]& 35 & 20 & 13 \\
Targeted signal $\langle {\vtarg_x}^2 \rangle_h^{1/2}$ [m\,s$^{-1}$]& 40 & 25 & 20 \\
\rule{0pt}{10pt}Noise $\sigma_z$ [m\,s$^{-1}$] & 3 & 13 & 133 \\
Inverted signal $\langle {\vakern_z}^2 \rangle_h^{1/2}$ [m\,s$^{-1}$]& 5 & 5 & 5 \\
Targeted signal $\langle {\vtarg_z}^2 \rangle_h^{1/2}$ [m\,s$^{-1}$]& 5 & 7 & 7 \\
\hline
\multicolumn{4}{c}{\rule{0pt}{5mm}10$^4$$~\times~T=6$~hours} \\
\hline
\rule{0pt}{10pt}Noise $\sigma_x$ [m\,s$^{-1}$] & 7 & 9 & 7 \\
Inverted signal $\langle {\vakern_x}^2 \rangle_h^{1/2}$ [m\,s$^{-1}$]& 41 & 25 & 19 \\
Targeted signal $\langle {\vtarg_x}^2 \rangle_h^{1/2}$ [m\,s$^{-1}$]& 40 & 25 & 20 \\
\rule{0pt}{10pt}Noise $\sigma_z$ [m\,s$^{-1}$] & 2 & 2 & 1 \\
Inverted signal $\langle {\vakern_z}^2 \rangle_h^{1/2}$ [m\,s$^{-1}$] & 5 & 6 & 4 \\
Targeted signal $\langle {\vtarg_z}^2 \rangle_h^{1/2}$ [m\,s$^{-1}$]& 5 & 7 & 7 \\
\hline
\end{tabular}
\tablefoot{Presented results are for inversions with cross-talk minimised and for two discussed cases: (1) inversion using travel-time maps averaged over few days and (2) averaging over many realisations of similar flow structure, each averaged over a short time (here we assume 10$^4$ realisations, each averaged over 6~hours).}
\end{table}

For the given $\mu$, we then choose $\nu$ so that the degree of cross-talk is less than 10$^{-5}$~Mm$^{-3}$. We observe that this constraint places the following upper bound on the magnitude of the cross-talk terms:
\begin{equation}
\frac{{\rm max} \left|\sum\limits_{\beta \ne \alpha} \cK^\alpha_\beta(\br, z; z_0)\right|}{{\rm max\ }\left|\cK^\alpha_\alpha(\br, z; z_0)\right|} < 0.05\ .
\end{equation}

The parameter $\epsilon$ controls the degree of spatial confinement of the inversion weights and has an impact on the misfit and error as well. If $\epsilon$ is too large, the weights will be highly localised in space around the central point, but a large misfit will result.

\begin{figure*}[!t]
\centering
\includegraphics[width=0.85\textwidth]{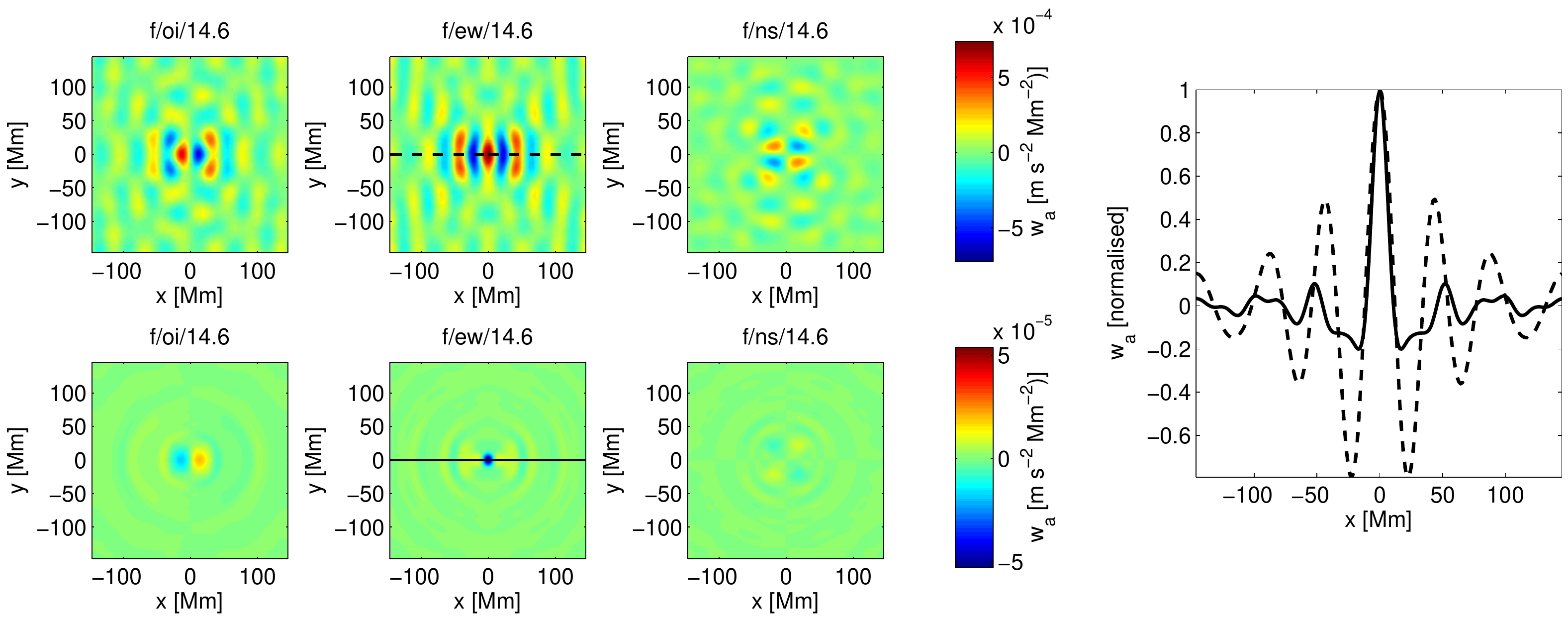}
\caption{Example inversion weights $w_a^x(x,y)$ for 1~Mm depth and effect of the minimisation of the spread $S$. A strong regularisation $\epsilon$ confines the inversion weights in the spatial domain (bottom row, $\epsilon=10^{0.5}$) compared to the case $\epsilon=0$ (top row). The cut at $y=0$ (horizontal black line) is displayed on the right-hand-side panel, where the values of $w^x_a(x,y=0)$ were scaled by $w^x_a(x=0,y=0)$.}
\label{fig:weights-vx}
\end{figure*}

For the particular cases that we study here, we find that localisation of weights is accomplished when $S<2\times10^{-3}~{\rm km}^2{\rm s}^{-2}$.  The impact of this regularisation term is illustrated in Fig.~\ref{fig:weights-vx}. For the case shown, weights are confined to within a disc of radius $\sim$50~Mm.

\subsection{The code}
The code implementing the above described procedure is written in {\sc Matlab}. {\sc Matlab} provides a compromise between computational efficiency and the availability of higher-level software constructs that make the code lucid, modular, and easy to modify. {\sc Matlab}-based code may be compiled into binary-executable form suitable for pipeline pseudo-automatic processing. The input is a text file, in which the user specifies all relevant parameters including the kernels to be used in the inversion and the set of trade-off parameters to be investigated. This allows for the development of user interfaces such as the web-based or graphical interface, which can serve as the user front-end of the inversion code. 

The code is parallel and scales linearly with number of parallel jobs. The execution is fast, e.g., a set of inversions for one flow component involving 240 kernels, 200$\times$200 spatial points, and a grid of 200 trade-off parameter values takes around 6 hours using 48 Opteron-2.3GHz CPUs. When many CPUs are involved, the extensive input-output load becomes a bottle-neck and affects the total processing time.

\section{Inversion for horizontal flow}
\label{sect:vx}
We applied the above-described method to known synthetic travel times in order to validate the set-up and the performance of the inversion procedure. We focus first on the horizontal flow components, i.e., $\alpha=x$ or $\alpha=y$. We compare two principal quantities: (1) the flow map that is obtained by convolving the known velocity field with the target function ($\vtarg_\alpha$), i.e., the best-case inversion scenario, and (2) the flow map actually resulting from the inversion ($\vinv_\alpha$). In this manner, we may investigate in detail different sources of bias in the results. 

The simulation convolved with the target function gives us $\vtarg_\alpha$, from which we estimate the expected magnitude of the flow that we want to invert for (see Table~\ref{tab:errors}). This places limits on the required noise level of the inversion so that the results have signal-to-noise ratios larger than 1. Requirements on the targeted error level fix the trade-off parameter $\mu$.

The choice of the target function depends on the discretion and needs of the user. Here we focus on layers in the top few Mm, in particular on depths $-z_0=1$, $3.5$, and $5.5$~Mm. This set-up was selected because similar flow inversions were also performed and discussed by \cite{2008SoPh..251..381J}, which makes it possible to compare the results of both methods. 
As discussed in the preceding sections, the outcome of the inversion is a set of weights $w^\alpha_a(\br)$, which are used to combine the travel-time maps in order to obtain estimates of velocity $\vinv_x(\br)$ and $\vinv_y(\br)$. Example weights for $\vinv_x$ are displayed in Fig.~\ref{fig:weights-vx}. 

All components of the averaging kernels $\cK^x_\beta(\br,z;z_0)$ are shown in Figs~\ref{fig:akerns-vx-1Mm-ES}--\ref{fig:akerns-vx-5.5Mm-ES} (available in the electronic supplement). We also show comparisons between inversions when the cross-talk is minimised and not minimised. The minimisation of the cross-talk with $\nu=100$ is very efficient. However, for depths of $3.5$ and $5.5$~Mm the minimisation of the cross-talk introduces some small artefacts in the inversion averaging kernel $\cK^x_x$, which are a small price to pay. 

\begin{figure}[!b]
\centering
\includegraphics[width=6cm]{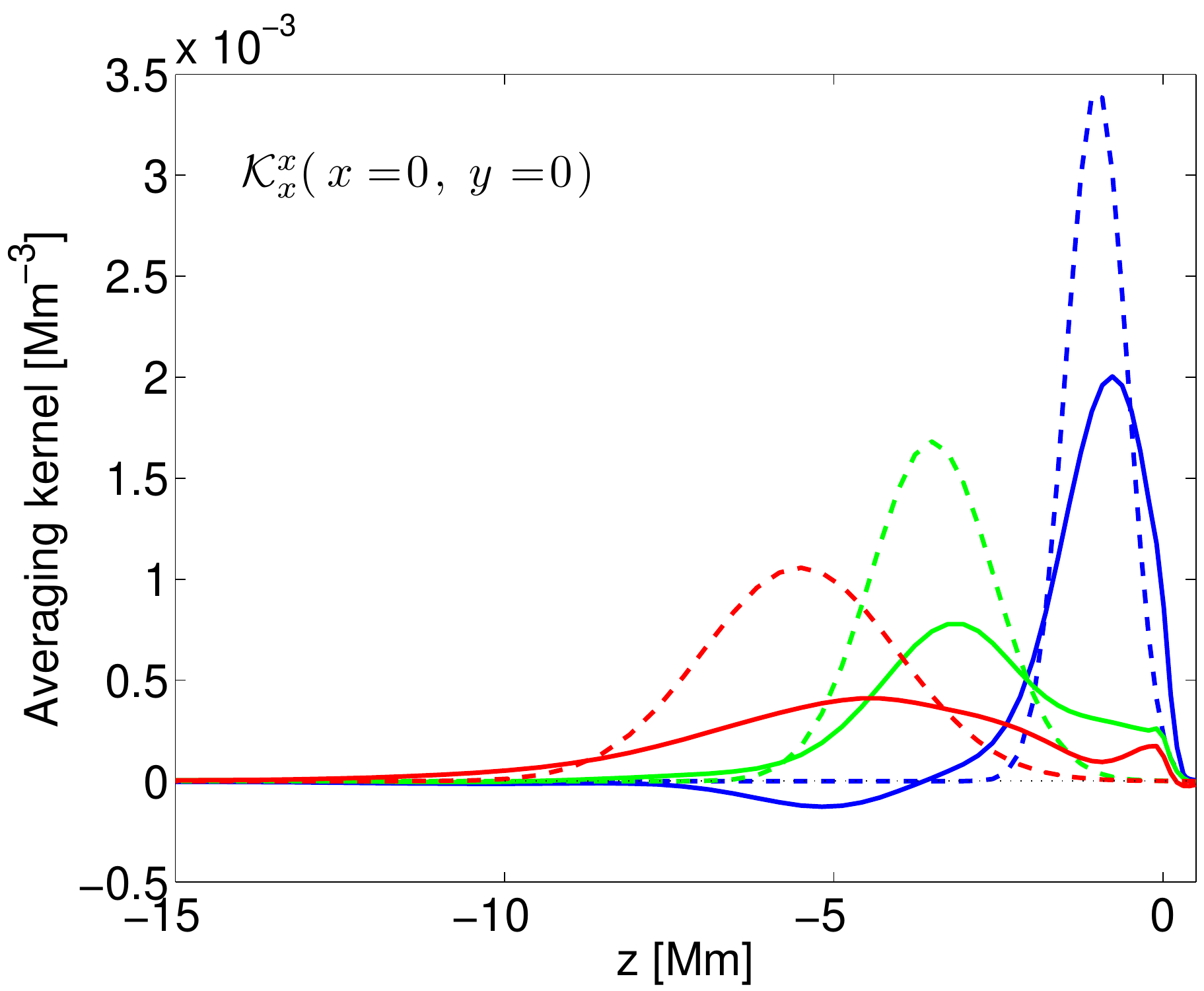}
\caption{The cut through the $x=y=0$ point of the averaging kernel (solid) and the respective target function (dashed) for $v_x$ inversion with minimised cross-talk, using travel times averaged over 4~days, at three discussed depths (1, 3{.}5, and 5{.}5~Mm).}
\label{fig:akerns-vx-1D}
\end{figure}

Fig.~\ref{fig:akerns-vx-1D} shows vertical cuts through the target functions and the averaging kernels when the cross-talk is minimised. Except for the target depth of $5.5$~Mm, the depth dependences of the averaging kernels $\cK^x_x$  resemble that of the target functions. The near-surface inversion at 1~Mm is dominated by the $f$-mode. The inversion at $3.5$~Mm has equal contributions from $f$ and $p_1$, with opposite signs, as shown in Fig.~\ref{fig:akern-modes-contributions} in the on-line supplement, where we plot the contribution of individual modes to the averaging kernels $\cK_x^x$ as a function of depth. 

\begin{figure*}[!t]
\centering
\sidecaption
\includegraphics[width=12cm]{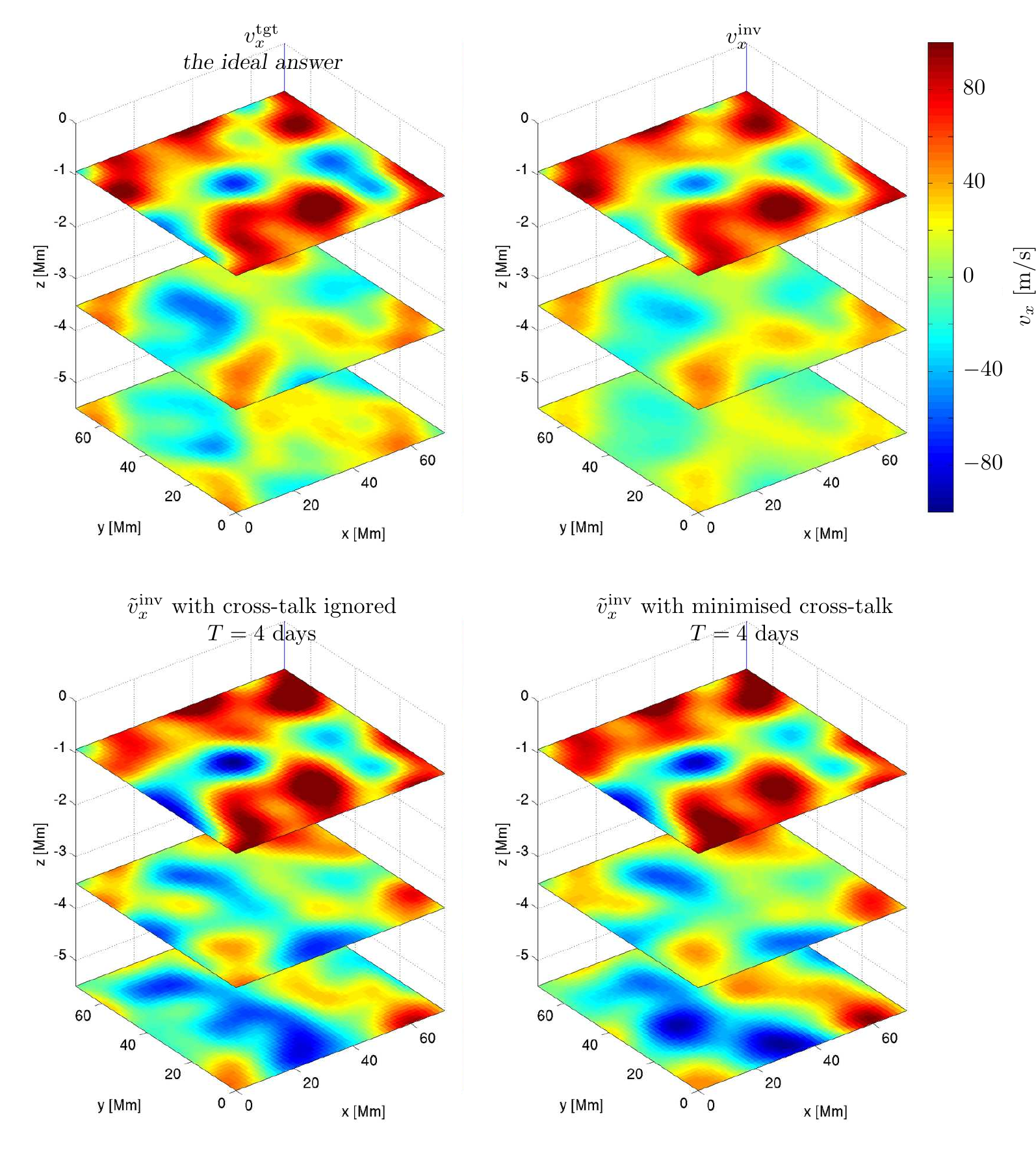}
\caption{Comparison of inverted $v_x$ with input data. Top-row panels show the input flow field convolved with the target function and the averaging kernel respectively. Bottom rows show inversion results in cases when cross-talk is ignored (left) and minimised (right). Random errors of the inversion are given in Table~\ref{tab:errors}.}
\label{fig:vx-validation}
\end{figure*}

The validation of the $v_x$ inversion is demonstrated in Fig.~\ref{fig:vx-validation}. Here we plot the desired $\vtarg_x$ at three different depths and the inverted $\vakern_x$ without noise contributions. These two are very close for the depths $1$ and $3.5$~Mm. The differences between $\vtarg_x$ and $\vakern_x$ at depth $5.5$~Mm are caused by an imperfect averaging kernel. When the random noise is added to the solution (bottom row of Fig.~\ref{fig:vx-validation}), we see that the inversions for $v_x$ at 1~Mm and $3.5$~Mm are still very good, while the inverted $v_x$ at depth $5.5$~Mm is dominated by noise. Minimising cross-talk does not improve the quality of the solution in this case.

\begin{figure*}[!t]
\sidecaption
\mbox{\includegraphics[width=6cm]{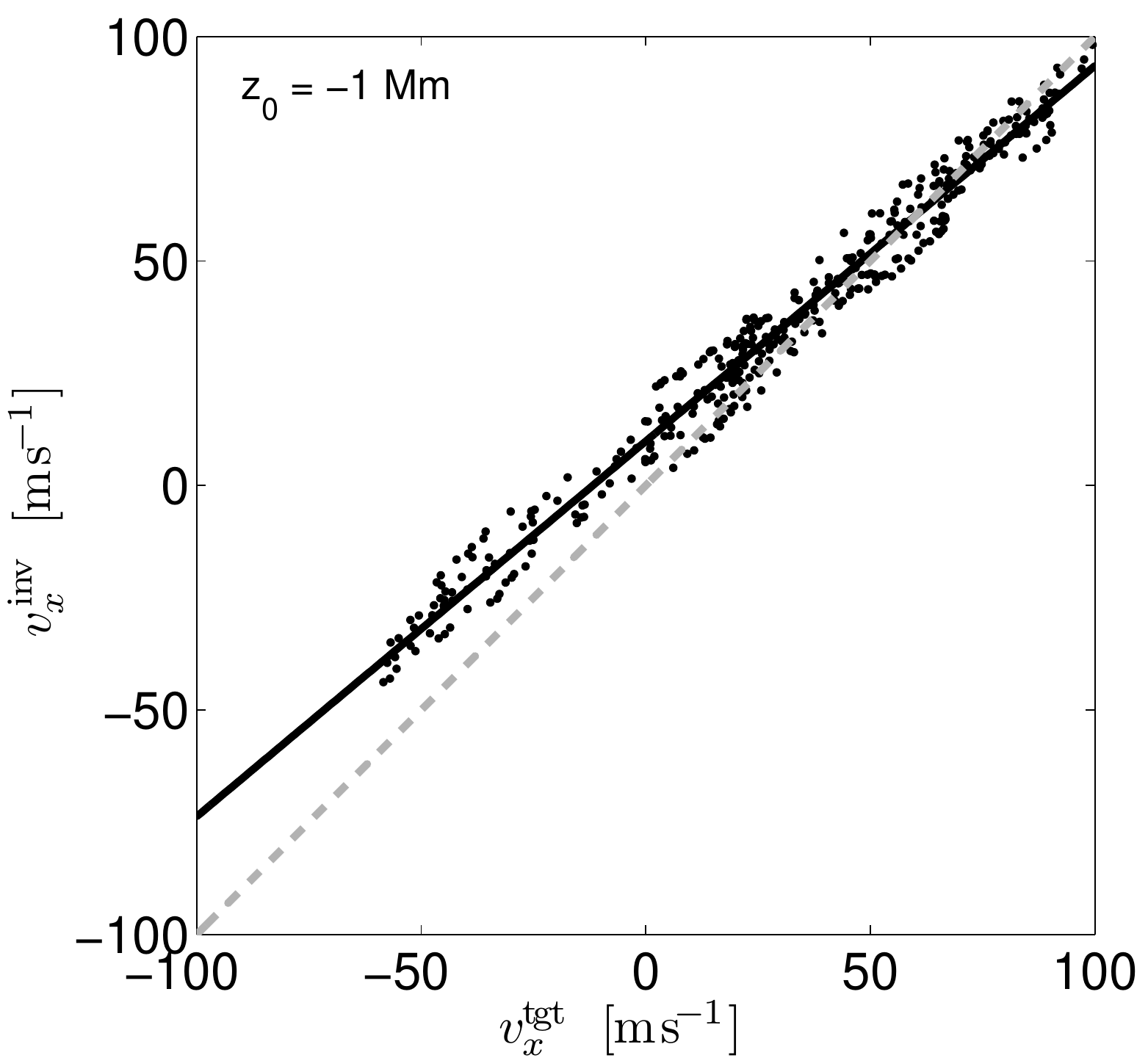}\includegraphics[width=6cm]{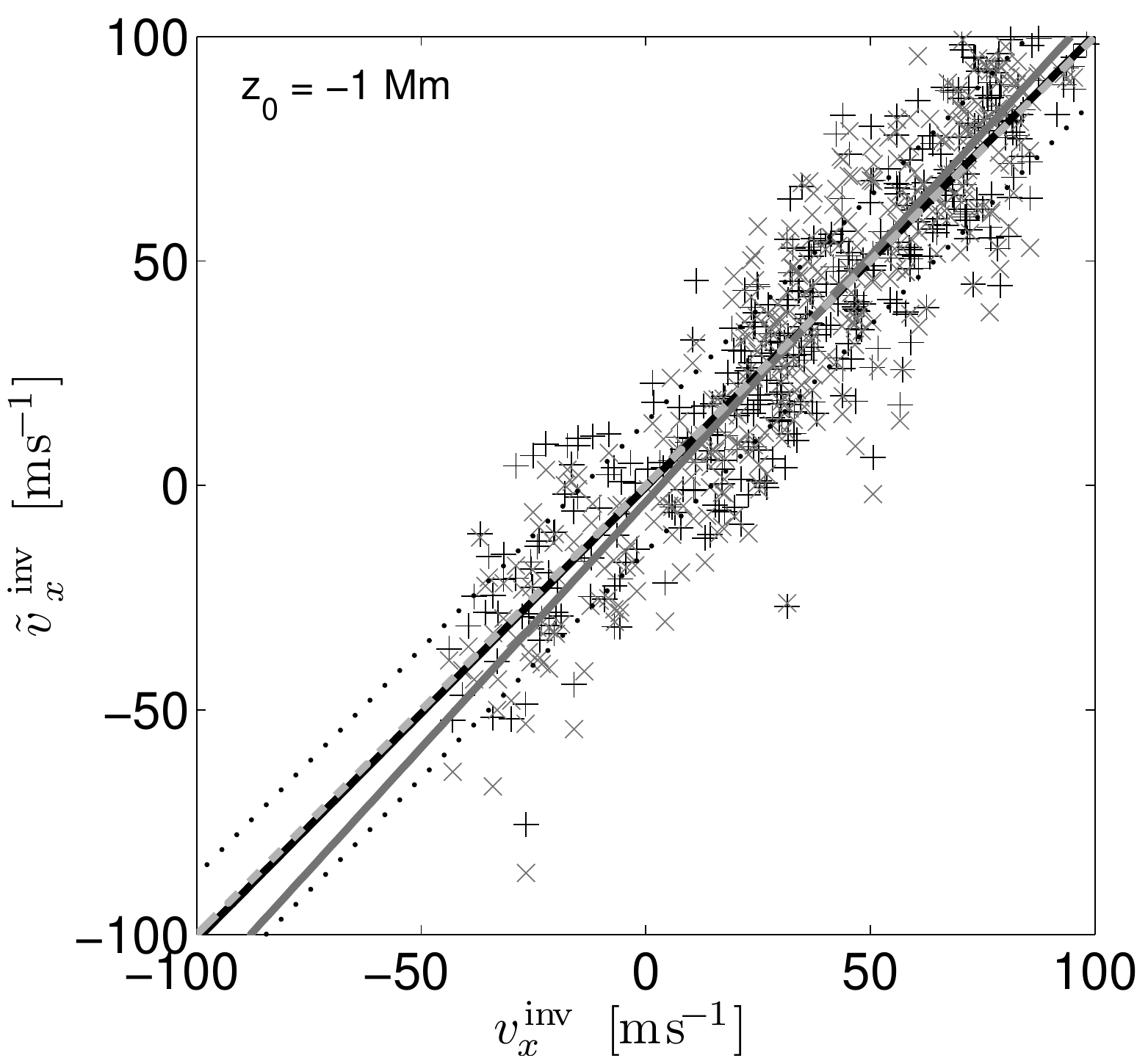}}
\caption{Inversion biases for $v_x$ at 1~Mm depth. Left: Noiseless $\vakern_x$ with cross-talk minimised versus the ideal $\vtarg_x$. The departure from slope unity (grey dashed line) is due to an imperfect match between $\cK$ and $\cT$. Right: Noisy $\vinv_x$ versus noiseless $\vakern_x$. The results are plotted in two cases: when the cross-talk is ignored (grey $\times$) and minimised (black $+$). The linear fit to the black crosses coincides with the dashed line of slope unity. The black dotted lines represent the predicted error of 14~\mps{}, which is consistent with the observed scatter of the black crosses.}
\label{fig:validation-vx-scatterplots}
\end{figure*}

In order to quantify the inversion biases, we compare directly the expected and inverted values for a set of spatial locations. We select points separated by $7.5$~Mm in each horizontal coordinate. This sampling interval is equal to half of the horizontal FWHM of the target function (thus the points are somewhat independent). In addition, to avoid possible edge effects, we cut the outer part of the horizontal plane so that only the central 200$\times$200~pixel patch is kept. The scatter plots comparing various inversion components are displayed in Fig.~\ref{fig:validation-vx-scatterplots} for the depth of 1~Mm, the plots for the depth of 3.5~Mm are qualitatively similar. We estimate that for the depth of 1~Mm, the imperfect averaging kernel leads to an average underestimation of the horizontal flow components by some 20\% (Fig.~\ref{fig:validation-vx-scatterplots} left). Furthermore we note that in the case of the inversion for the horizontal flow components, the bias caused by the cross-talk is not important and the random noise level corresponds to the predicted value (Fig.~\ref{fig:validation-vx-scatterplots} right). The results for the depth $3.5$~Mm are similar, while the results for the depth $5.5$~Mm are dominated by the random noise. 

\begin{figure*}[!t]
\includegraphics[width=6cm]{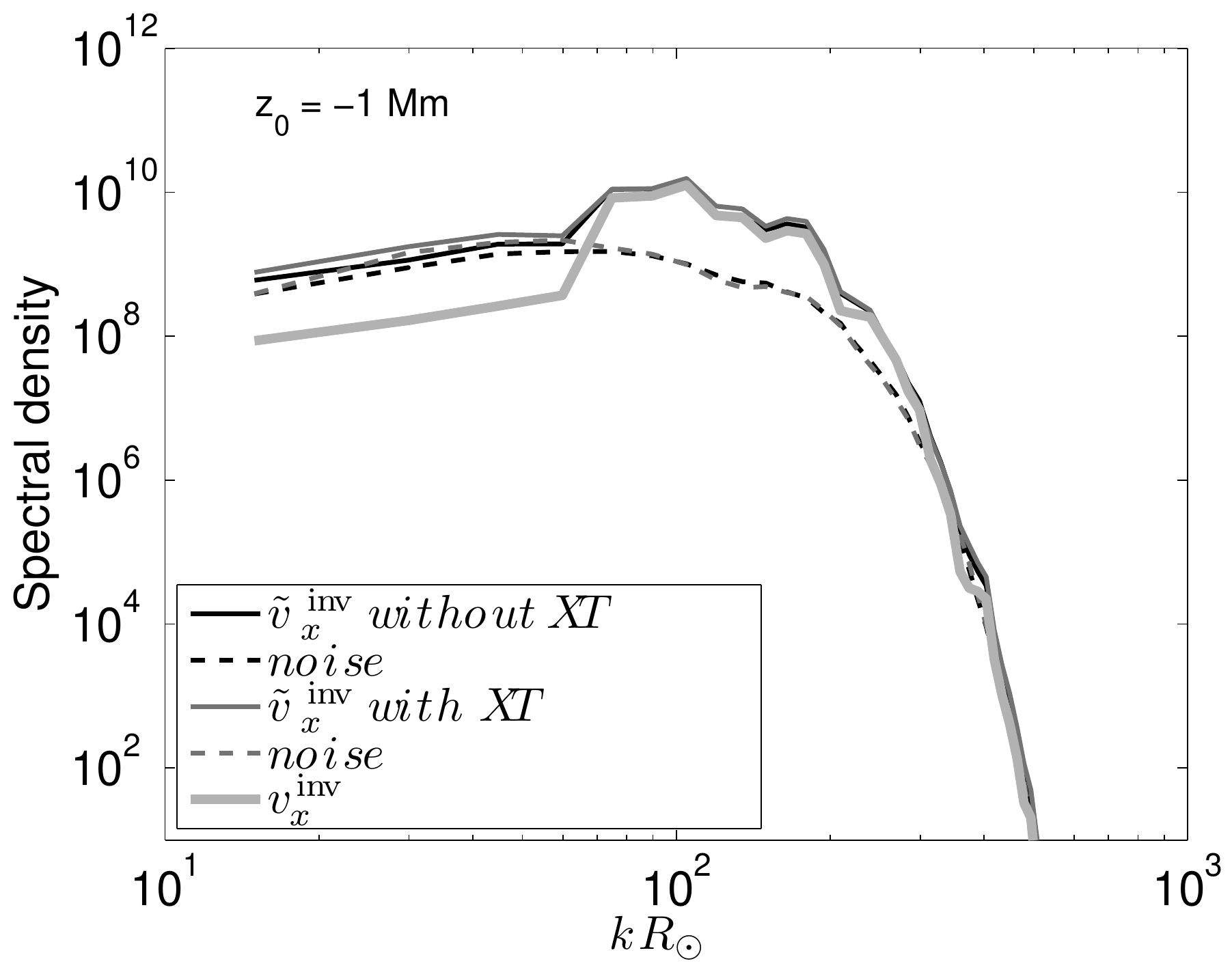}
\includegraphics[width=6cm]{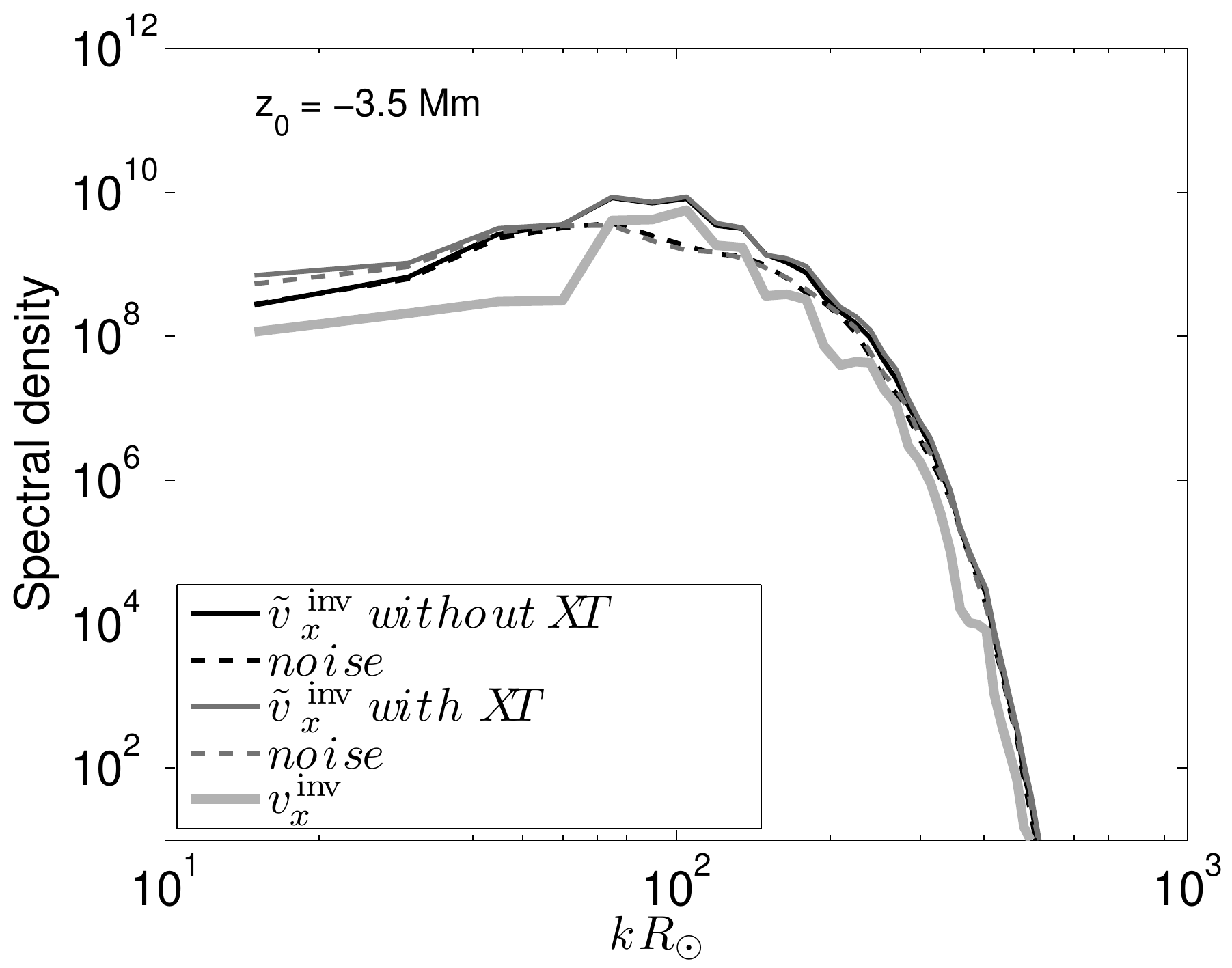}
\includegraphics[width=6cm]{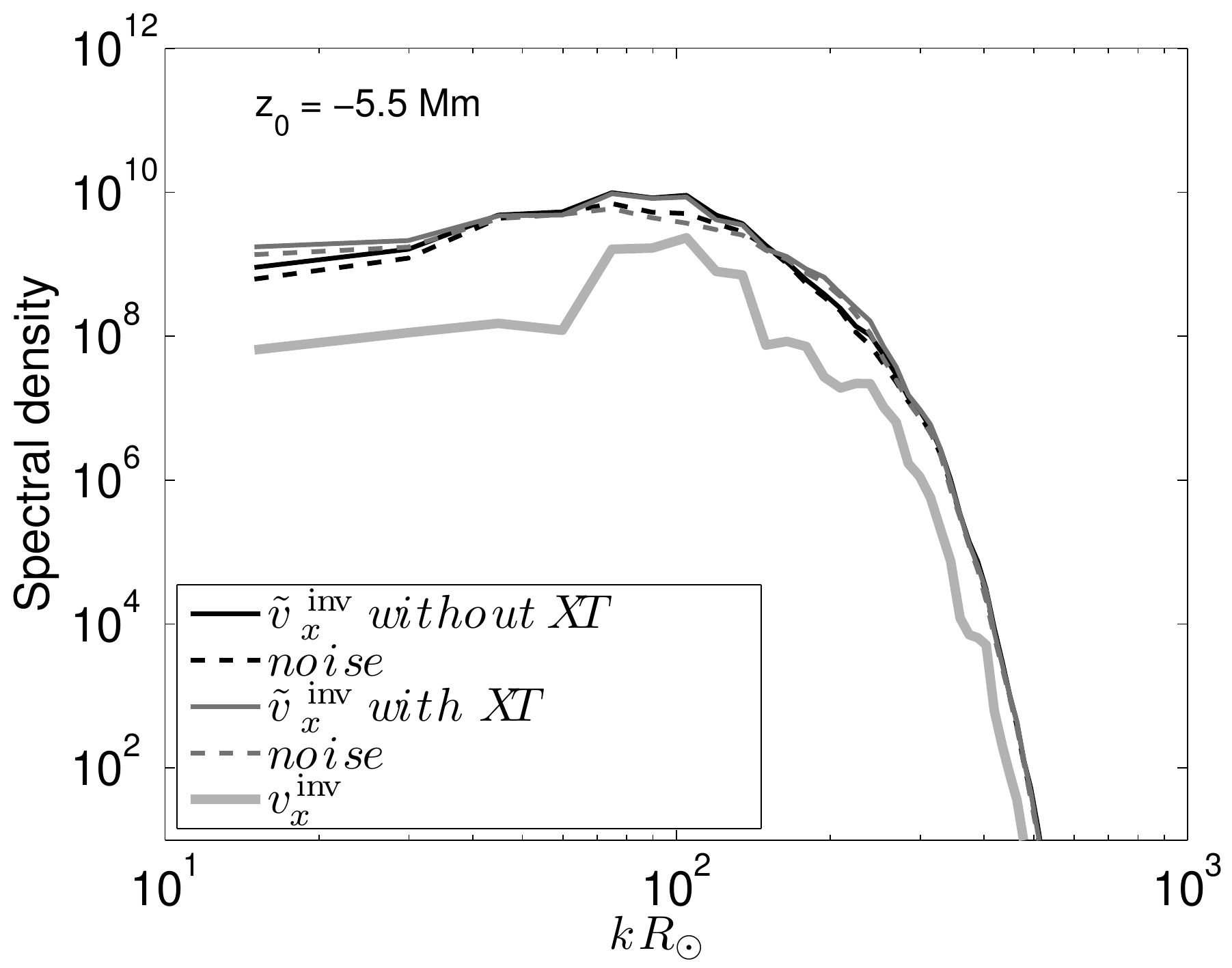}
\caption{The azimuthally-averaged power spectra of various components of the $v_x$ inversions using travel times averaged over 4~days. For reference, we plot the power spectrum of $\vakern_x$ (thick grey solid line). We plot also the power spectrum of $\vinv_x$ (solid line) and the power spectrum of the noise (i.e., the power spectrum of $\vinv_x-\vakern_x$; dashed line) for the inversion where the cross-talk is minimised (black) and ignored (grey).}
\label{fig:snr-vx-k}
\end{figure*}

Inversions for $v_x$ at the depth of $5.5$~Mm are dominated by noise; this is evident from the plot of the azimuthally averaged power spectra of the random noise and of the signal as a function of $k\,R_\odot$ (Fig.~\ref{fig:snr-vx-k} right). For comparison, similar plots for the depths 1~Mm and 3{.}5~Mm are given also in Fig.~\ref{fig:snr-vx-k}, where the power spectrum of the signal is well above the random noise near supergranular spatial scales. The decrease in power of the signal at low $k R_\odot$ in Fig.~\ref{fig:snr-vx-k} is because the convection simulation does not contain these scales.

We showed that it is possible to retrieve $v_x$ and $v_y$ in the top 3.5~Mm without noticeable bias and within the predicted noise level of $\sim$25~m/s (for observing time $T=4$~days) with 240 different travel-time measurements for ridges $f$ to $p_4$. In Table~\ref{tab:summary}, we summarise our findings. We state the statistical quantities (correlation coefficient and the slope of the linear fit) comparing $\vinv_\alpha$ and the corresponding $\vakern_\alpha$. We also show the signal-to-noise ratio of the inverted $\vinv_\alpha$.

The cross-talk is unimportant in inversions for the horizontal flow components. The cross-talk could come only from $v_z$, which is weak. We estimate that the influence of the cross-talk is less than 5\%. The inversion at 5.5~Mm is already dominated by noise. If we drop the requirement on horizontal resolution, inversions at a depth of 5.5 Mm are also possible (e.g., with a FWHM of $s_h=25$~Mm). 

\begin{table}[!b]
\caption{Correlation analysis of various inversion components using synthetic travel times averaged over 4~days. }
\centering
\label{tab:summary}
{\it Crosstalk ignored}\\
\begin{tabular}{l|ccc}
\rule{0pt}{5mm}Depth [Mm] & 1 & 3.5 & 5.5 \\ 
\hline
\rule{0pt}{5mm}${\rm corr}(\vakern_x, \vinv_x)$ & 0.93 & 0.75 & 0.54 \\
${\rm slope}(\vakern_x, \vinv_x)$ & 1.03 & 1.04 & 1.17 \\
$SNR(\vinv_x)$ & 2.63 & 1.10 & 0.54 \\
\hline
\rule{0pt}{5mm}${\rm corr}(\vakern_z, \vinv_z)$ & $-$0.63 & $-$0.02 & 0.06 \\
${\rm slope}(\vakern_z, \vinv_z)$ & $-$1.98 & $-$0.07& 1.67\\
$SNR(\vinv_z)$ & 2.01 & 0.42 & 0.04 \\
\end{tabular}\\
\rule{0pt}{5mm}

{\it Improved inversion, cross-talk minimised}\\
\begin{tabular}{l|ccc}
\rule{0pt}{5mm} Depth [Mm] & 1 & 3.5 & 5.5 \\ 
\hline
\rule{0pt}{5mm}${\rm corr}(\vakern_x, \vinv_x)$ & 0.92& 0.73 & 0.51\\
${\rm slope}(\vakern_x, \vinv_x)$ & 1.02 & 1.05 & 1.29 \\
$SNR(\vinv_x)$ & 2.40 & 1.03 & 0.46 \\
\hline
\rule{0pt}{5mm}${\rm corr}(\vakern_z, \vinv_z)$ & 0.82 &0.31& 0.05\\
${\rm slope}(\vakern_z, \vinv_z)$ & 0.92& 0.91 & 1.41\\
$SNR(\vinv_z)$ & 1.67& 0.35& 0.03 \\
\end{tabular}
\end{table}

It is possible to perform inversions with less temporal averaging, but to obtain a reasonable signal-to-noise ratio, we would have to relax the demand on the match between the averaging kernel and the target function. Typically, this leads to side-lobes in the averaging kernel (especially in the $z$-direction), which may make interpreting results more difficult. Another possibility would be to include a larger number of independent travel-time measurements (and therefore more sensitivity kernels). These issues are being worked on.

  \section{Inversion for vertical flow} 
\subsection{Specificity of inversions for vertical flow}
\label{sect:vz-issue}
Inversions for $v_z$ require a different methodology because sensitivity kernels $K_z^a$ have zero horizontal integrals at each depth (and therefore zero total integral) implying (see Eq.(\ref{eq:akerndefinition}))
\begin{equation}
\int\limits_{-\infty}^{+\infty} \id^2\br\, \cK_z^z(\br,z';z_0) = 0,\ \foral z'\ .
\label{eq:zero_vz_sensitivity}
\end{equation}
Consequently, with measurements discussed here, it is impossible to retrieve horizontal average $\langle v_z \rangle$ of vertical flow and we may only invert for fluctuations $v_z - \langle v_z \rangle$. Equation (\ref{eq:zero_vz_sensitivity}) implies

\begin{equation}
\int_\odot \id^2\br\,\id z' \, \cK^z_\beta(\br,z';z_0) = 0\ {\rm for\ } \beta=z\  
\end{equation}
and thus equation (\ref{eq:constraint}) cannot be written for $\alpha=z$. This implies in turn that the matrix equation (\ref{eq:problem-zerok}) for $\bk=\bzero$ may not be written either. 

\subsection{Ignoring $\bk=0$}
\label{sect:ignorek}
A solution to this issue is simply to replace Eq.~(\ref{eq:problem-zerok}) by 
\begin{equation}
\bar{w}^z_a(\bzero) \equiv 0\ \foral a\ .
\label{eq:problem-zerok-new}
\end{equation}
Since the averaging kernel $\cK_z^z$ must have zero integral, the averaging kernel will be offset by a small negative constant away from the central peak. Luckily, regularization by the term (\ref{eq:confinement}) will ensure that the averaging kernel will drop to zero at the edge of the inversion box. However, this solution is not quite satisfactory, because we do not have very much control over this extended negative surrounding sidelobe.  

\subsection{Target function with zero mean}
\label{sect:tzeromean}
It is more elegant to select a target function $\cT(\br-\br_0,z;z_0)$ with vanishing horizontal integral at each depth. If this target function is to peak around $\bx_0=(\br_0,z_0)$, then this peak must be compensated by negative side-lobes in horizontal directions. These side-lobes have to be constructed in such a way that they do not lead to significant biases. We suggest to choose
\begin{equation}
\cT(\br,z;z_0)=H(\br) \frac{\sqrt{4\ln2}}{\sqrt{\pi} s_z} \exp{\left[-\frac{4\ln2}{s_z^2}{(z-z_0)^2}\right]}\ , 
\end{equation}
where
\begin{eqnarray}
H(\br) & = & \frac{4\ln2}{\pi s_h^2} \exp{\left[-\frac{4\ln2}{s_h^2}{\|\br\|^2}\right]} \nonumber \\
& & - \frac{1}{c} \frac{4\ln2}{\pi (ns_h)^2} \exp{\left[-\frac{4\ln2}{(ns_h)^2} \left(\|\br\|-\frac{s_h}{2\sqrt{2\ln2}}\right)^2\right]}\ .
\label{eq:improved_target1}
\end{eqnarray}
The constant 
\begin{equation}
c=\exp{\left( -\frac{1}{2n^2} \right)}+\sqrt{\frac{\pi}{2n^2}}\left[1+{\rm erf}{\left(\frac{1}{n\sqrt{2}} \right)} \right]
\end{equation} 
ensures that the horizontal integral of $\cT$ is zero. The horizontal part of the target function is constructed from a Gaussian peaked at $\br_0=0$, from which a wide surrounding Gaussian annulus is subtracted. The free parameter $n$ balances the width of a side-lobe compared to the width of the central peak. 

The inversion is performed for each $\bk \ne \bzero$ with this new target function, together with $\bar{w}^z_a(\bk=\bzero)=0$. 

\begin{figure}[!b]
\centering
\includegraphics[width=0.4\textwidth]{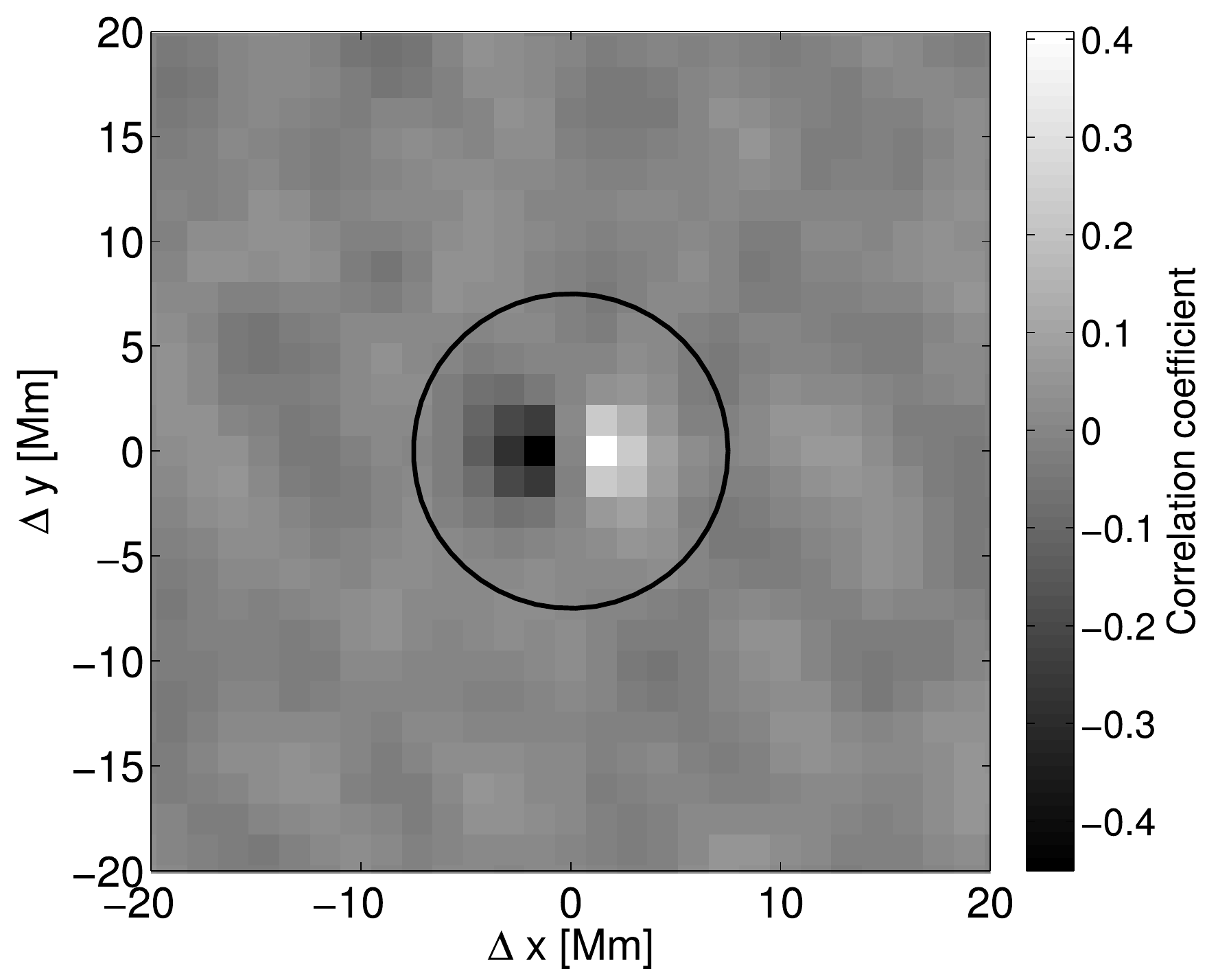}
\caption{Horizontal spatial cross-correlation between $\vsim_x$ and $\vsim_z$ from the simulation in the top 1~Mm. These flow correlations together with a non-vanishing cross-talk averaging kernel $\cK_x^z$ may lead to a systematic bias of the inferred $v_z$. The circle of radius of 15~Mm shows the width of the desired target function.}
\label{fig:vx-vz-correlations}
\end{figure}

\subsection{Both target functions provide similar answers}
We compared the results obtained using the two target functions proposed above and found two conclusions.

The two solutions described above provide results which are very close (see, e.g., Fig.~\ref{fig:zeroT-ES} in the electronic supplement). However, the solution with the target function having zero horizontal average is elegant and provides more control over the solution averaging kernel. 

By comparing the results obtained with various sizes of the negative annulus surrounding the central peak we found that the bias in the inverted flow caused by this negative sidelobe is negligible if $n > 3$ in Eq.~(\ref{eq:improved_target1}). 

The results presented in the following sections are obtained using the first formalism. 

\begin{figure}[!t]
\centering
\includegraphics[width=6cm]{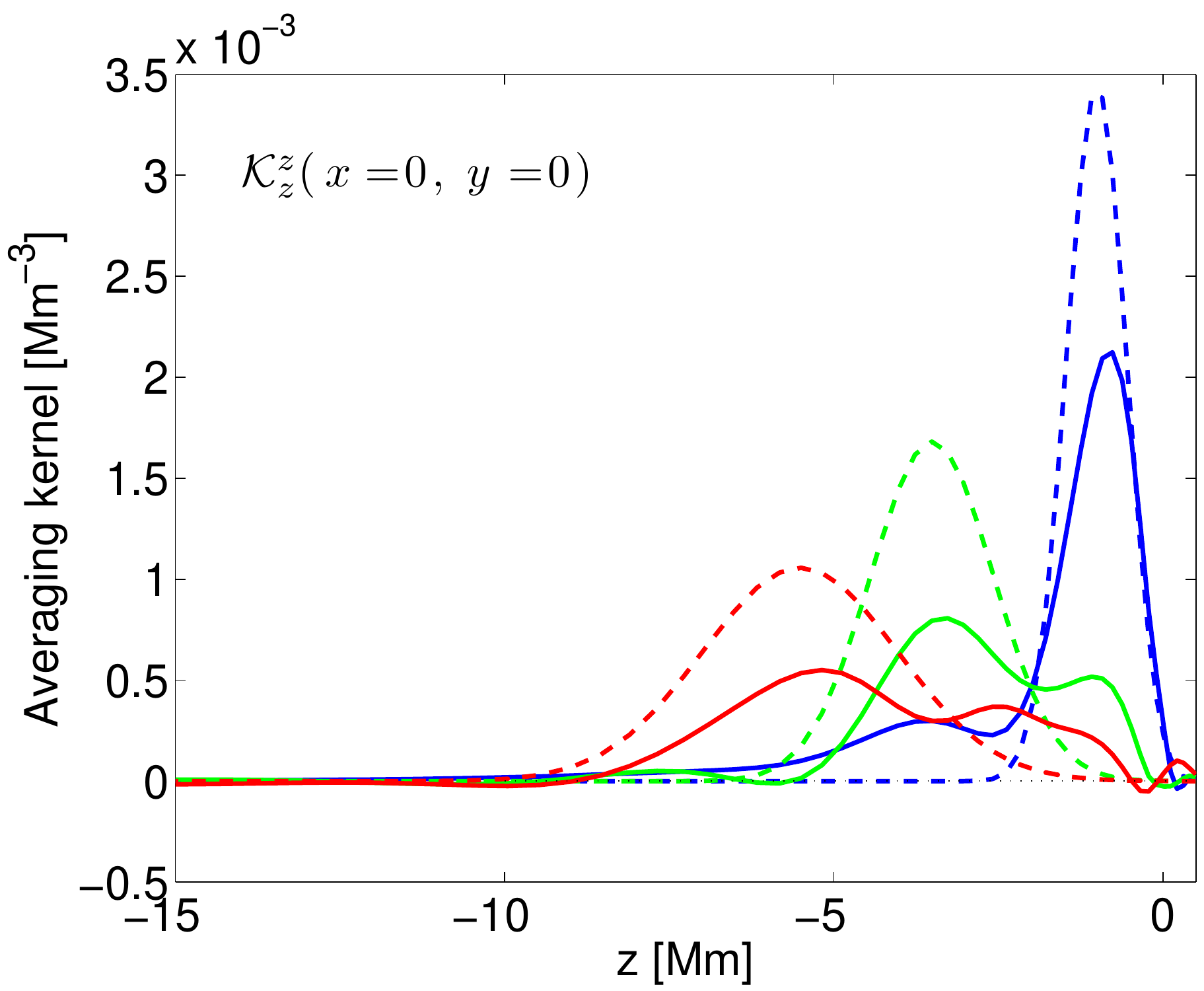}
\caption{The cut through the $x=y=0$ point of the averaging kernel (solid) and the respective target function (dashed) for $v_z$ inversion using travel times averaged over 4~days. Compare with Fig.~\ref{fig:akerns-vx-1D}.}
\label{fig:akerns-vz-1D}
\end{figure}
\subsection{Validation of vertical flow inversion}
\begin{figure*}[!t]
\centering
Inversion with no cross-talk regularisation\\
\includegraphics[width=0.8\textwidth]{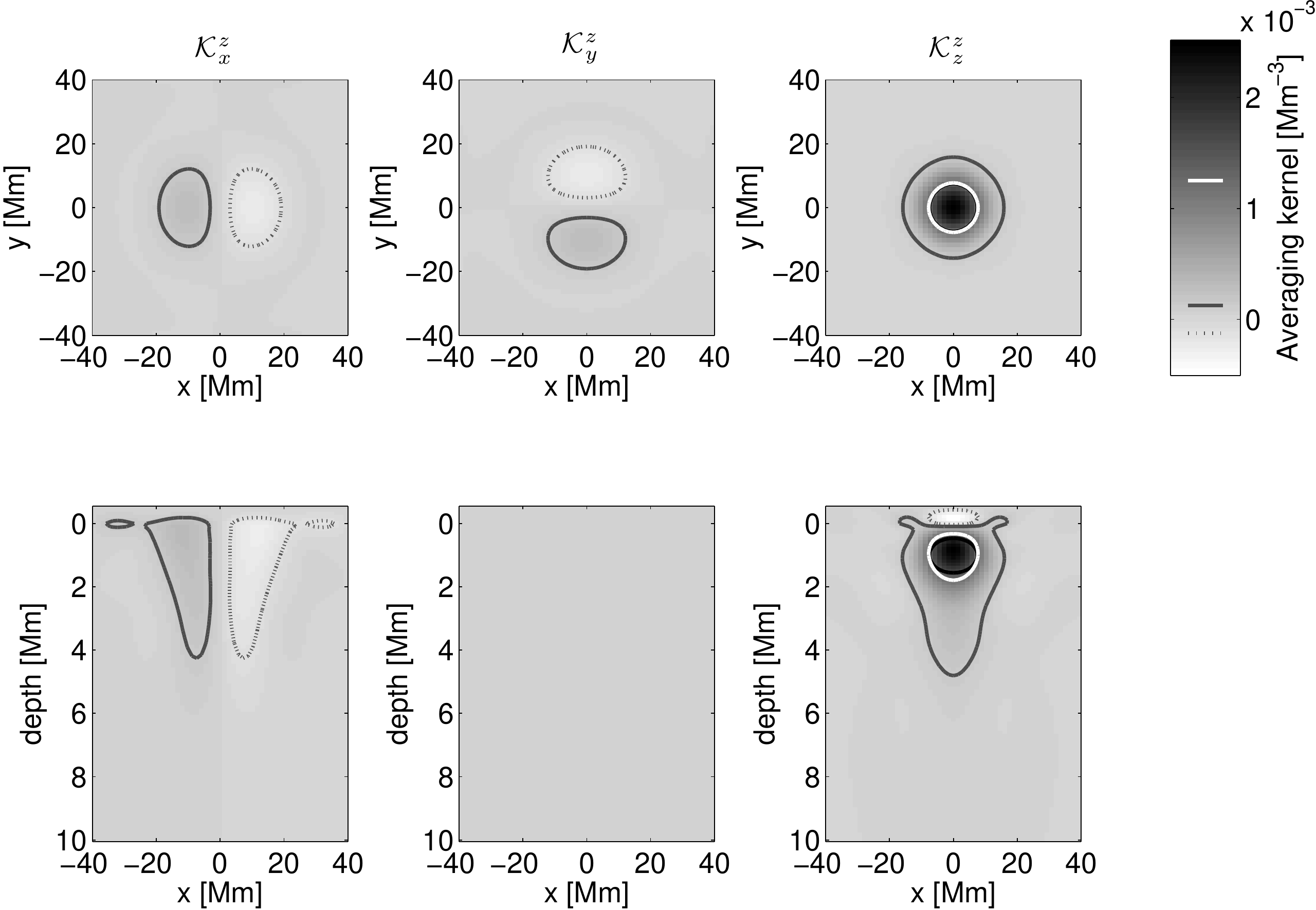}\\
\rule{0.8\textwidth}{1pt}\\
Improved inversion\\
\includegraphics[width=0.8\textwidth]{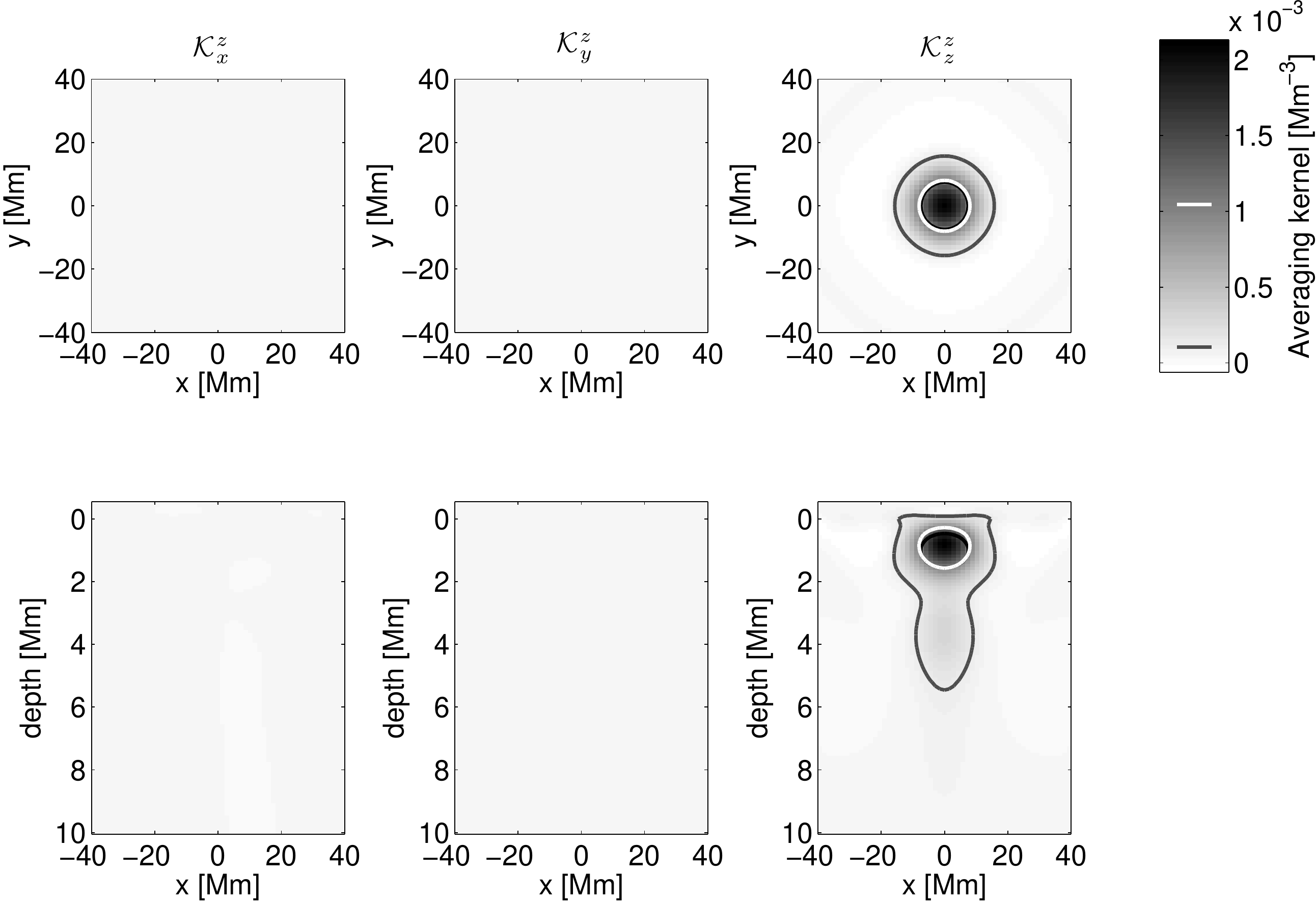}\\
\caption{All components of the averaging kernel for $v_z$ inversion at 1~Mm depth with a FWHM of $s_z=1.1$~Mm and $s_h=15$~Mm. Bottom row: with cross-talk minimised, top row: cross-talk is ignored. The cross-talk is presented in the form of $\cK_x^z$ and $\cK_y^z$ averaging kernel components. Random error of the results is 3~\mps{} when assuming data averaged over 4~days. Over-plotted contours, which are also marked on the colour bar for reference, denote the following: half-maximum of the kernel (white), half-maximum of the target function (black), and by grey lines $\pm 5$\% of the maximum value of the kernel (solid and dotted, respectively). }
\label{fig:akerns-vz-1Mm}
\end{figure*}

\begin{figure*}[!t]
\sidecaption
\includegraphics[width=12cm]{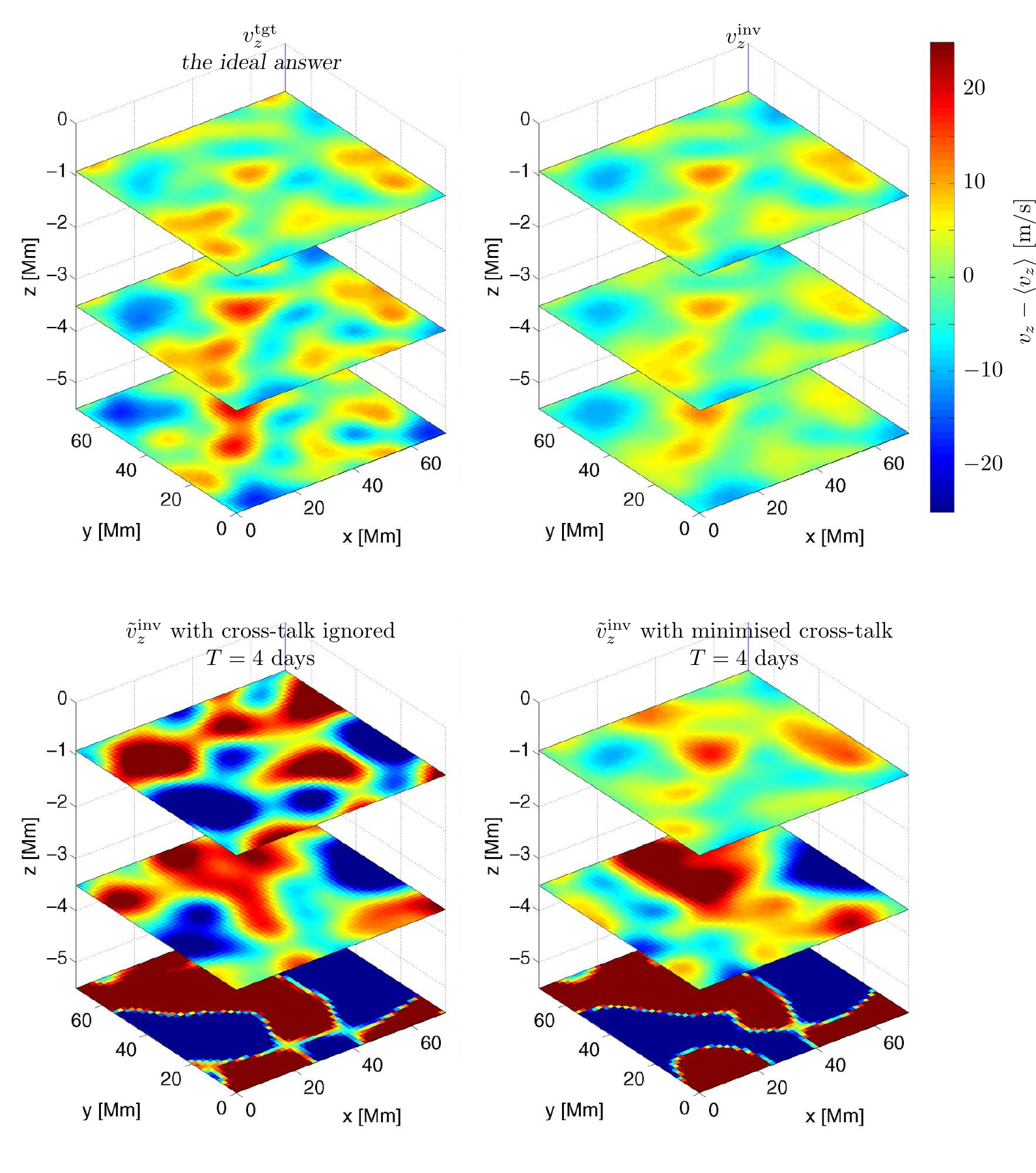}
\caption{Comparison of inverted $v_z$ with input data, compare with Fig.~\ref{fig:vx-validation}. $\left< v_z \right> =11$~\mps{}. }
\label{fig:vz-validation}
\end{figure*}
 
As is evident already, inversions for vertical flow are not as seamless as in the case of horizontal components. This is mostly because the vertical flow is much weaker on supergranular scales than the horizontal flow. As shown in  Table~\ref{tab:errors}, the expected RMS of the vertical flow in the top layers is of the order of 5~\mps. Therefore, the predicted noise of the inverted $v_z$ has to be set to a much smaller value than for the horizontal velocities. As a result, the match between the desired target function and the resulting averaging kernel will have to be poorer.

Furthermore, it is absolutely crucial to minimise the cross-talk to avoid the leakage of the large horizontal velocities into the small inferred vertical velocity. Even an apparently negligible cross-talk averaging kernel could in the end cause a significant bias in the results. Minimising the cross-talk is especially important because of the natural correlations between the vertical and horizontal flow components in the mass-conserving flow of the supergranules (see Fig.~\ref{fig:vx-vz-correlations}). In the upper layers, horizontal outflows are associated with upflows. The structure of the cross-talk averaging kernel $\cK^z_x$ as shown in the top half of Fig.~\ref{fig:akerns-vz-1Mm} in the case when cross-talk is not minimised then implies a negative bias in $v_z$ due to $\vakern_z{}^{(x)}$ and $\vakern_z{}^{(y)}$.  

Fig.~\ref{fig:akerns-vz-1D} shows vertical cuts through the target functions and the averaging kernels when the cross-talk is minimised. The averaging kernels $\cK^z_z$ for the inversions at depth $3.5$ and $5.5$~Mm have sidelobes towards the surface. All components of the averaging kernel $\cK^z_\beta(\br,z;z_0)$ for $v_z$ are displayed in Fig.~\ref{fig:akerns-vz-1Mm} (and Figs.~\ref{fig:akerns-vz-3.5Mm-ES}--\ref{fig:akerns-vz-5.5Mm-ES} available in the electronic supplement), again comparing the cases when the cross-talk is and is not minimised. The action of the cross-talk minimisation term is very efficient. 

The validation of the $v_z$ inversion is demonstrated in Fig.~\ref{fig:vz-validation}. Here we plot $\vtarg_z$ at three different depths and the inverted $\vakern_z$ without noise contributions. These two are very similiar at the depth of $1$~Mm. Small differences between $\vtarg_z$ and $\vakern_z$ caused by an imperfect averaging kernel are visible at the depth of $3.5$~Mm, become more significant at depth $5.5$~Mm. When random noise is added to the solution (bottom row of Fig.~\ref{fig:vz-validation}), we see that the inversion for vertical flow is possible at 1~Mm depth only if cross-talk is minimised. Inversions at depths $3.5$ and $5.5$~Mm are buried in random noise, where minimising cross-talk does not help. The magnitude of cross-talk at a depth 1~Mm is much larger than that of the vertical flow; further, cross-talk is highly anti-correlated with the vertical flow, which makes the correlation coefficient between $\vtarg_z$ and $\vinv_z$ close to $-1$. Note that a similarly high anti-correlation was measured by \cite{2007ApJ...659..848Z}.
\begin{figure*}[!t]
\centering
\includegraphics[width=0.8\textwidth]{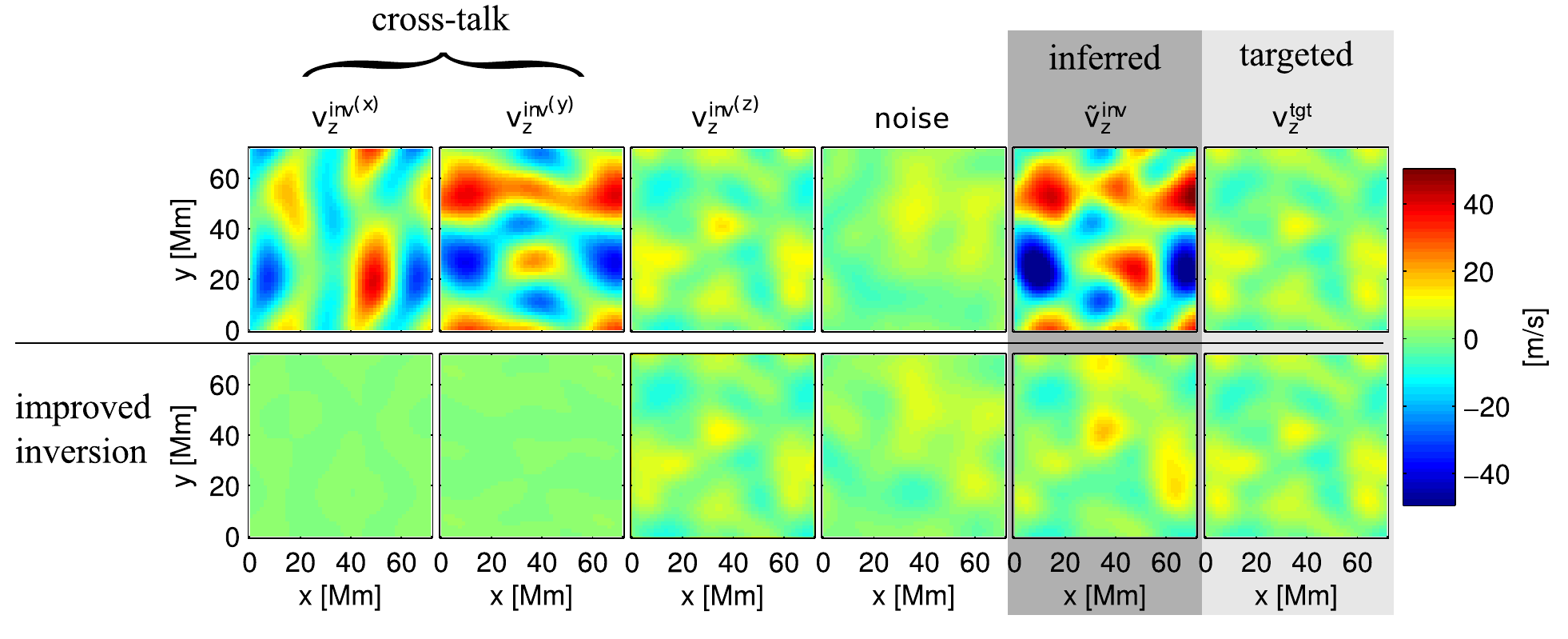}
\caption{All components of the $v_z$ inversion at 1~Mm depth.  Top row with the cross-talk ignored, bottom row with the cross-talk minimised. We demonstrate that if cross-talk is not addressed, horizontal components will leak into the inverted $v_z$ and cause a bias.}
\label{fig:vz-inversion-components}
\end{figure*}

\begin{figure*}[!t]
\sidecaption
\includegraphics[width=6cm]{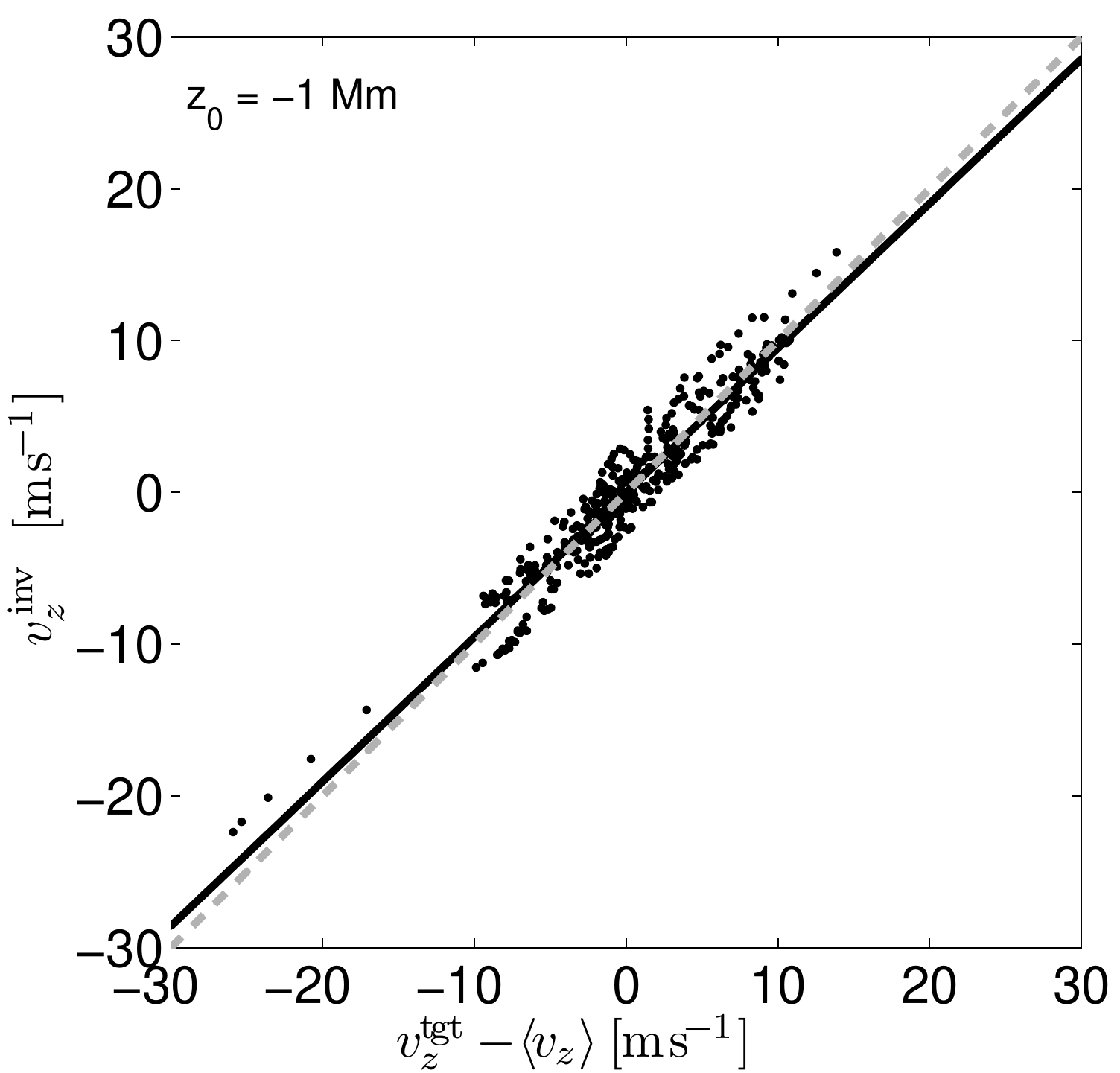}
\includegraphics[width=6cm]{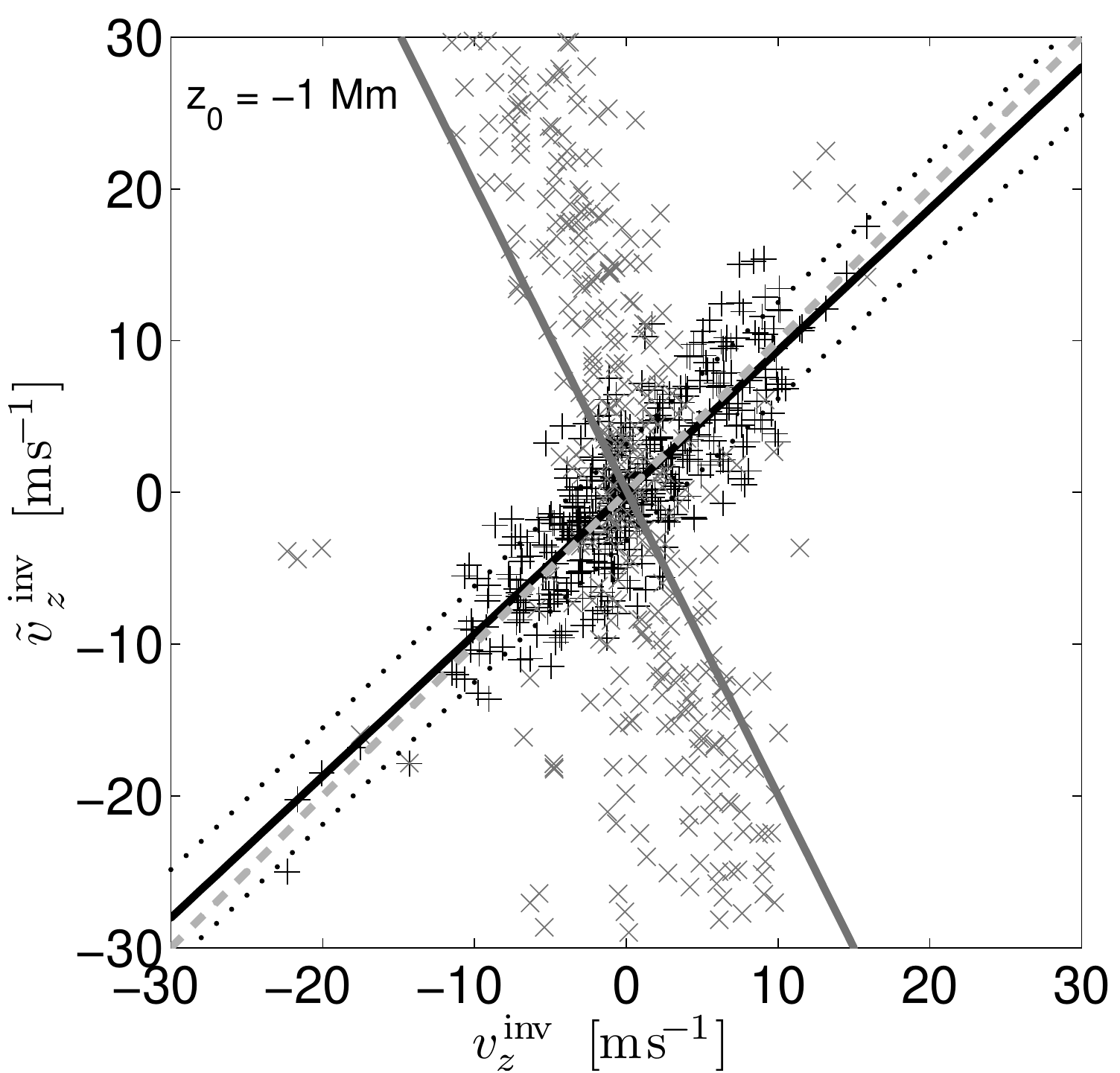}
\caption{Inversion biases for $v_z$ at 1~Mm depth, similar to Fig.~\ref{fig:validation-vx-scatterplots}. While the black line (with minimised cross-talk) almost coincides with the dashed line of slope unity, the grey one (cross-talk ignored) indicates a horrible bias. The predicted error is 3~\mps{}. The inversion is not sensitive to the horizontally averaged horizontal flow $\left< v_z \right> =11$~\mps{}, therefore it was subtracted from $\vtarg_z$ in this comparison.}
\label{fig:validation-vz-scatterplots}
\end{figure*}

\begin{figure*}[!t]
\includegraphics[width=6cm]{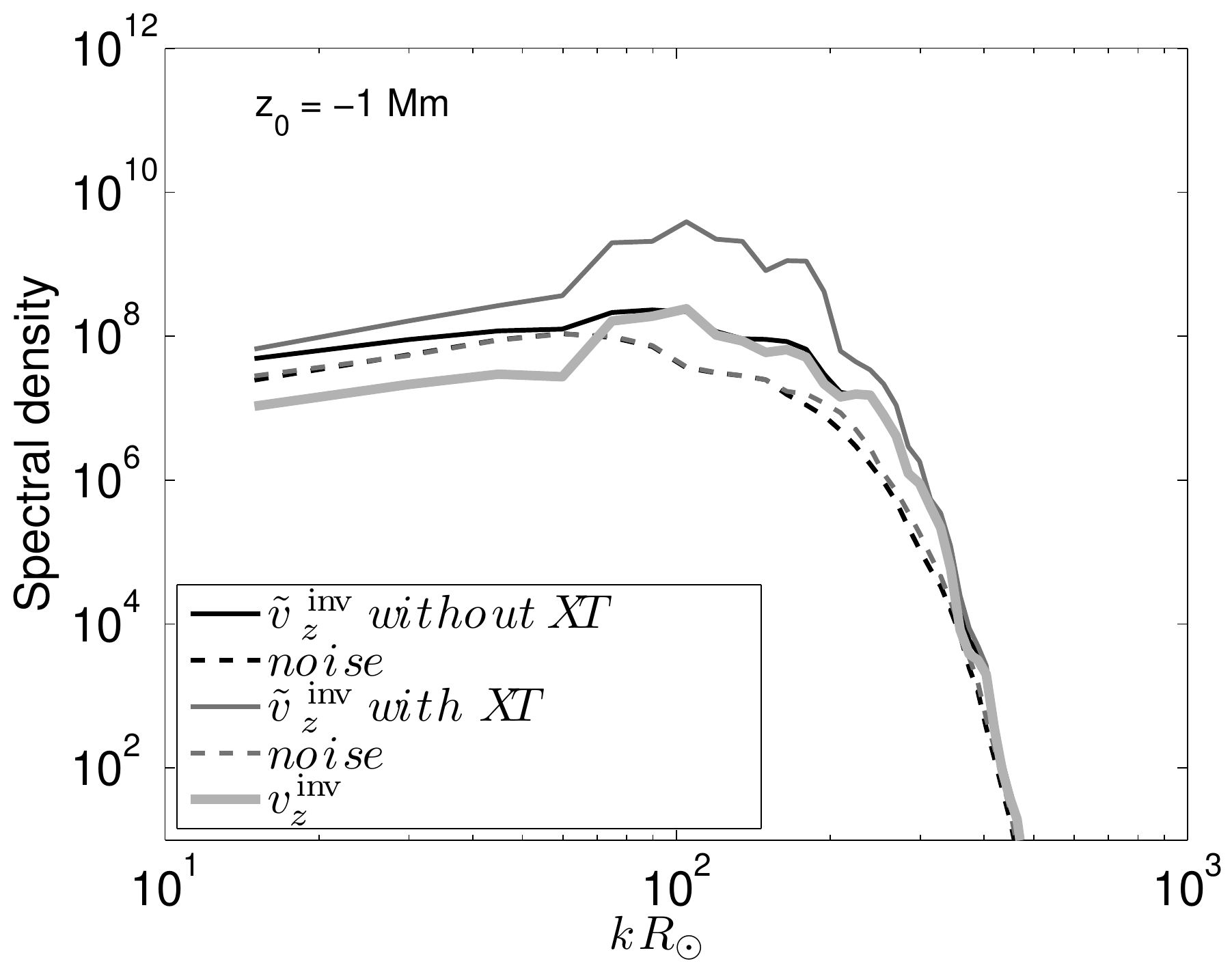}
\includegraphics[width=6cm]{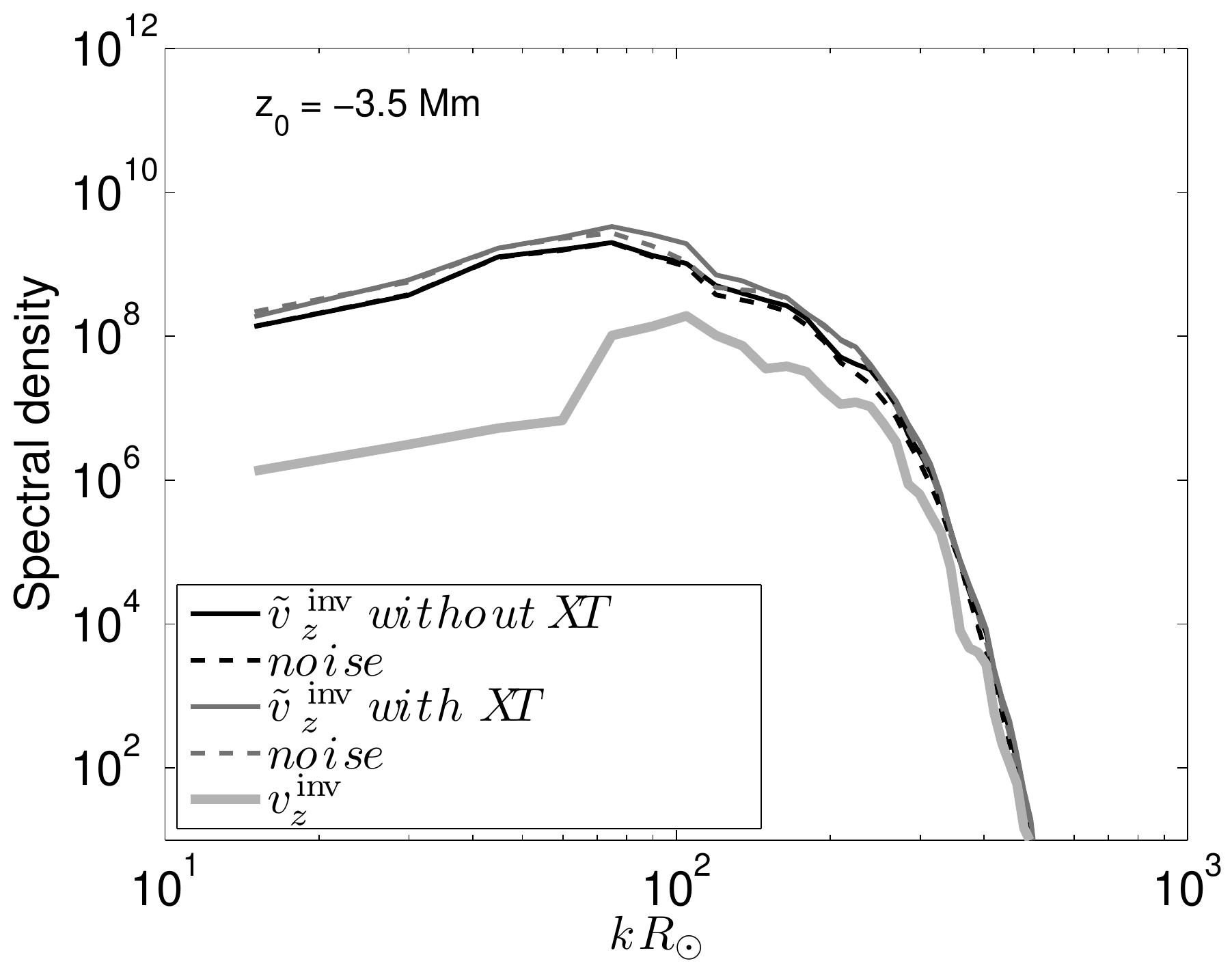}
\includegraphics[width=6cm]{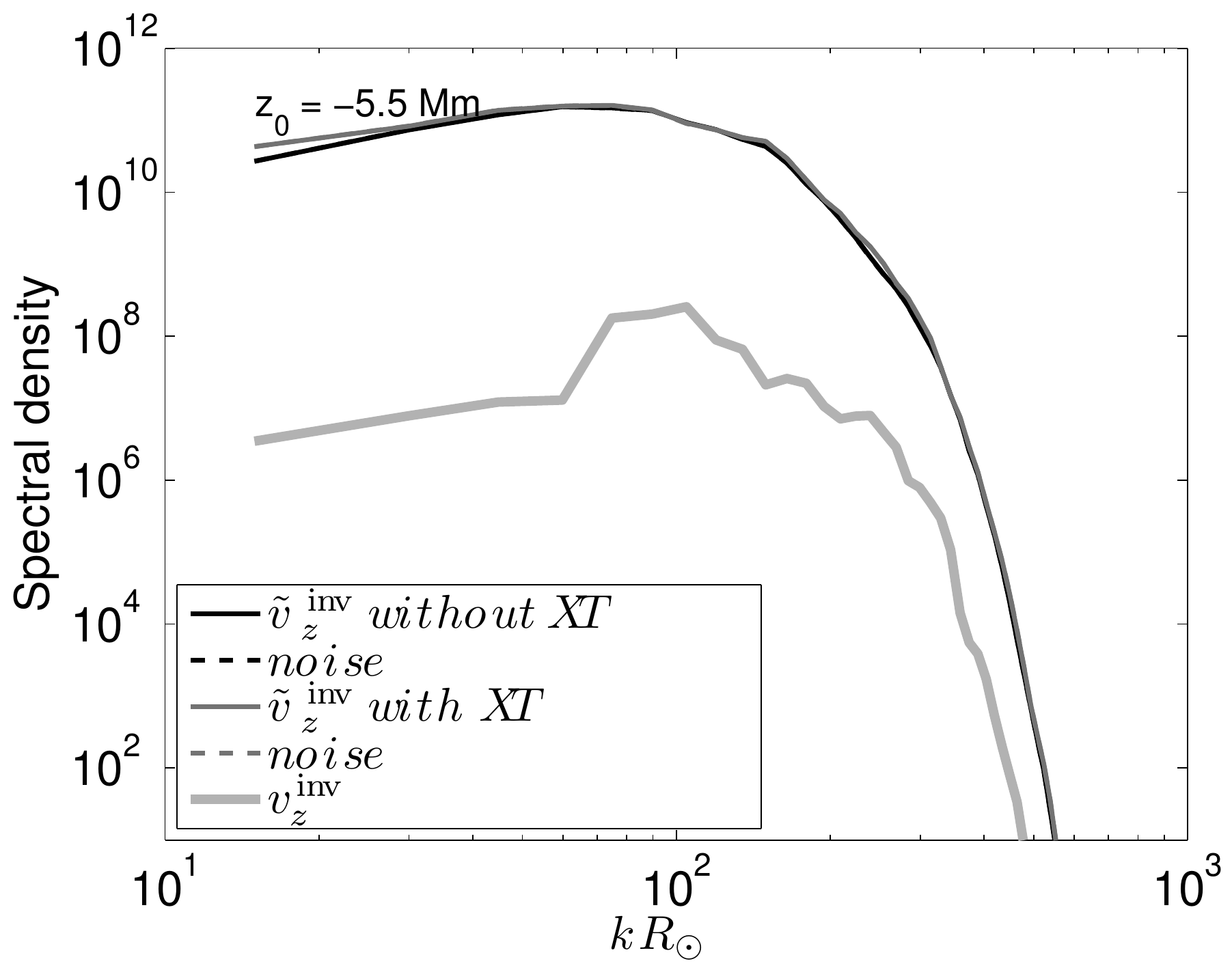}
\caption{The azimuthally-averaged power spectra of various components of the $v_z$ inversions using travel times averaged over 4~days. Compare with Fig.~\ref{fig:snr-vx-k} for $v_x$ inversions. Notice that in the case of $\vinv_z$  containing the cross-talk at 1~Mm depth, the excess in power around $k R_\odot \sim 150$ is not caused by the random noise, but by the bias coming from the cross-talk (not plotted separately). In the other two cases the random noise is a main cause why these inversions are not possible.}
\label{fig:snr-vz-k}
\end{figure*}

All components $\vakern_z{}^{(\beta)}$ and the random noise component in the inversion for $v_z$ at 1~Mm depth are displayed in Fig.~\ref{fig:vz-inversion-components}. We see in the top row that the leakage of the horizontal components ($\vakern_z{}^{(x)}$ and $\vakern_z{}^{(y)}$) covers up completely the weak signal of the vertical flow when the cross-talk is not minimised. 

Biases in the $v_z$ inversion at a depth of 1~Mm may be quantified by directly comparing expected and inverted values for a set of spatial locations in the horizontal plane (Fig.~\ref{fig:validation-vz-scatterplots}), as in the case of the $v_x$ inversion (Section~\ref{sect:vx}). The imperfect averaging kernel in this case does not cause any significant bias (Fig.~\ref{fig:validation-vz-scatterplots} left). The $v_z$ inversion is not sensitive to the horizontally averaged vertical flow $\left< v_z \right>$, which is therefore subtracted from $\vtarg_z$ in the corresponding plot. The effect of minimising cross-talk is shown in Fig.~\ref{fig:validation-vz-scatterplots}. The inverted $\vinv_z$ is anti-correlated with $\vakern_z$ expected from the inversion due to the leakage of the horizontal mass-conserving flow components into the vertical one.

Meaningful inversions for the vertical flow are possible in shallow near-subsurface layers of the convection zone only when the cross-talk between the vertical and the horizontal components is minimised. Vertical flow inversions on supergranular horizontal scales at depths greater than $\sim$1~Mm performed using $f$ to $p_4$ modes and travel time maps averaged over 4 days are dominated by random noise and the signal of the vertical flow may not therefore be inferred at all (see summary in Table~\ref{tab:summary}). This fact is also demonstrated in Fig.~\ref{fig:snr-vz-k}, where we plot power spectra of individual inversion components as a function of the spatial scale. At depths larger than 1~Mm, the power of the signal is much less than the power of the random noise regardless of the spatial scale. 

\section{Beating the noise: statistical averaging}
\begin{figure*}[!t]
\centering
\sidecaption
\includegraphics[width=12cm]{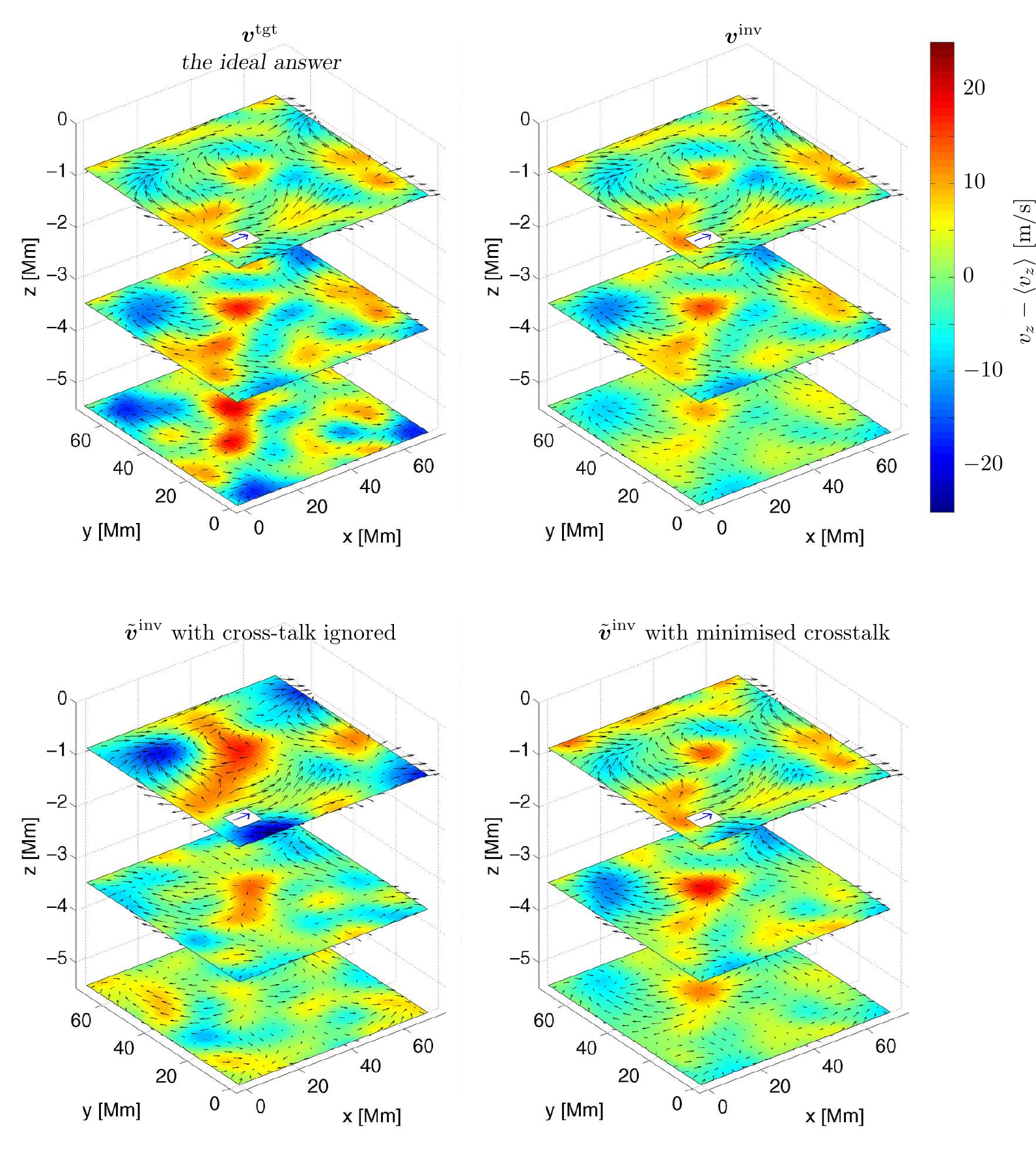}
\caption{Comparison at three depths of statistical averages of velocity-vector over many flow realisations. The horizontal flow is displayed by arrows while the vertical flow is color-coded. The reference arrow indicates 100 \mps{}, the random errors of the inversions are given in Table~\ref{tab:errors}. Compare to Figs.~\ref{fig:vx-validation} and \ref{fig:vz-validation}.}
\label{fig:3dflow-validation-largeaveraging}
\end{figure*}

As demonstrated in previous sections, it is very difficult to have a meaningful inversion for solar flows on supergranular scales averaged over a short time, even in the shallow near-subsurface layers of the solar convection zone. The signal is overwhelmed by random noise. It is not feasible to average over longer times, because the expected lifetime of convection on supergranular scales is on the order of a day. By averaging over longer time, a significant portion of the scientifically useful information would be lost.

It seems feasible to solve the issue by averaging over many realisations of similar flow structures each averaged over short time (therefore noisy). This concept was already used by other authors \citep[e.g.,][]{2006ApJ...646..553D}. As an example we note the possibility to measure the flows in many individual supergranules and to average over this sample in order to obtain the typical flow structure in an ``average supergranule'', similar to \cite{2010ApJ...725L..47D}. The predicted error in the results after averaging scales as $1/\sqrt{\cal{N}}$, where $\cal{N}$ is the size of the ensemble of independent flow realisations. This scaling law allows us to relax the noise constraint on the inversion and therefore regularise more strongly about the misfit and cross-talk terms. See Table~\ref{tab:errors} for required random error of inversions results.

For instance, let us assume that we average over 10$^4$ independent inverted flows, each obtained by considering travel-time maps averaged over 6 hours. Individual flow maps are noisy and therefore contain little useful information about underlying flows. This selection allows us to relax the constraint on the estimated error level by a factor of a 100 and therefore obtain a much better fit to the target function (see Fig.~\ref{fig:akerns-1D-largeaveraging} in the electronic supplement). With this set-up, the validation displayed in Fig.~\ref{fig:3dflow-validation-largeaveraging} shows almost perfect correspondence between expected $\bvtarg$ and inverted $\bvinv$ flows in the top 5{.}5~Mm of the convection zone. Inversion for $v_z$ at 5.5~Mm with a FWHM of $s_h=15$~Mm is an exception, because the selected target function cannot be matched by the averaging kernel. Different inversions for $v_z$ at this depth are possible with greater averaging, e.g., with $s_h=25$~Mm. It is still crucial to minimise cross-talk in the case of $v_z$ inversions to retrieve the correct answer. Power spectra of  inversion components at all discussed depths are displayed in Fig.~\ref{fig:snr-vz-k-10000sgs-ES} in the electronic supplement.

  \section{Conclusions}

We improved and validated an inversion algorithm based on a SOLA inversion approach. The formalism, algorithm and the code is universal and can, in principle, be used to invert for any quantity describing inhomogeneities in the solar plasma, provided that the corresponding sensitivity kernels are available. The code is also ready to be used for application to real measurements. It will become part of the helioseismic pipeline running at German Data Center for SDO. We plan to use it to analyse all available SOHO/MDI and SDO/HMI data in order to routinely provide a  tomographic image of the structure of the solar upper convection zone. 

The code is absolutely scalable allowing to include more independent measurements (and therefore more sensitivity kernels) in order to further refine the precision of the results. Thanks to the decoupling, the problem will still be solvable even using nowadays computers. 

We improved the inversions by introducing additional terms that allow to control and minimise some sources of bias in the results. Most importantly, we minimise the cross-talk between individual flow components, which is crucial especially for $v_z$ inversions. In principle, the formalism allows to minimise cross-talk between any selected quantities during the general inversion. The validation performed here silently assumes that the sensitivity kernels and noise covariance matrices are perfect. For the validation of the method and the code it is important if the sensitivity kernels and noise covariance matrices are solar-like. This allows us to study different sources of biases we may expect in the Sun. It is also very important that the sensitivity kernels used in the inversion contain all the details of the travel-time measurements, including the instrumental function affecting the solar oscillation power spectrum. 

We found that by considering $f$ to $p_4$ frequency-averaged modes and supergranular spatial scales it is possible to perform reliable and trustworthy flow inversions of the travel-time maps averaged over a few days in the top 3.5~Mm layer of the convection zone in the case of horizontal $v_x$ and $v_y$ components, and in the top 1~Mm in the case of vertical $v_z$ component. Based on our experiment, we expect that using travel-time maps averaged over 4~days, it is possible to measure 3-D velocities as weak as 10~\mps{} at the surface and horizontal velocities having amplitude 20~\mps{} in the top 5~Mm. We estimate, that using the travel-times averaged over 1~day, it should still be possible to measure all components of the supergranular flow at the surface and its horizontal components in the top 5~Mm. The cross-talk minimisation is crucial in order to measure the correct vertical velocity. Its presence in the results may explain the opposite sign of the vertical flow inversion discovered by \cite{2007ApJ...659..848Z}. 

By considering many flow realisations and statistical averaging we might go deeper in the convection zone to learn about the horizontal flow components and perform vertical flow inversions for the top few Mm depths. 


\subsection*{Acknowledgements}                                                                                                              
This study was supported by the European Research Council under the European Community's Seventh Framework Programme (FP7/2007--2013)/ERC grant agreement \#210949,  ``Seismic Imaging of the Solar Interior'', to PI L. Gizon (Milestone \#5). Authors would like to thank J.~Jackiewicz for providing us with the noise covariance matrices. The flow sensitivity kernels were computed using the code written by A.~C.~Birch deployed within the HELAS project at {\tt http://www.mps.mpg.de/projects/seismo/NA4}. M.\v{S}. acknowledges a partial support through the Grant Agency of Academy of Sciences of the Czech Republic under grant IAA30030808.


\Online 
\onecolumn
\onlfig{16}{
\begin{figure*}[!h]
\centering
\includegraphics[width=0.49\textwidth]{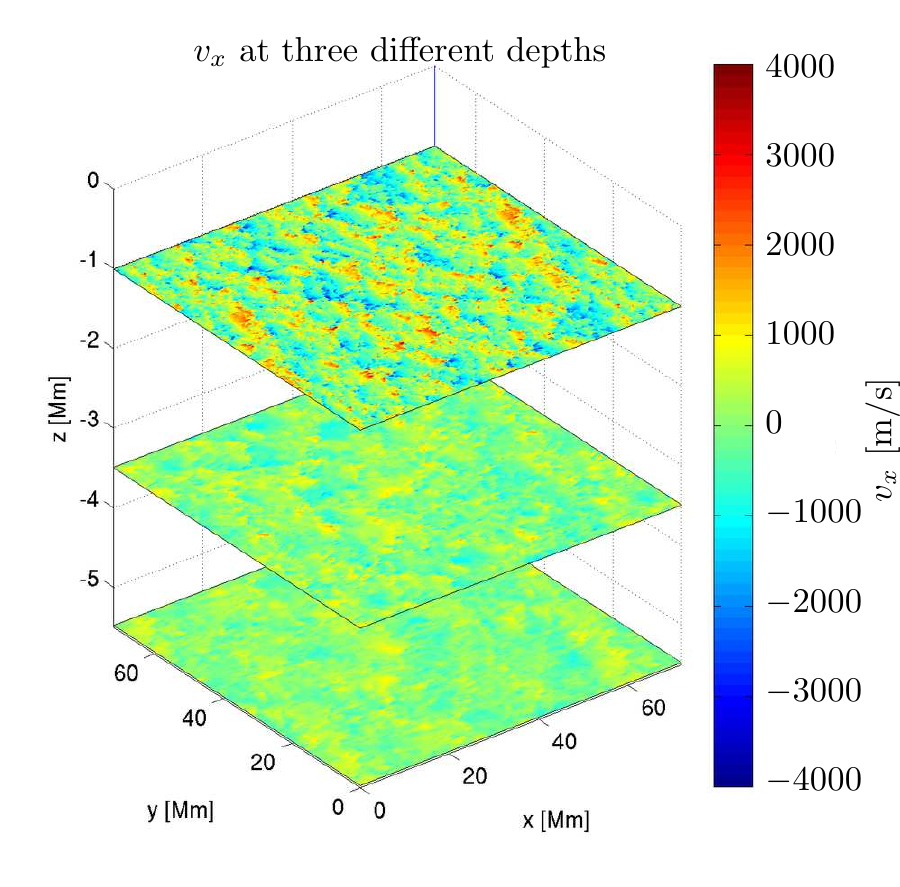}
\includegraphics[width=0.49\textwidth]{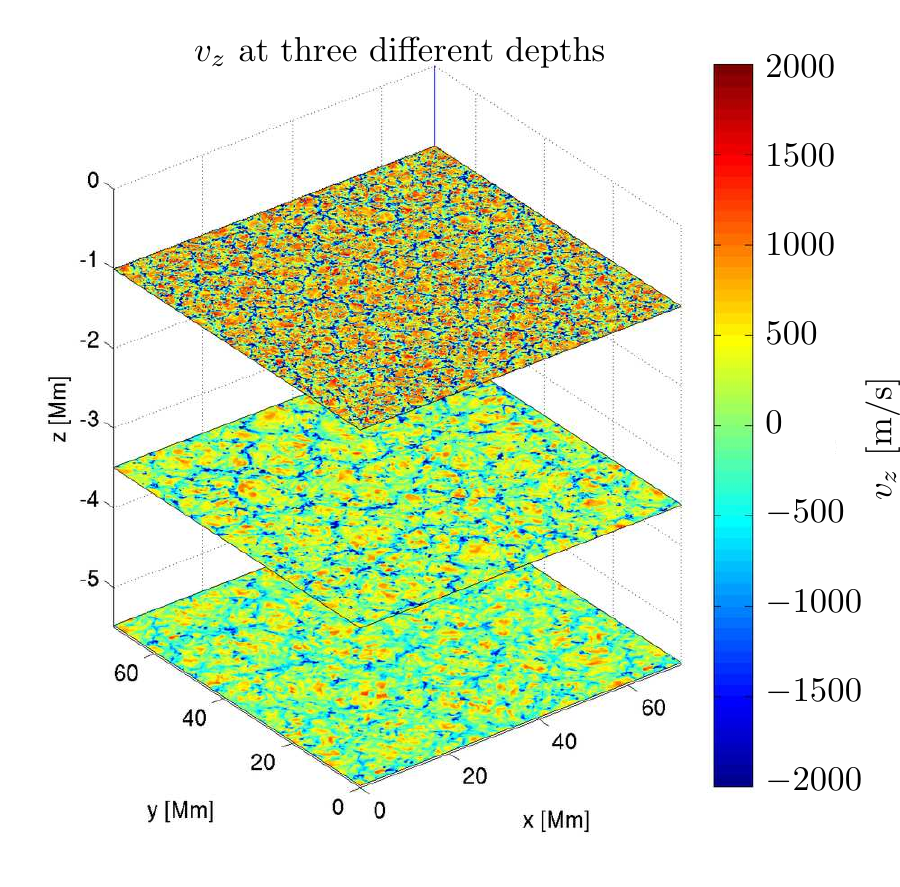}
\caption{Slices through the $x$ and $z$ components of the simulated flow $\bvsim$ at depths 1~Mm, $3.5$~Mm, and $5.5$~Mm.}
\label{fig:simulation}
\end{figure*}
}

\onlfig{17}{
\begin{figure}[!h]
\sidecaption
\includegraphics[width=12cm]{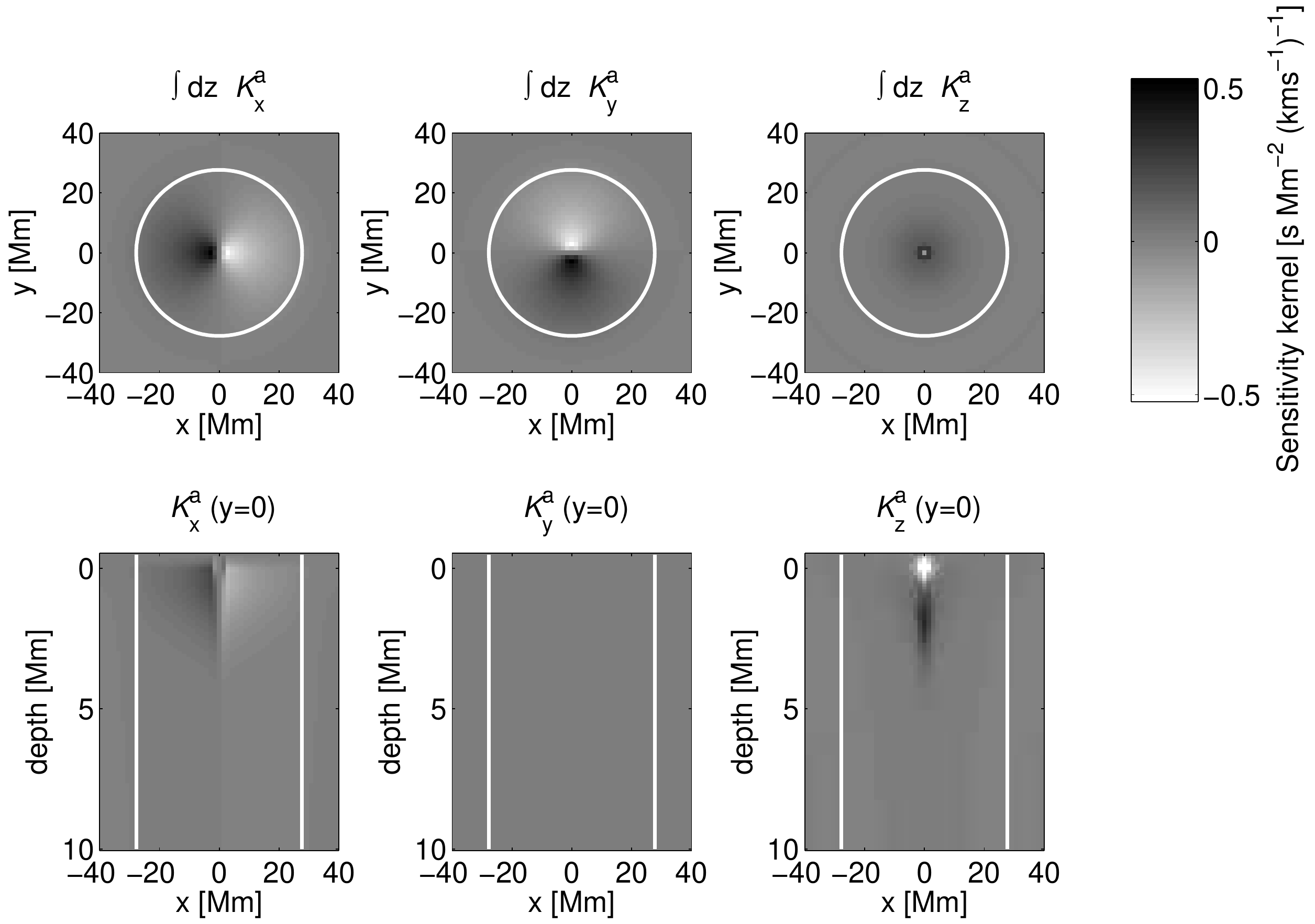}
\caption{The point-to-annulus sensitivity kernels for three flow components computed for the $f$-mode, distance 28~Mm and outward-inward geometry. The white circle represents the location of the averaging annulus. }
\label{fig:sensitivitykernel-ES}
\end{figure}
}

\onlfig{18}{
\begin{figure}[!h]
\sidecaption
\includegraphics[width=0.35\textwidth]{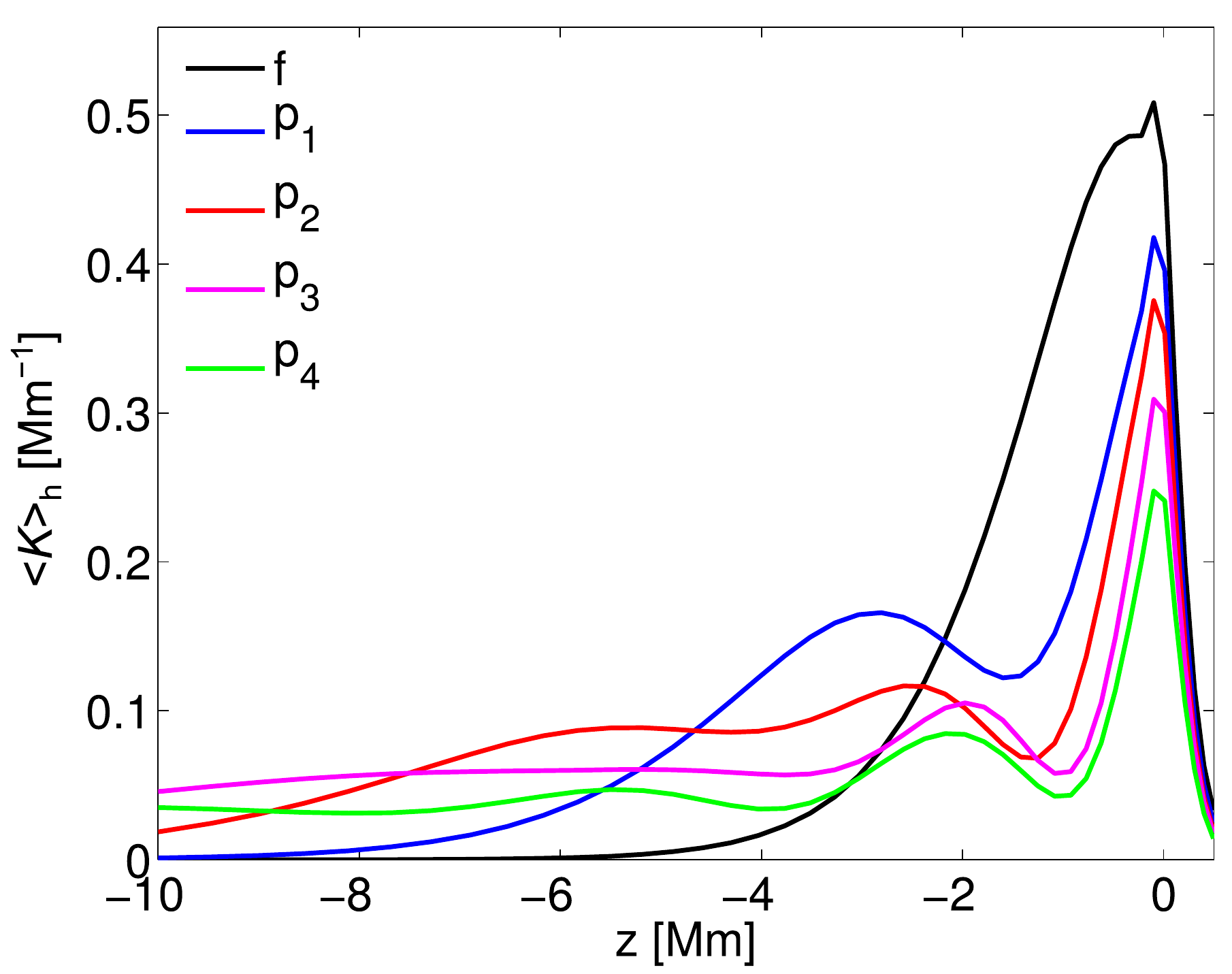}
\caption{Horizontal averages of sensitivity kernels may serve useful when estimating which depths are easier to target. The trend for each mode/ridge was obtained by taking $\int K_x^a(\br;z) \id^2\br$ and averaging over all $a$s within the given mode.}
\label{fig:1Dsensitivity}
\end{figure}
}

\onlfig{19}{
\begin{figure}[!h]
\sidecaption
\includegraphics[width=6cm]{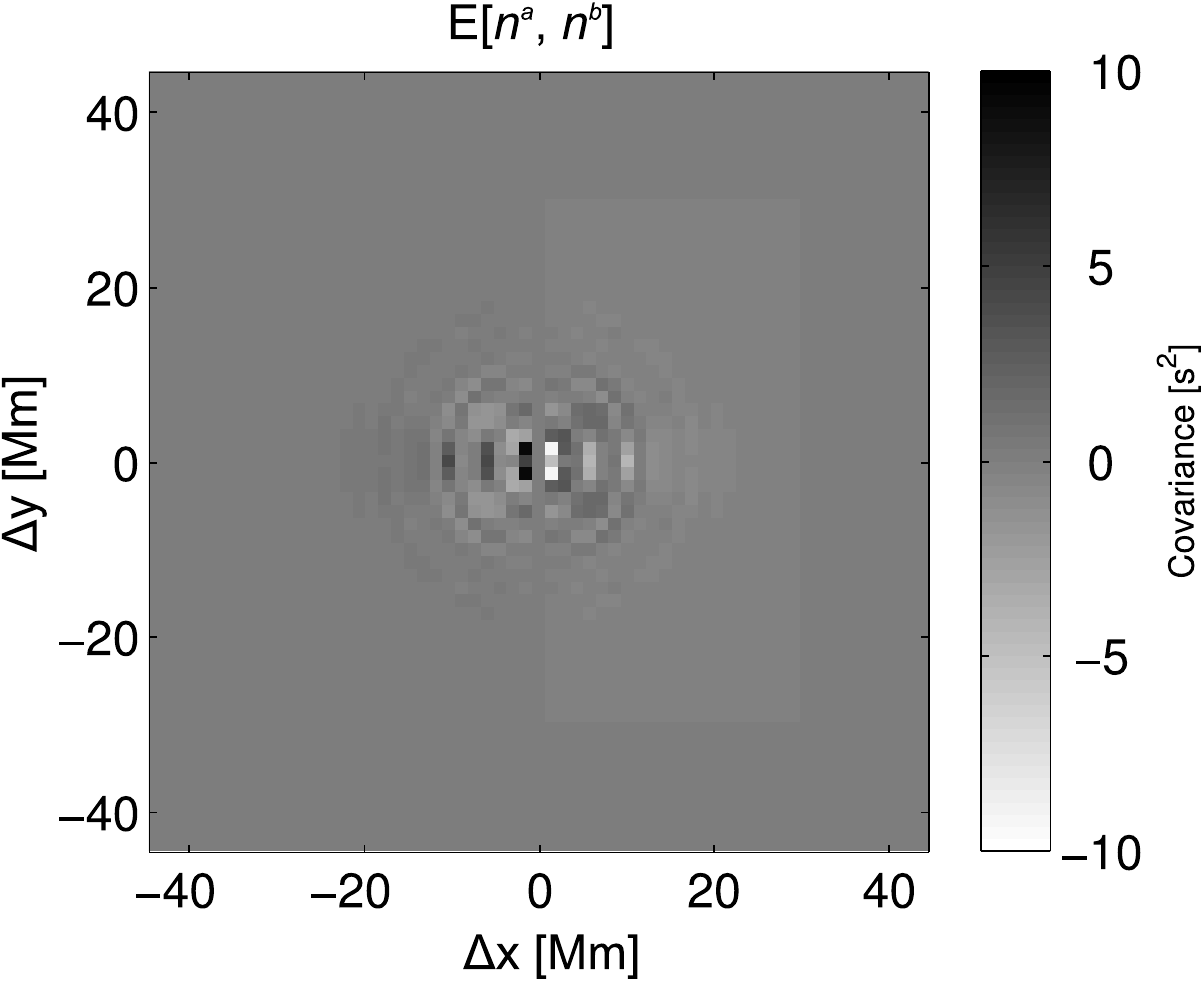}
\caption{An example noise-covariance matrix for $f$-mode travel times averaged over 6 hours. In this plot, $a$ stands for the combination of the $f$-mode, oi geometry, and annulus radius of 7.3~Mm, $b$ stands for the combination of $f$-mode, we geometry, and annulus radius of 8.8~Mm.}
\label{fig:covariancematrix-ES}
\end{figure}
}

\onlfig{20}{
\begin{figure}[!h]
\centering
Inversion with no cross-talk regularisation\\
\includegraphics[width=0.8\textwidth]{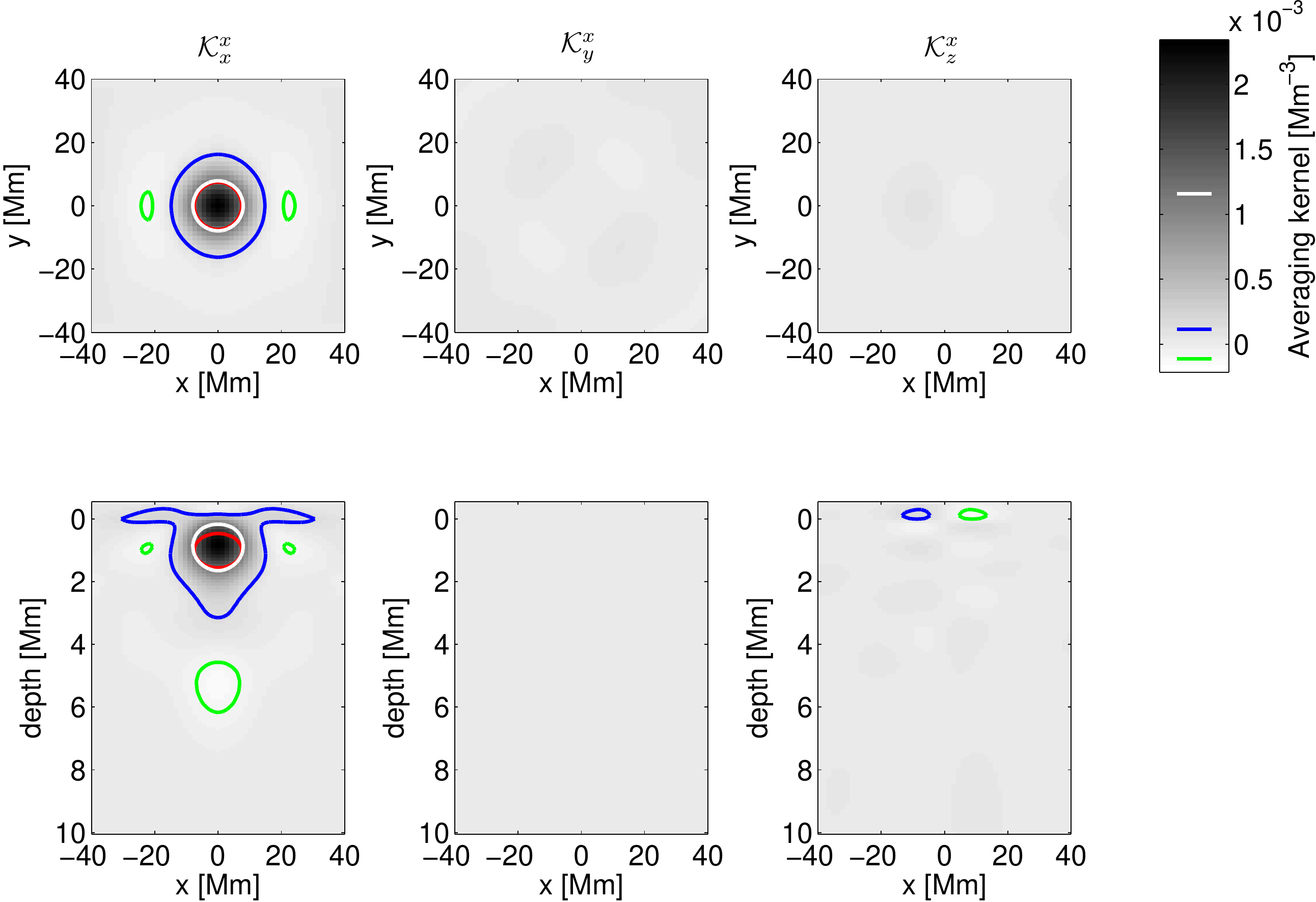}\\
\rule{0.8\textwidth}{1pt}\\
Improved inversion\\
\includegraphics[width=0.8\textwidth]{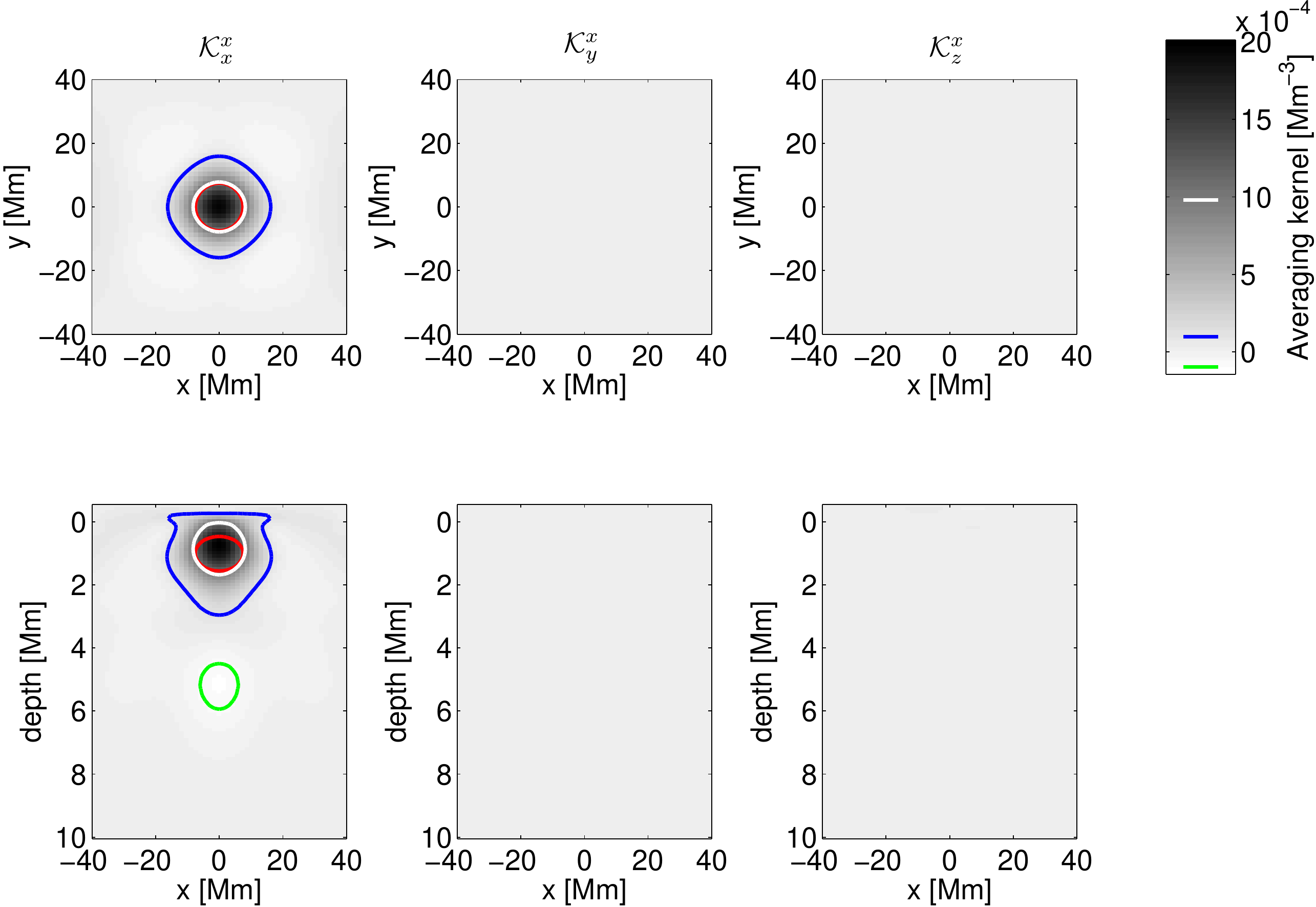}\\
\caption{All components of the averaging kernel for $v_x$ inversion at 1~Mm depth with a FWHM of $s_z=1.1$~Mm and $s_h=15$~Mm. Bottom row: with cross-talk minimised, top row: cross-talk is ignored. Random error of the results is 14~\mps{} when assuming data averaged over 4~days. Over-plotted contours, which are also marked on the colour bar for reference, denote the following: half-maximum of the kernel (white), half-maximum of the target function (red), and $\pm 5$\% of the maximum value of the kernel (blue and green, respectively).}
\label{fig:akerns-vx-1Mm-ES}
\end{figure}
}

\onlfig{21}{
\begin{figure}[!h]
\centering
Inversion with no cross-talk regularisation\\
\includegraphics[width=0.8\textwidth]{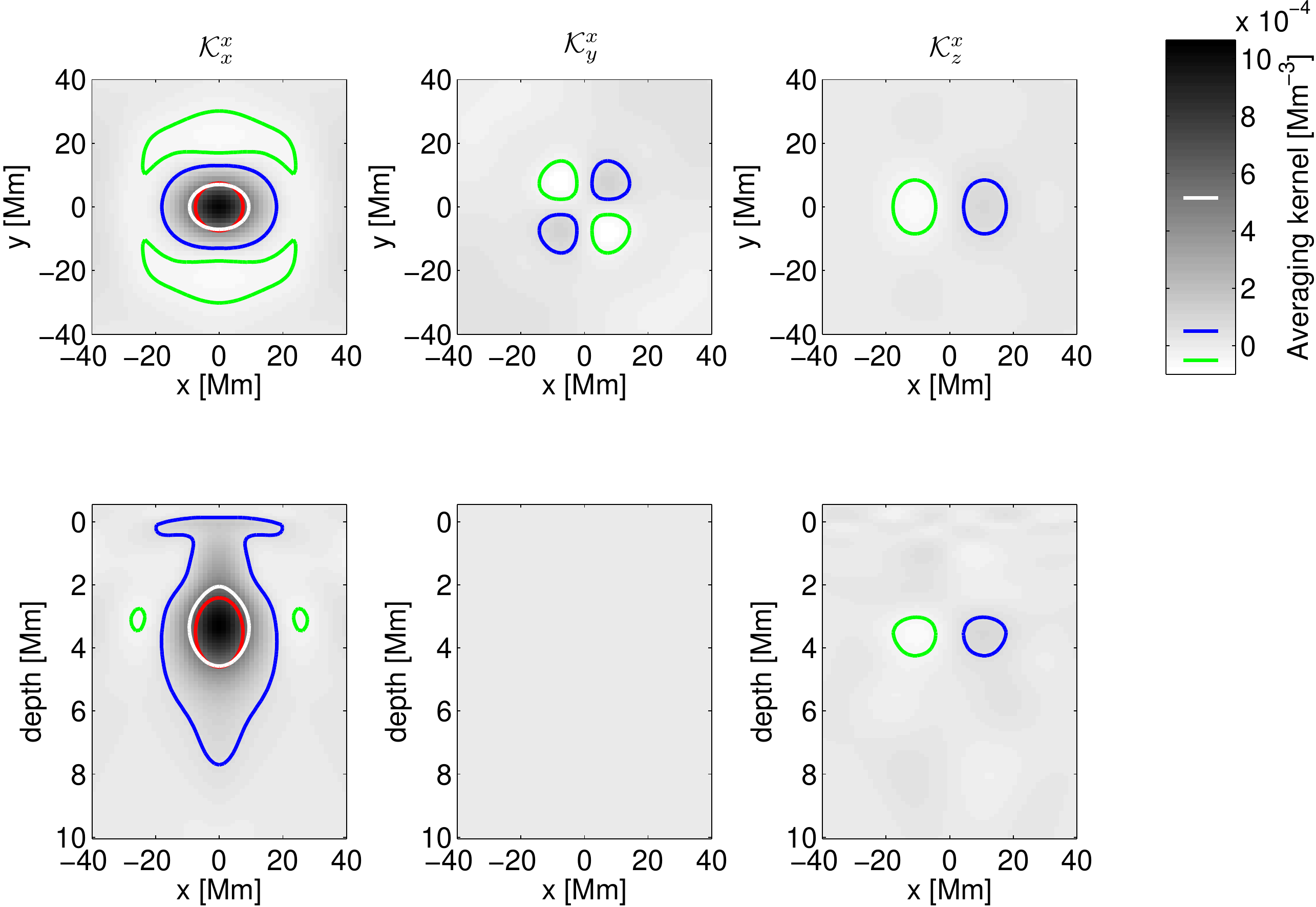}\\
\rule{0.8\textwidth}{1pt}\\
Improved inversion\\
\includegraphics[width=0.8\textwidth]{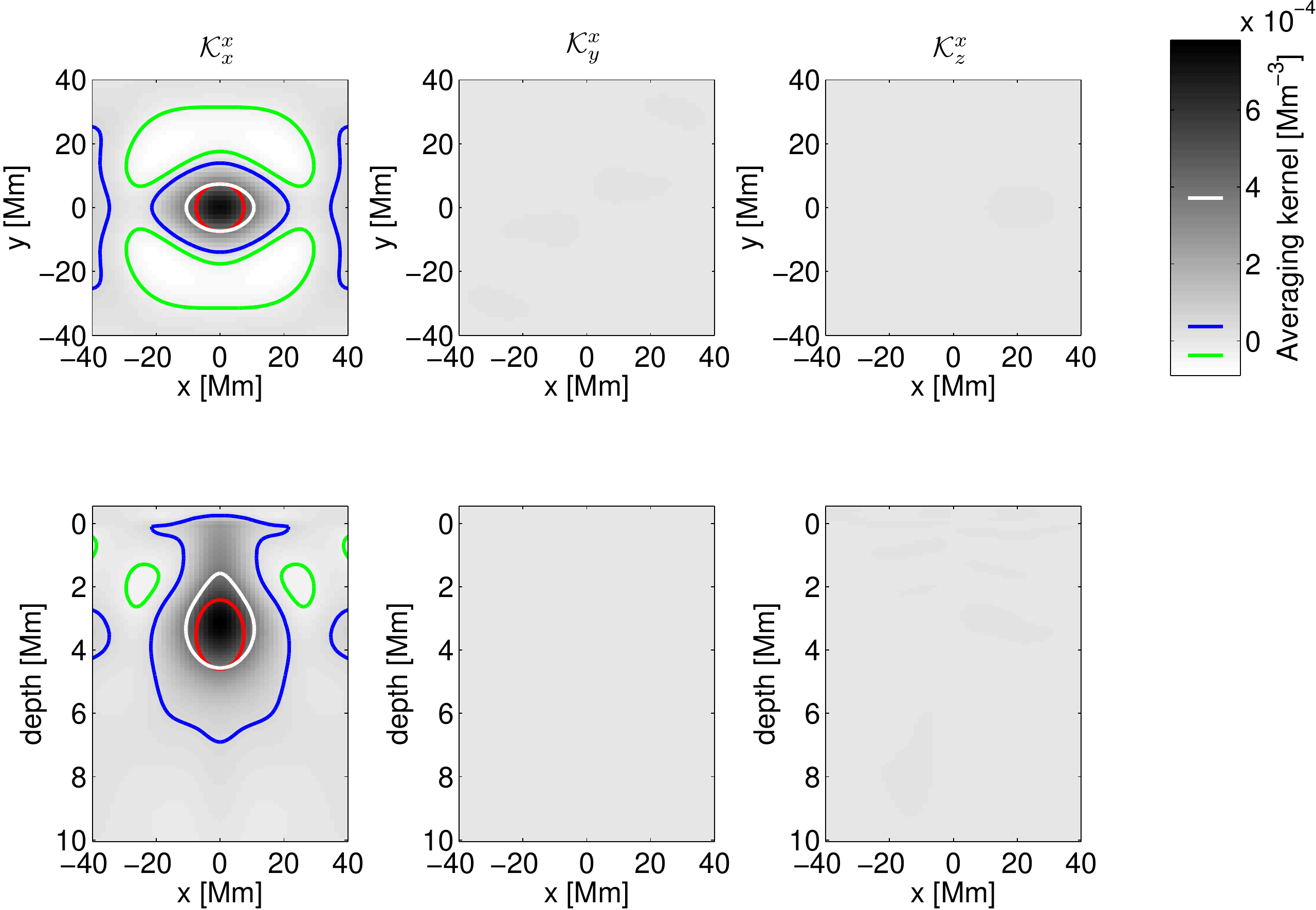}\\
\caption{All components of the averaging kernel for $v_x$ inversion at 3.5~Mm depth with a FWHM of $s_z=2.2$~Mm and $s_h=15$~Mm. Random error of the results is 20~\mps{} when assuming data averaged over 4~days. For details see  Fig.~\ref{fig:akerns-vx-1Mm-ES}.}
\label{fig:akerns-vx-3.5Mm-ES}
\end{figure}
}

\onlfig{22}{
\begin{figure}[!h]
\centering
Inversion with no cross-talk regularisation\\
\includegraphics[width=0.8\textwidth]{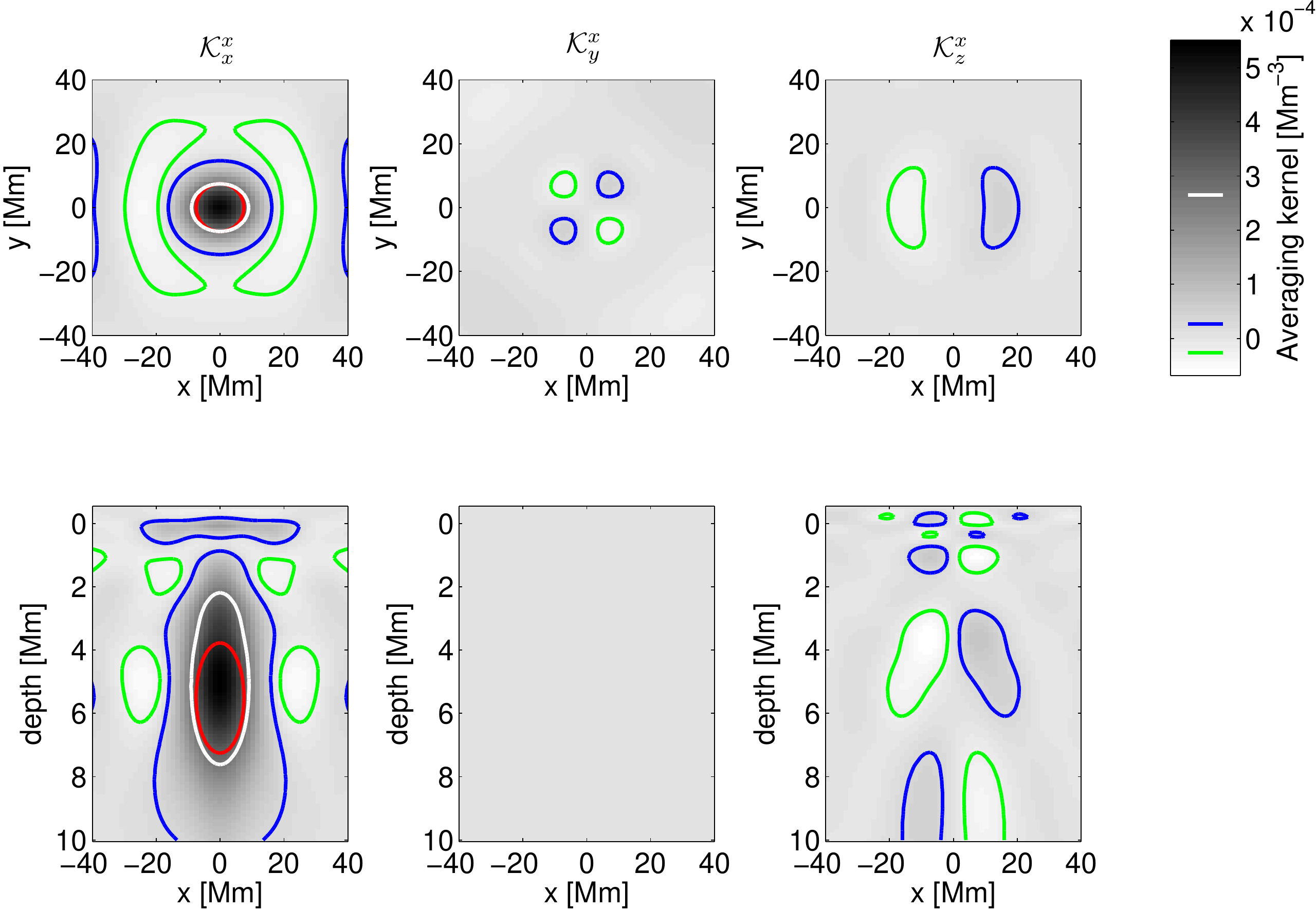}\\
\rule{0.8\textwidth}{1pt}\\
Improved inversion\\
\includegraphics[width=0.8\textwidth]{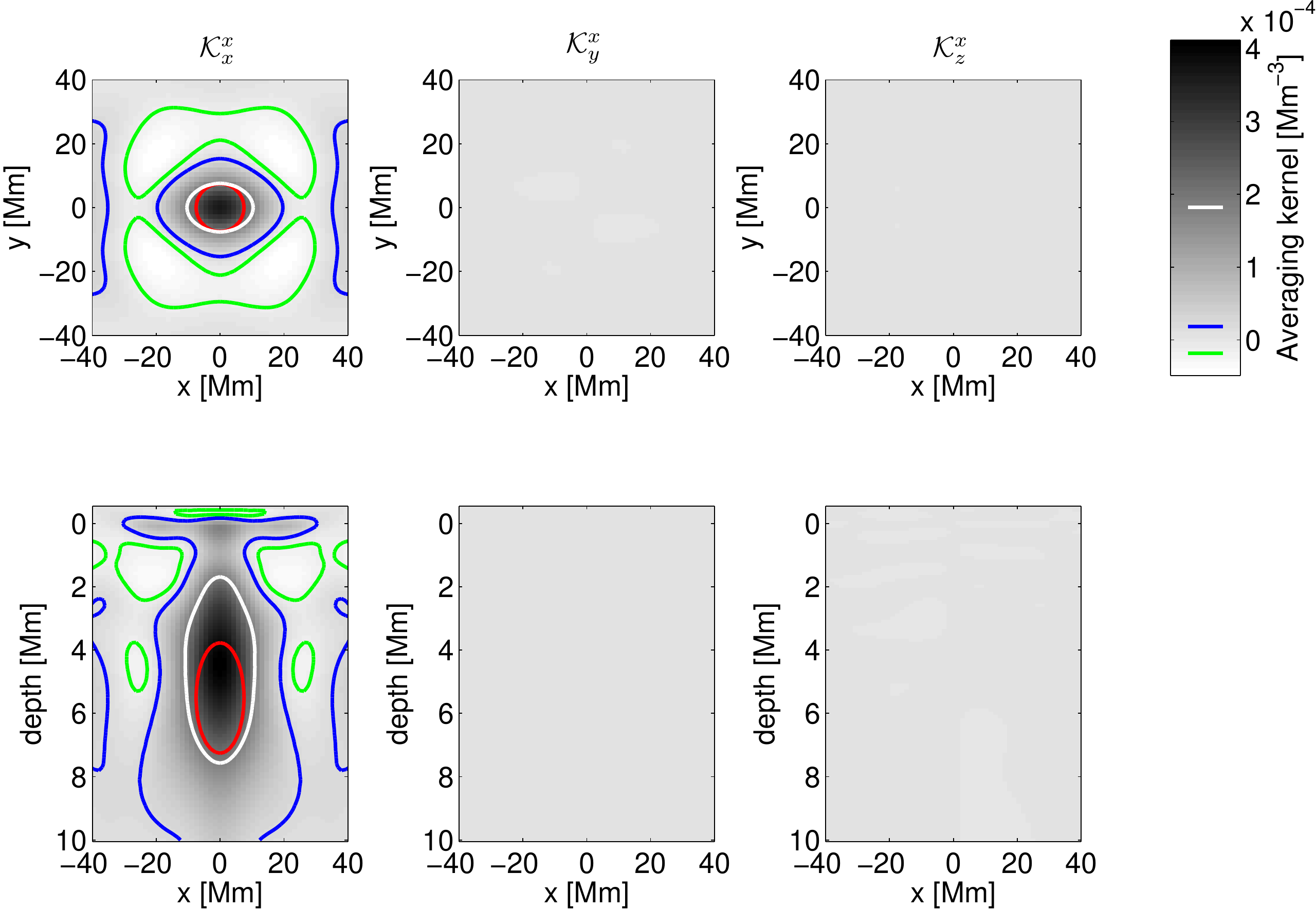}\\
\caption{All components of the averaging kernel for $v_x$ inversion at 5.5~Mm depth with a FWHM of $s_z=3.5$~Mm and $s_h=15$~Mm. Random error of the results is 28~\mps{} when assuming data averaged over 4~days. For details see  Fig.~\ref{fig:akerns-vx-1Mm-ES}.}
\label{fig:akerns-vx-5.5Mm-ES}
\end{figure}
}

\onlfig{23}{
\begin{figure*}[!h]
\sidecaption
\includegraphics[width=6cm]{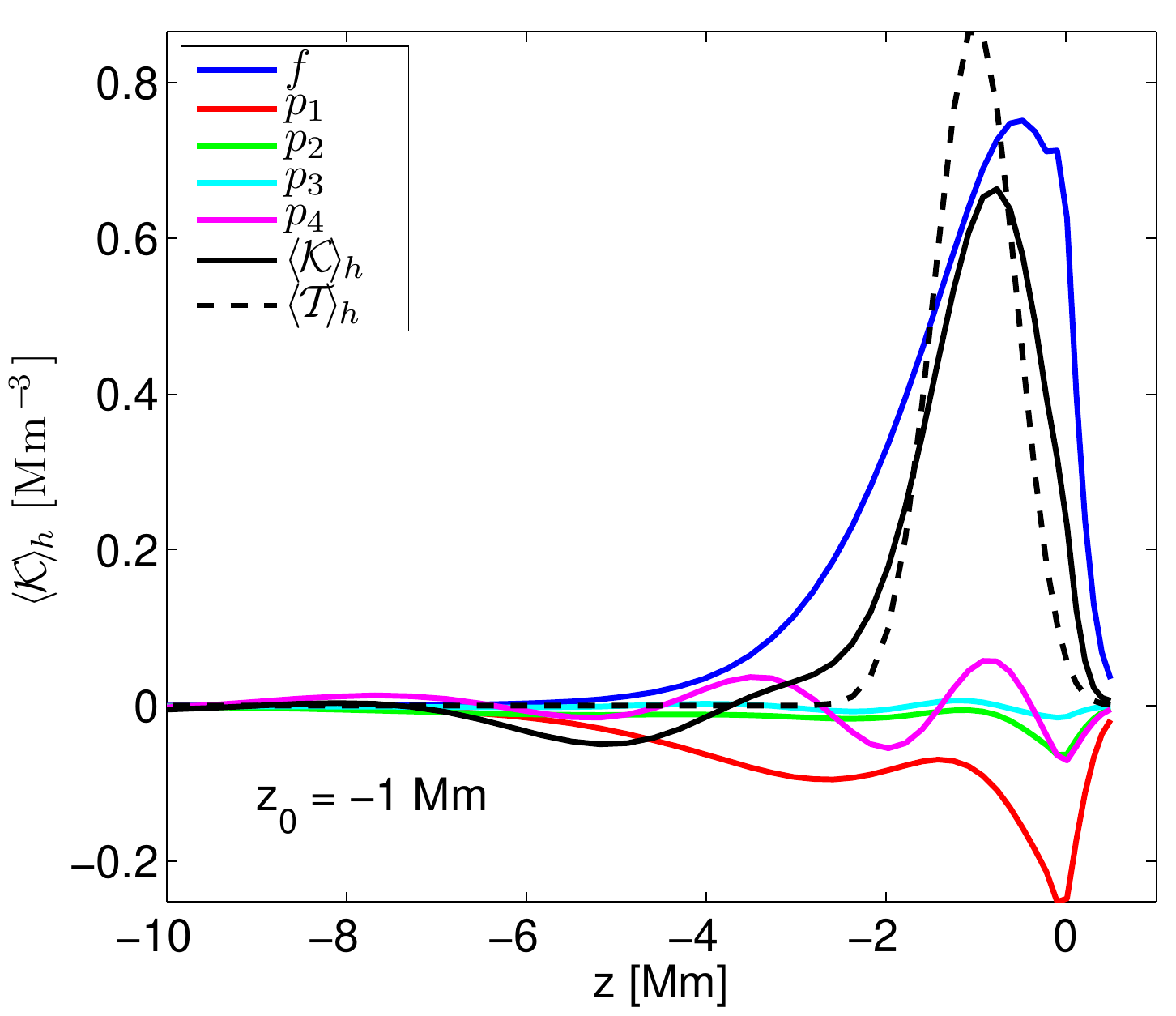}
\includegraphics[width=6cm]{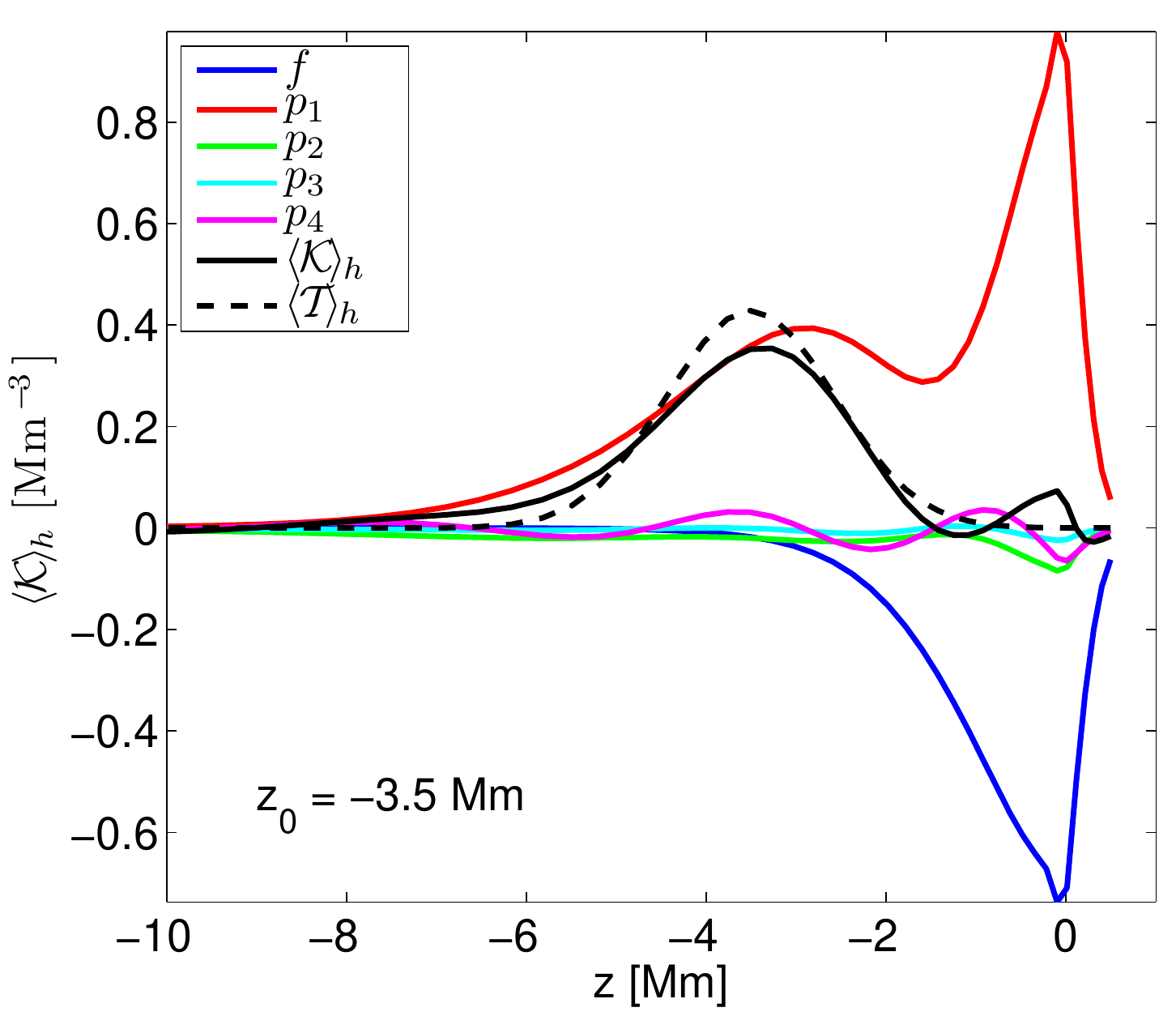}
\caption{The contributions of particular modes to the horizontally averaged averaging kernel for $v_x$ inversions using travel times averaged over 4~days for depths 1 and 3.5~Mm. We do not display the inversion for the depth of 5.5~Mm, because it is heavily dominated by noise.}
\label{fig:akern-modes-contributions}
\end{figure*}
}

\onlfig{24}{
\begin{figure}[!h]
\sidecaption
\includegraphics[width=6cm]{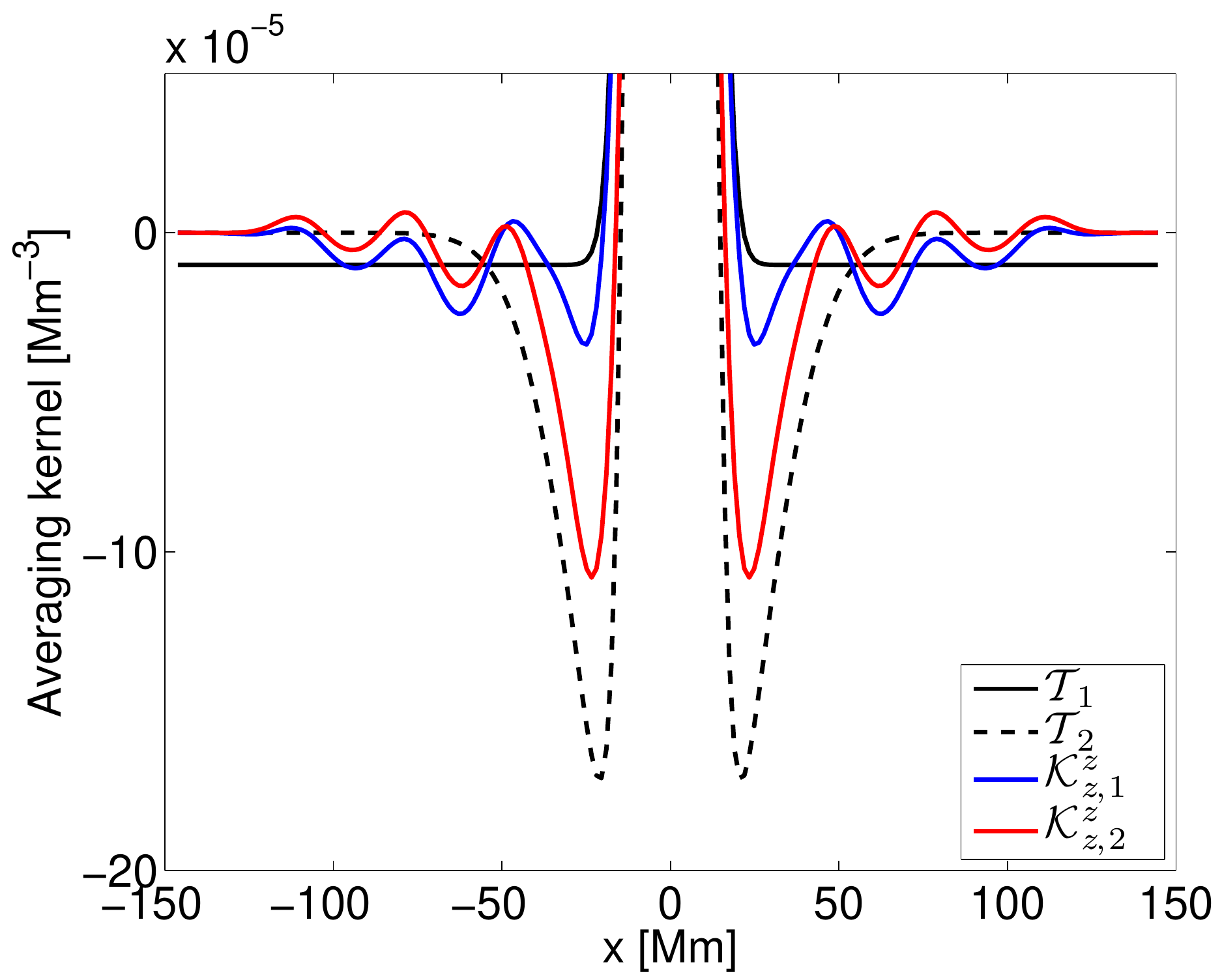}
\caption{To solve the peculiarity of the $v_z$ inversion, we introduced two formalisms in Section~\ref{sect:vz-issue}. Here we plot performance of those. In black, the magnified section a $y=0$ and $z=z_0$ of different target functions are displayed, one with removed mean (1) and one constructed with negative side-lobes (2). The resulting averaging kernels are also plotted. It is evident that the resulting averaging kernels are qualitatively very similar even when  different formalisms were used to compute them.  }
\label{fig:zeroT-ES}
\end{figure}
}  

\onlfig{25}{
\begin{figure}[!h]
\centering
Inversion with no cross-talk regularisation\\
\includegraphics[width=0.8\textwidth]{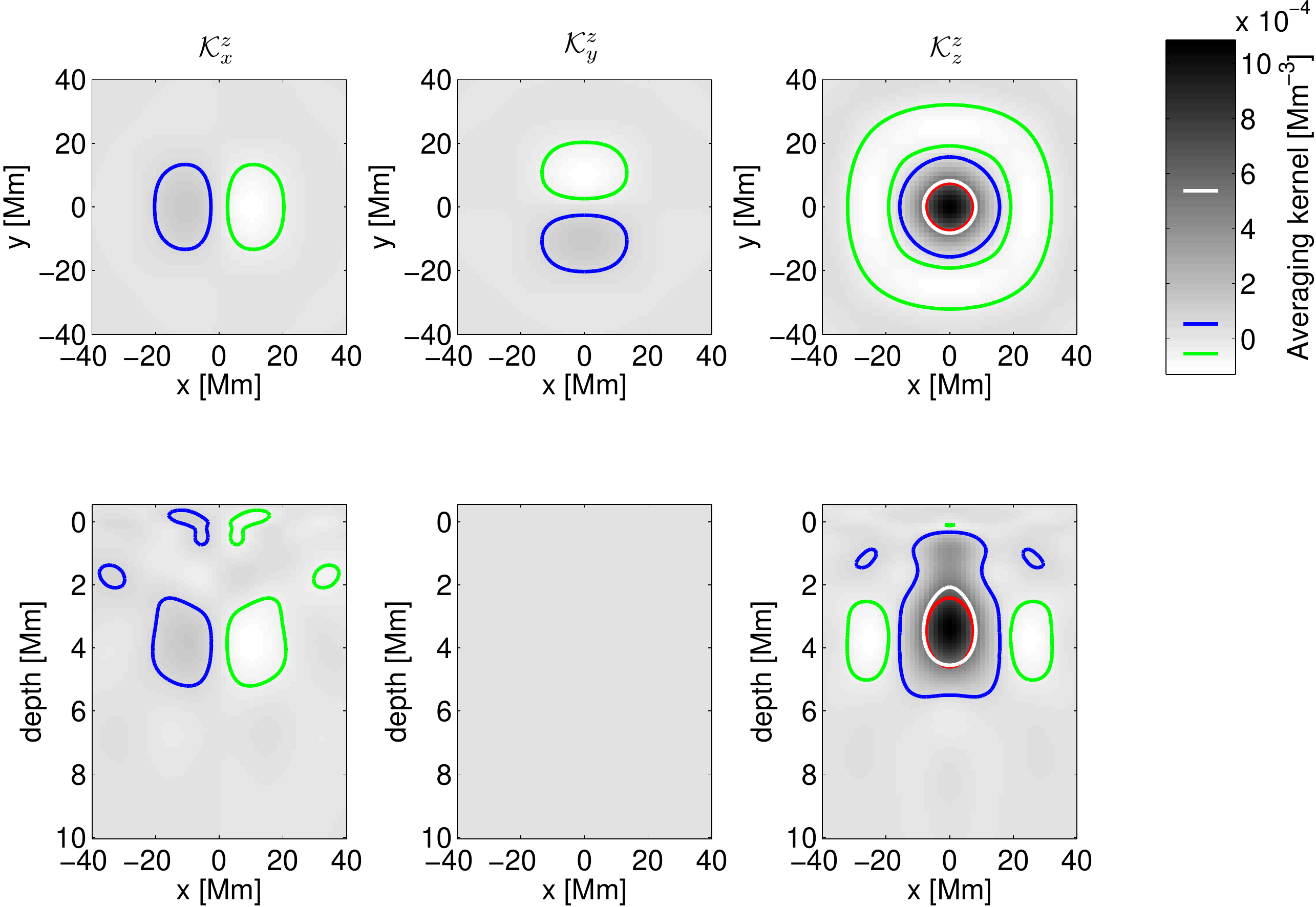}\\
\rule{0.8\textwidth}{1pt}\\
Improved inversion\\
\includegraphics[width=0.8\textwidth]{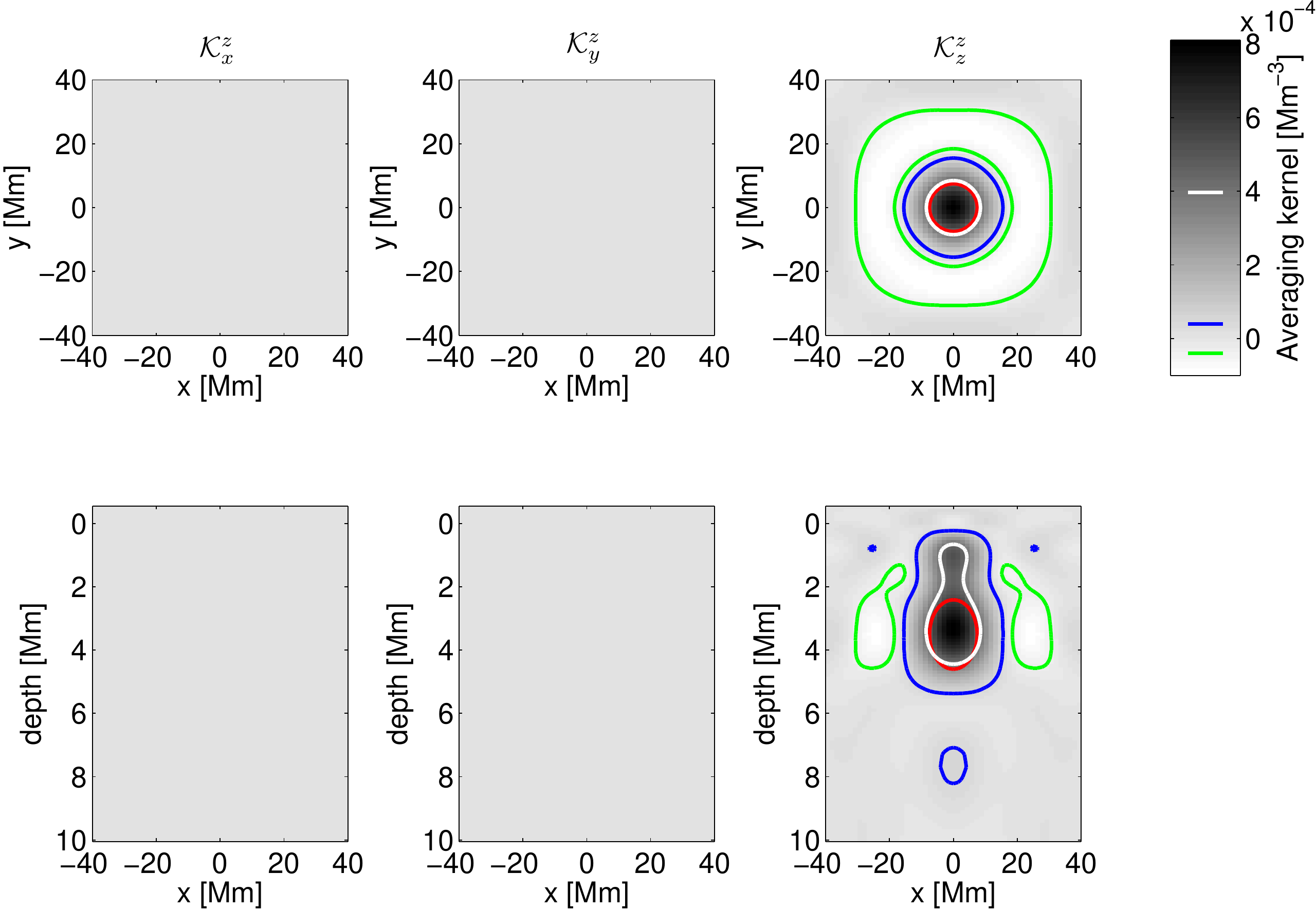}\\
\caption{All components of the averaging kernel for $v_z$ inversion at 3.5~Mm depth with a FWHM of $s_z=2.2$~Mm and $s_h=15$~Mm. Random error of the results is 13~\mps{} when assuming data averaged over 4~days. For details see  Fig.~\ref{fig:akerns-vx-1Mm-ES}.}
\label{fig:akerns-vz-3.5Mm-ES}
\end{figure}
}

\onlfig{26}{
\begin{figure}[!h]
\centering
Inversion with no cross-talk regularisation\\
\includegraphics[width=0.8\textwidth]{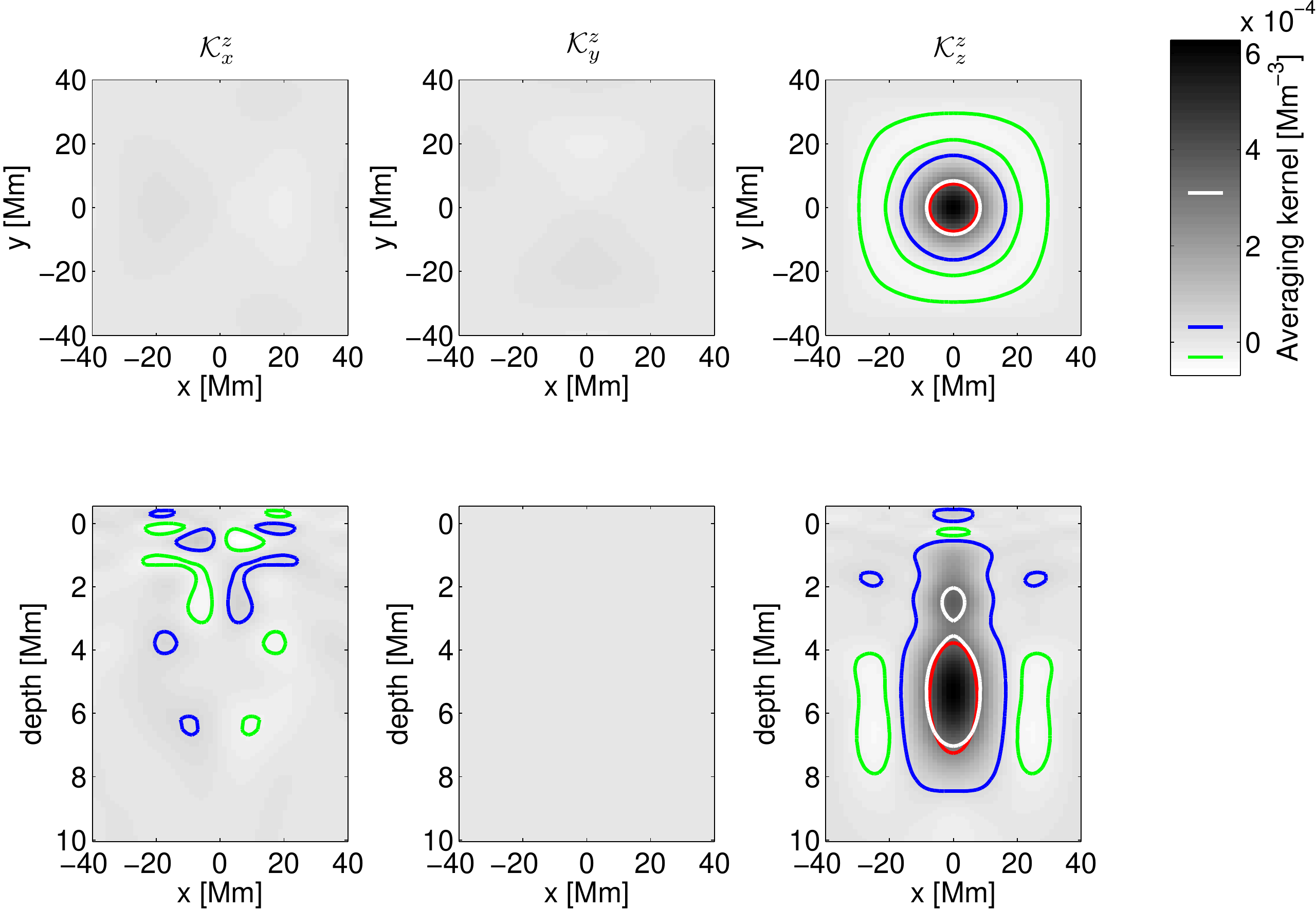}\\
\rule{0.8\textwidth}{1pt}\\
Improved inversion\\
\includegraphics[width=0.8\textwidth]{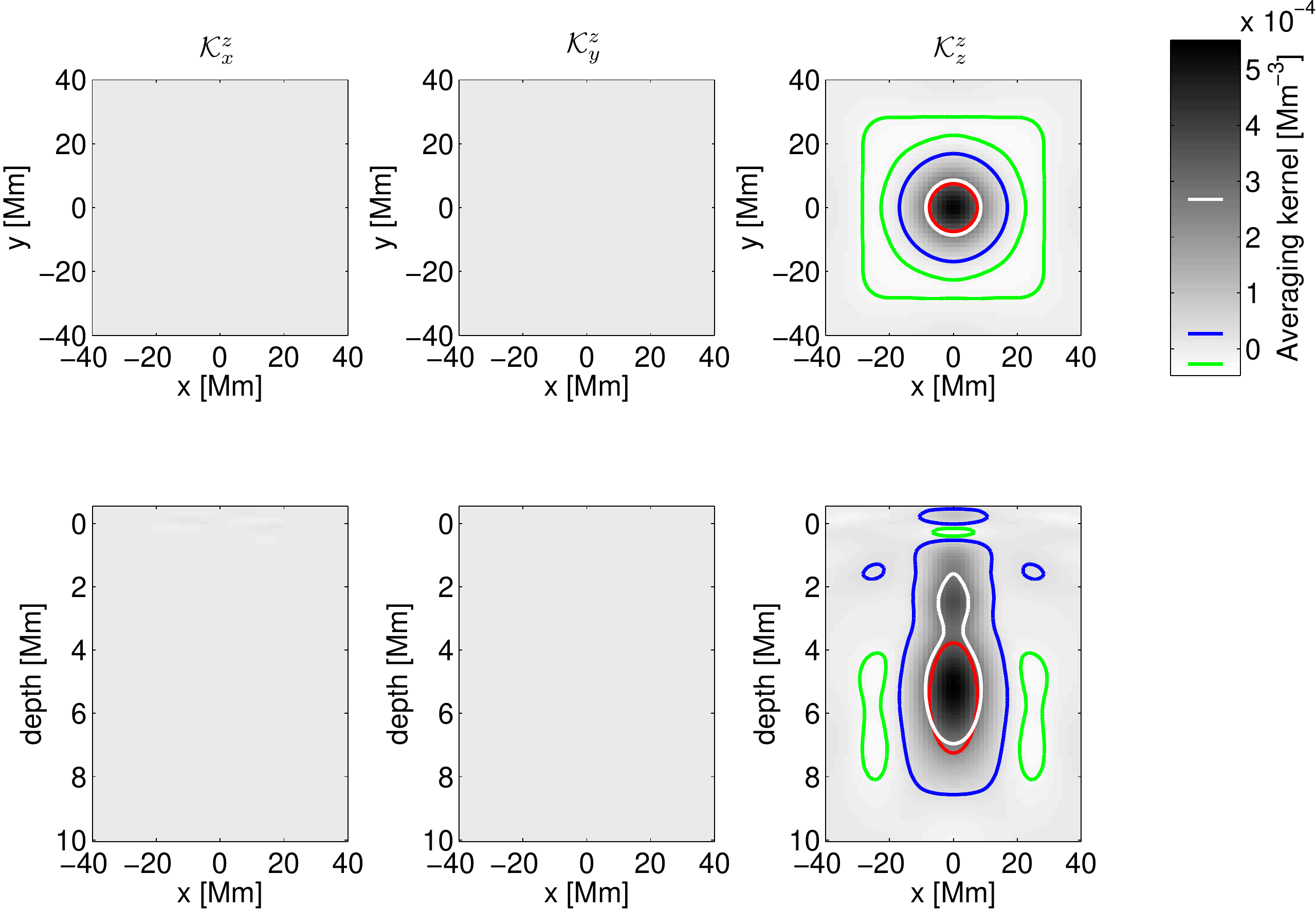}\\
\caption{All components of the averaging kernel for $v_z$ inversion at 5.5~Mm depth with a FWHM of $s_z=3.5$~Mm and $s_h=15$~Mm. Random error of the results is 133~\mps{} when assuming data averaged over 4~days. For details see  Fig.~\ref{fig:akerns-vx-1Mm-ES}.}
\label{fig:akerns-vz-5.5Mm-ES}
\end{figure}
}

\onlfig{27}{
\begin{figure*}[!h]
\sidecaption
\includegraphics[width=6cm]{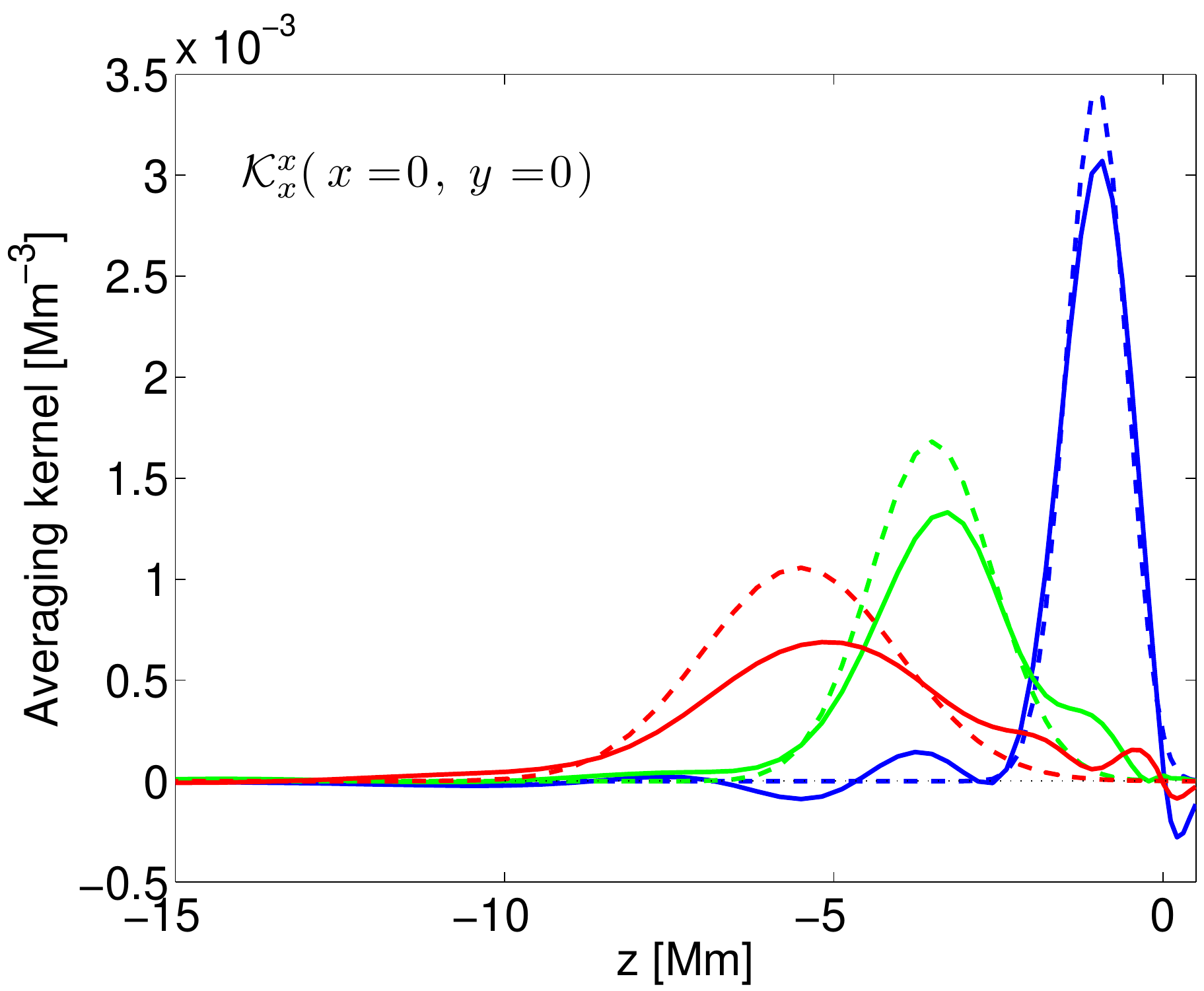}
\includegraphics[width=6cm]{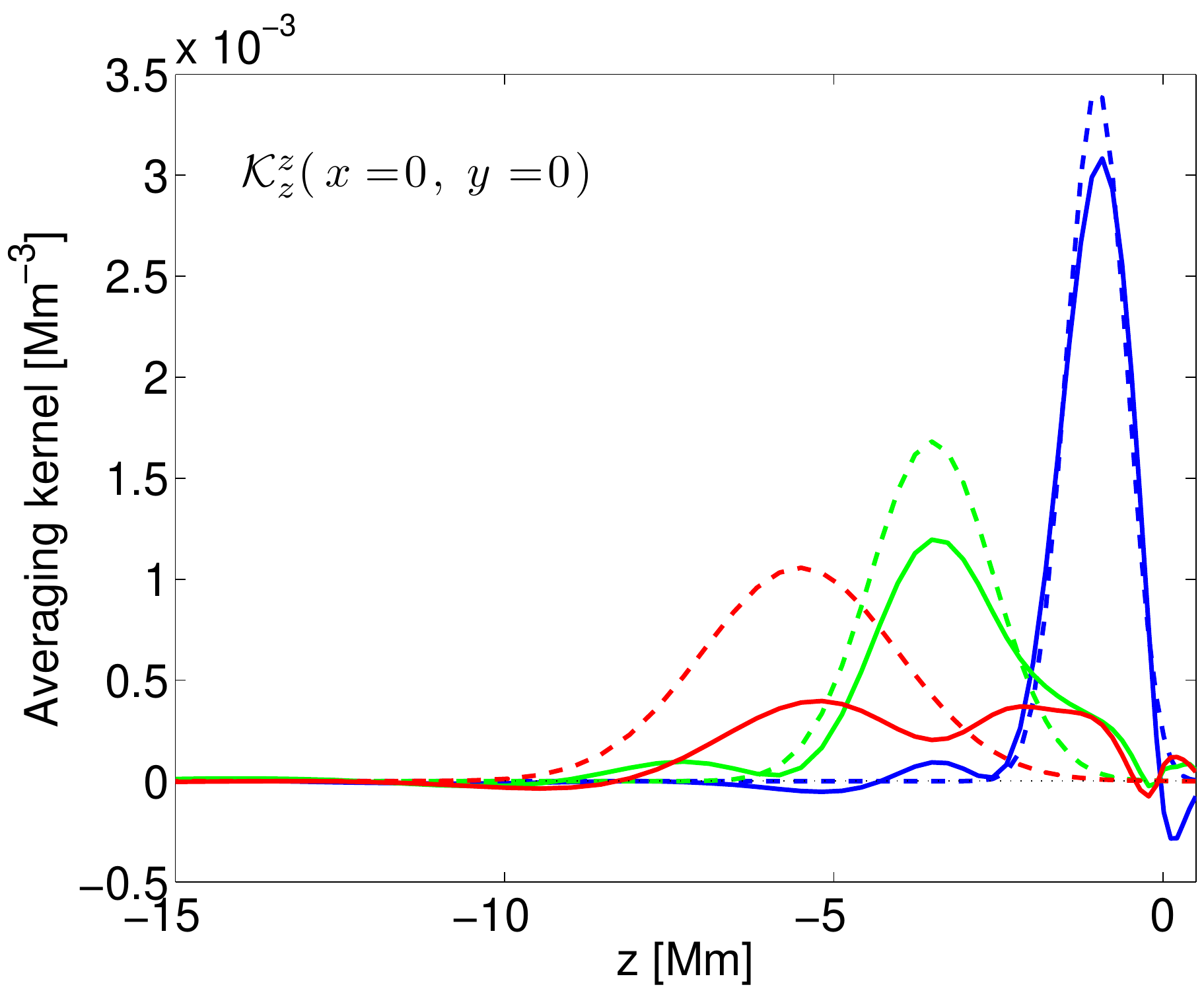}
\caption{The cut through the $x=y=0$ point of the averaging kernel (solid) and the respective target function (dashed) for the $v_x$ (left) and $v_z$ (right) inversions using averaging over many flow realisations plotted along with the corresponding target functions at three discussed depths (1~Mm in blue, 3.5~Mm in green, and 5.5~Mm in red). Compare to Figs.~\ref{fig:akerns-vx-1D} and \ref{fig:akerns-vz-1D} where the resemblance of the target functions is worse. The random error of the results is given in Table~\ref{tab:errors}.}
\label{fig:akerns-1D-largeaveraging}
\end{figure*}
}

\onlfig{28}{ 
\begin{figure}[!h]
\includegraphics[width=6cm]{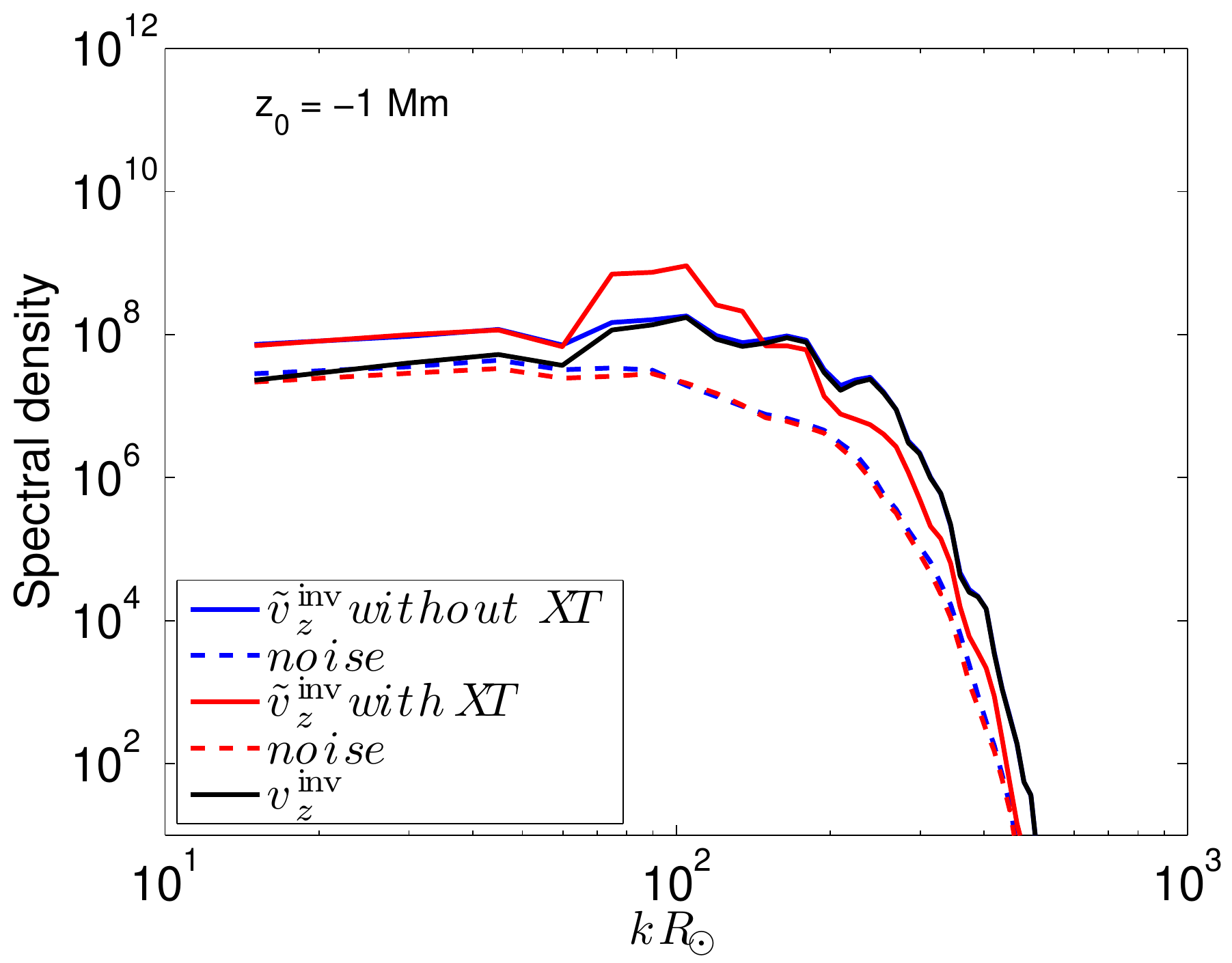}
\includegraphics[width=6cm]{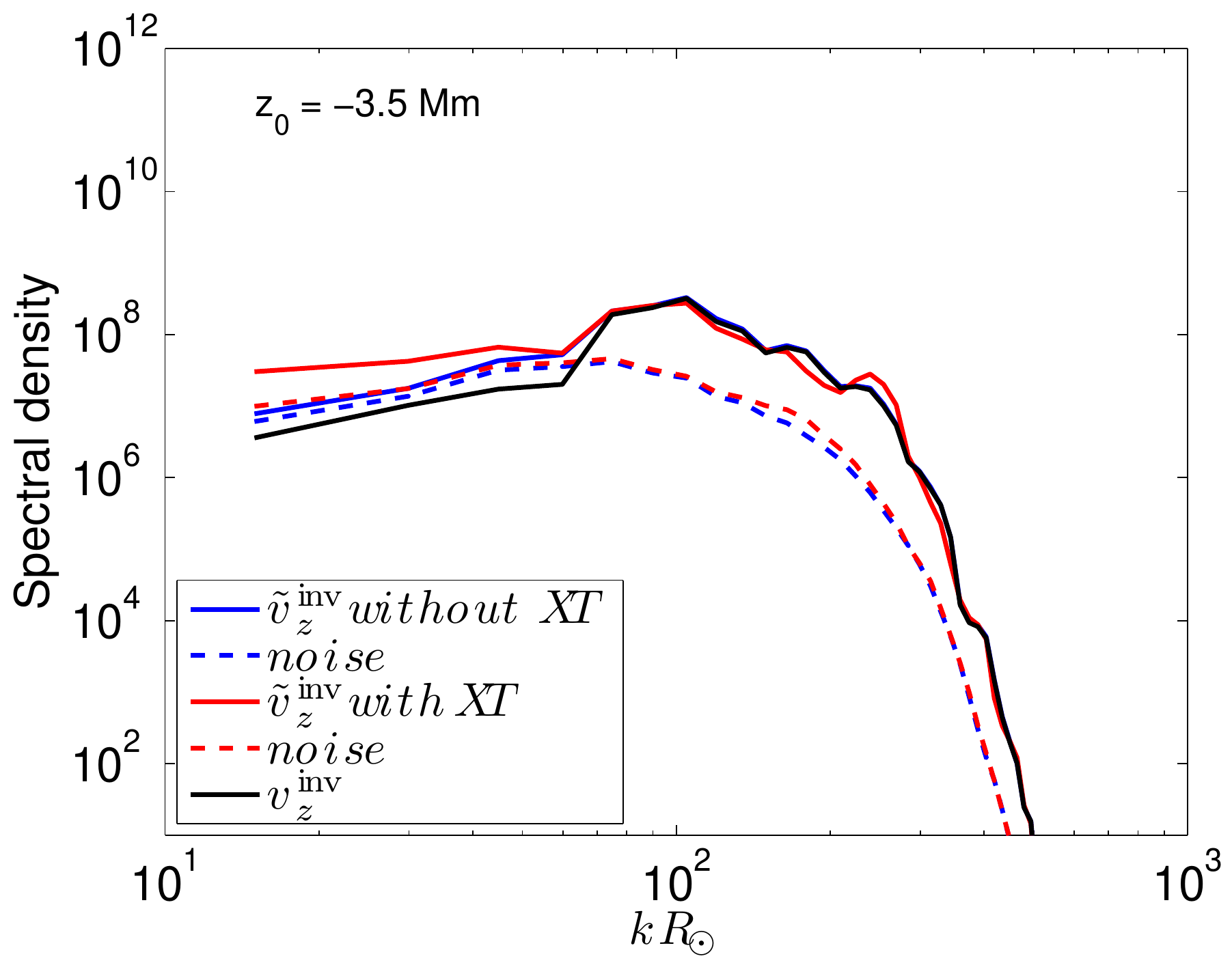}
\includegraphics[width=6cm]{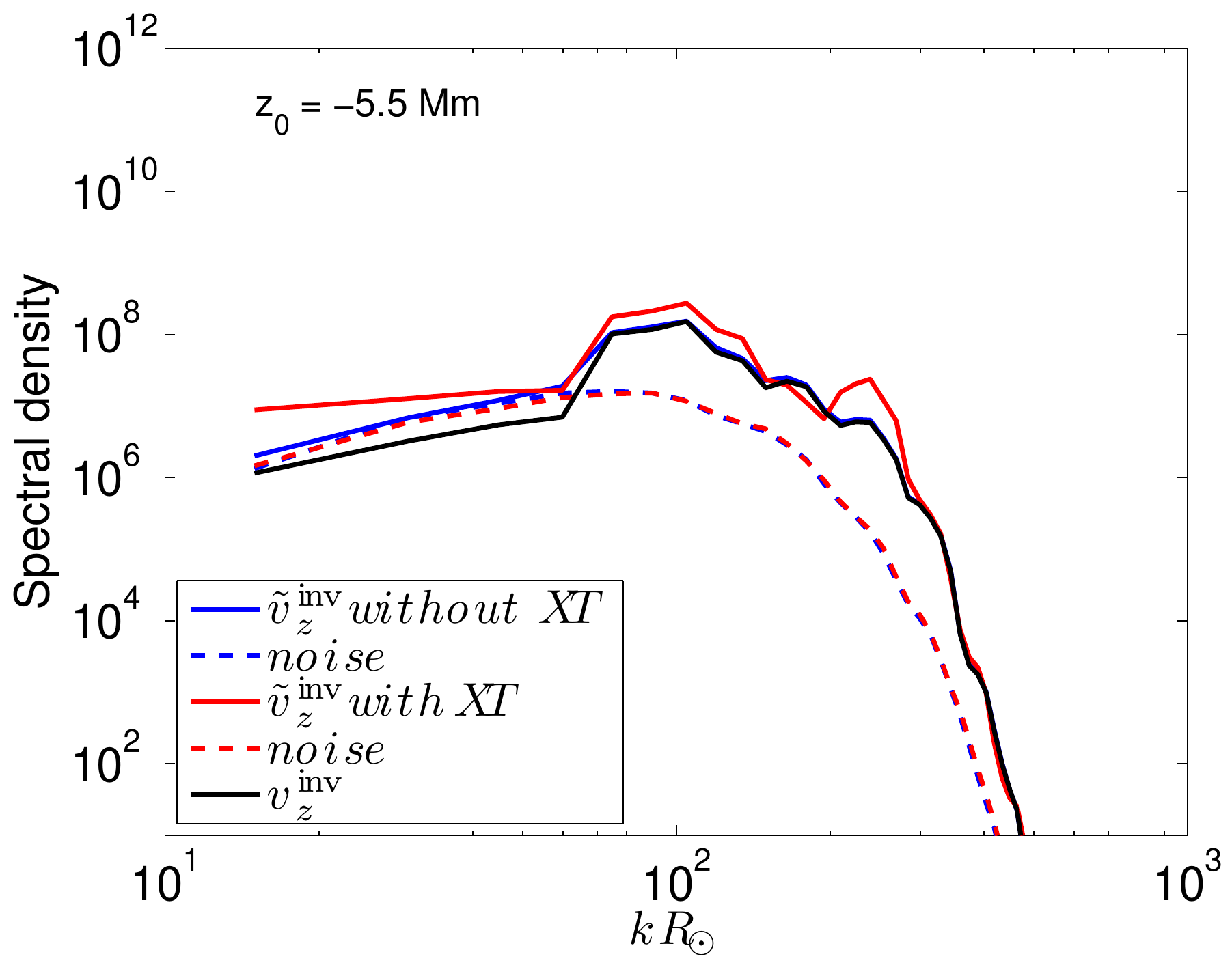}
\caption{The azimuthally-averaged power spectra of the $v_z$ inversion components at depths of 1, 3.5 and 5.5~Mm for averaging over many flow representations. For reference, we plot the power spectrum of $\vakern_z$ using the black solid line. Then we plot the power spectrum of $\vinv_z$ (solid line) and power spectrum of the noise (i.e., the power spectrum of $\vinv_z-\vakern_z$; dashed line) for the inversion where the cross-talk is minimised (blue) and ignored (red). Compare to Fig.~\ref{fig:snr-vz-k}.}
\label{fig:snr-vz-k-10000sgs-ES}
\end{figure}
}
 
\end{document}